\documentstyle[12pt,twoside,epsf]{article}
\jot\medskipamount
\begin{document}
%\documentstyle[12pt]{article}          %%% alternative, also 11pt. 

% for proof-reading
%\renewcommand{\baselinestretch}{1.2}
%\parskip 2ex
%\addtolength{\textheight}{0cm}
%\addtolength{\evensidemargin}{-4cm}
%\addtolength{\oddsidemargin}{-2.5cm}

% This is a nice title page including abstract ....
%

\thispagestyle{empty}
\noindent\hspace*{\fill}  FAU-TP3-97/1 \\
\noindent\hspace*{\fill}  NIKHEF-97/013 \\
%\noindent\hspace*{\fill}  nucl-th/9703019\\
\noindent\hspace*{\fill}  Feb. 27, 1997 \\

\begin{center}\begin{Large}\begin{bf}
Stochastic Methods\\
for\\
Zero Energy Quantum Scattering\\
\end{bf}\end{Large}\vspace{.75cm}
 \vspace{0.5cm}
Justus H. Koch$^{\,a,b}$, Hubertus R. Mall$\,^a$, Stefan Lenz$\,^a$\\
 \vspace{0.5cm}
$^a$Institut f\"ur Theoretische Physik III,
Universit\"at Erlangen - N\"urnberg \\
Staudtstra{\ss}e 7, D-91058 Erlangen, Germany\\

 \vspace{0.5cm}
$^b$National Institute for Nuclear Physics and High Energy Physics (NIKHEF)\\
Postbus 41882, 1009 DB Amsterdam, The Netherlands\\
and Institute for Theoretical Physics, University of Amsterdam\\
email: justus@nikhef.nl
\vspace{1.0cm}
 \vspace{1.0cm}
\end{center}
\vspace{1cm}\baselineskip=25pt
\begin{abstract} \noindent
We investigate the use of stochastic methods for zero energy quantum scattering
based on a path integral approach.
With the application to the scattering of a projectile from a nuclear many body 
target in mind, we use the potential scattering of a 
particle as a test for the accuracy and efficiency of several methods. To 
be able to deal with 
complex potentials, we introduce a path sampling action and a modified scattering
observable. The approaches considered are the random walk, where the points of a
path are sequentially generated, and the Langevin algorithm, which updates an
entire path. Several improvements are investigated.
A cluster algorithm for dealing with scattering problems is finally proposed, which shows the best accuracy and stability.
\end{abstract}
\newpage\baselineskip=18pt

\section{Introduction}

The most frequently used method to study small scale composite systems is
quantum scattering. While the interaction of the projectile with an isolated 
constituent of the target may be well known and can be described accurately, 
the scattering from the bound system can in general only be solved with 
several simplifying assumptions. These approximations concern both the 
interaction of the projectile with the target as well as the internal target 
dynamics during the scattering process.

While a better understanding of the scattering from composite systems is of 
interest for several fields of physics, we focus here specifically on the 
scattering of a strongly interacting particle from nuclei at zero energy. There 
exist several often used recipes for the description of low enegy nuclear 
scattering, mainly in the form of effective single particle potentials for the 
projectile - nucleus interaction. However, the inherent approximations have not 
been tested in a conclusive way. That there are surprises if one goes beyond the
standard approximations has been shown recently in the framework of the frozen
nucleus picture, where all nucleons are held fixed during the scattering
process. By evaluating the scattering amplitude with Monte Carlo methods for a
target with a large number of scattering centers, it was found that the 
projectile gets 'trapped' by clusters of target constituents \cite{StoLe}, 
\cite{StoLe2}. This is an effect 
which is absent in any of the approximate descriptions. Clearly it is important
to investigate methods how to solve in practice the scattering from a many-body
target exactly, such that also the target dynamics can be taken into
account. 
  
A formalism that has hardly ever been applied to scattering is the path
integral description. In a recent study \cite{SLen95b} this approach was
applied to the potential scattering of a particle and stochastic methods were
used to evaluate the multi-dimensional integrals one encounters. 
It was then shown that these methods can in principle be extended in a 
straightforward fashion to calculate the exact scattering amplitude for
scattering from nuclei with the full inclusion of the target dynamics. 
Before this method can be applied to a detailed study of nuclear scattering, 
which requires a considerable computational effort, it is necessary to find
more efficient and numerically stable stochastic methods to obtain the 
scattering amplitude. This is the purpose of this paper where we examine
different stochastic algorithms that can be applied to the scattering 
problem. To be able to assess the accuracy of these methods we apply 
them to the potential scattering of a low energy particle, where the exact 
solution can be obtained by standard numerical methods from the 
single particle Schr\"odinger equation.
Keeping in mind the application to problems such as the interactions 
of low energy anti-protons with nuclei, we include complex potentials
in our investigation of stochastic methods for scattering, which has not been 
considered before. 

In several fields of physics path integrals are a well established 
approach to obtain the ground state properties of a system and there
exist stochastic methods to evaluate the resulting high-dimensional integrals. 
Examples are spin systems \cite{Swe87}, \cite{Wol89} and quantum 
fields on the lattice \cite{Cre92}. However, the subject of these 
studies are localized, bound systems. The application of these methods to 
scattering, where the states  are not normalizable and the spectrum is 
continous, is therefore not straightforward. Another complication is due to the
use of complex potentials, which have been already used for 
bound states {\it e.g.}
in quantum chemistry \cite{Dol88, Mak88} or in the calculation of correlation 
functions of nuclear ground states \cite{Car90}. A complex action makes 
it impossible to interpret the distribution of paths as a 
probability distribution and we examine how to deal with this problem
in scattering. For completeness we mention here also the possibility of 
obtaining the scattering length of a system from the 
volume dependence of its energy spectrum \cite{Lu86}. The feasibility of this
approach has been shown for some special cases of quantum field theory on 
the lattice: the non-linear $\sigma$ 
model in two dimensions \cite{Lu90} and the Ising model in four dimensions
\cite{Mo87}. The generalization of this method to nuclear scattering 
is certainly not straightforward and we therefore focus below on more direct 
path integral methods.

In Section 2, we briefly outline the path integral approach to 
non-relativistic quantum scattering and define a 'modified observable' 
appropriate for scattering calculations with a complex action. Different 
stochastic methods to evaluate the resulting multi-dimensional integrals
are discussed in the subsequent chapters.
We examine their efficiency and accuracy in order to identify the method 
that is most suitable for the application to many-body targets. 
The random walk algorithm, the most straightforward implementation 
of the path integral concept, is discussed in Section 3.
Langevin and hybrid methods, which update entire paths, are 
studied in Section 4. An efficient new approach, a cluster algorithm for 
quantum mechanical scattering, is developed in Section 5. 
A summary and our conclusions are presented in Section 6.

\section{Potential Scattering}
\subsection{Scattering formalism}
\label{POTSCAT}

In this chapter we briefly describe the theoretical framework for solving
a scattering problem by using the path integral approach. We discuss only
the potential scattering for a single particle. It is used in this paper as 
a test case for finding a numerical method that is suitable for applications
to scattering from a composite many-particle target. Some modifications of 
standard methods in the literature are necessary since we allow for complex
potentials. 

The Hamiltonian, which will be used throughout this
paper, describes the nonrelativistic scattering of a particle from an complex,
local potential

\begin{eqnarray}
  \hat{H}  &=& \frac{\hat{p}^2}{2\mu}+ \hat{V} =  \hat{K} + \hat{V} 
\mbox{,}\label{hamil1}\\
  \hat{V}  &=& \hat{U} + i\;\hat{W}
\mbox{,}\label{hamil2}
\end{eqnarray}
where $\hat{p}$ is the momentum operator and $\mu$ the reduced mass of
projectile - target system. We are interested in stationary scattering states, 
corresponding to a real energy $E(k) = \frac{\vec{k}^2}{2\mu}$. Due to the 
complex potential the Hamiltonian $\hat{H}$ is not hermitian. Therefore one has
two solutions to the Lippmann-Schwinger equations, which belong to
$\hat{H}$ and $\hat{H}^{\dagger}$, respectively \cite{Fes85}:
\begin{eqnarray}
  \mid\psi^{+}_{\vec{k}}\rangle =
  \mid\chi_{\vec{k}}\rangle + \frac{1}{E^{+}-\hat{H}} \; \hat{V}
  \mid\chi_{\vec{k}}\rangle\;,
 &&
  \mid\psi^{-}_{\vec{k}}\rangle =
  \mid\chi_{\vec{k}}\rangle +\frac{1}{E^{-}-\hat{H}^{\dagger}}\;
  \hat{V}^{\dagger} 
  \mid\chi_{\vec{k}}\rangle\;,
  \label{biorth1}\nonumber\\
  \mid\tilde{\psi}^{-}_{\vec{k}}\rangle = 
  \mid\chi_{\vec{k}}\rangle +\frac{1}{E^{-}-\hat{H}}\;\hat{V}
  \mid\chi_{\vec{k}}\rangle\;, 
&&
  \mid\tilde{\psi}^{+}_{\vec{k}}\rangle =
  \mid\chi_{\vec{k}}\rangle +\frac{1}{E^{+}-\hat{H}^{\dagger}}\;
  \hat{V}^{\dagger} \mid\chi_{\vec{k}}\rangle
\quad\mbox{.} 
\label{biorth2}
\end{eqnarray}
\noindent
We assume the existence of a complete biorthogonal basis for the
Hamiltonian, $\hat H$  \cite{Nai68}, \cite{Ste72}.

The scattering wave functions $\mid\hat\psi_0^{\pm}\rangle$ for zero 
projectile energy can be obtained from the propagator $\hat{\rho}(\beta)$, 
\begin{eqnarray}
  \hat{\rho}(\beta) &=& \frac{1}{{\cal{N}}_{\beta}}
  \exp{\left[-\beta\hat{H}\right]}
\mbox{,}\label{project}\\
   {\cal{N}}_{\beta} &=& \left(\frac{2\pi\mu}{\beta}\right)^{\frac{3}{2}}
\label{projectnorm}
\mbox{,}
\end{eqnarray}
in the limit of large Euclidean times $\beta$. This operator contains the 
complete physical information about the evolution of the system. Its spectral
representation in the biorthogonal basis given by Eq. (\ref{biorth2}) is
\begin{eqnarray}
  \hat{\rho}( \beta )
 &=&
   \frac{1}{{\cal{N}}_{\beta}}
   \int{\,d^3k\; \exp{\left[-\beta E(k) \right]} \;\;
   \mid\psi_{\vec{k}}^{\pm}\rangle\langle\tilde\psi_{\vec{k}}^{\pm}\mid}
\;\mbox{.}\label{projectspec}
\end{eqnarray}
\noindent
In the limit $\beta\rightarrow\infty$ one is left with
\begin{eqnarray} 
  \lim_{\beta\rightarrow\infty} \hat{\rho}( \beta ) &=&
  \mid\psi_{0}^{\pm}\rangle\langle\tilde\psi_{0}^{\pm}\mid
\label{grundwell}\mbox{,}
\end{eqnarray}
which projects onto the ground state of the system. This projector can be 
used to calculate the {\it square} of the scattering amplitude,
\begin{eqnarray}
   f^{2}_{0}   
  &=&
   \left[ (2\pi)^2\mu\right]^2  \;
   \langle\chi_{0}\mid \hat{V} \mid\psi_{0}^{+}\rangle
   \langle\psi_{0}^{-}\mid \hat{V} \mid\chi_{0}\rangle
\label{phyquant}\mbox{.}
\end{eqnarray}
\noindent
The wavefunctions $\mid\chi_0\rangle$ are plane wave solutions, which in the
zero energy limit are constant. The wave functions 
$\mid\tilde{\psi}_{\vec{k}}^{+}\rangle$ and $\mid\psi_{\vec{k}}^{-}\rangle$
differ with respect to their boundary conditions. 
However, since no spatial direction is distinguished in the limit 
$E(k)\rightarrow0$, Eq. (\ref{phyquant}) therefore can be written as
\begin{eqnarray}
    f_{0}^{2} 
  &=&
   \left[ (2\pi)^2\mu\right]^2  \;
   \langle\chi_{0}\mid \hat{V} \mid\psi_{0}^{+}\rangle
   \langle\tilde\psi_{0}^{+}\mid \hat{V} \mid\chi_{0}\rangle
    \label{f2}\nonumber\\
  &=&
    \left[(2\pi)^2\mu\right]^2 \;
    \lim_{\beta\rightarrow\infty} 
    \langle\chi_{0}\mid \hat{V} \;\hat\rho (\beta)\;
    \hat{V} \mid\chi_{0}\rangle 
    \label{f3}
    \mbox{,}
\end{eqnarray}
or in $r$-space representation
\begin{equation}
    f_0^2=
    \left[(2\pi)^2\mu\right]^2 \;
    \lim_{\beta\rightarrow\infty} \int\,d\vec{r}_1\,d\vec{r}_{N+1}\;
    \chi_0(\vec{r}_{N+1}) \, V(\vec{r}_{N+1}) \,
    \rho(\vec{r}_{N+1},\,\vec{r}_1; \beta) \; V(\vec{r}_1) \,\chi_0(\vec{r}_1)
    \;
    \mbox{.}\label{f3a}
\end{equation}
\noindent
Note that, in contrast to other discussions of scattering in the literature,
the quantity $f_0^2$ is calculated instead of $\mid f_0 \mid^2$.
It contains information not only about the absolute value, but also the
phase of the zero energy scattering amplitude. 

In the following chapters, we will investigate different stochastic methods to 
solve Eq. (\ref{f3a}). In all subsequent applications, the shape of $\hat{V}$ 
will be assumed to be Gaussian,
\begin{eqnarray}
  V(\vec{r}) &=& (U_0\,+iW_0\,)\;v(\vec{r}) \,\equiv \, V_c\;v(\vec{r})\; 
\mbox{,}\label{potkonf}\\
  v(\vec{r})  &=& \exp{\left( -\frac{1}{2b^2}\vec{r}^{\,2} \right)}
\label{gausspot}\mbox{,}
\end{eqnarray}
\noindent
but this assumption is not crucial for the methods discussed here. 
Keeping in mind applications to the scattering length of antiprotons 
\cite{Koh86}, \cite{Bat89}, we take the real part of the potential
$U_0$ in the range of $-0.5\,\mbox{fm}^{-1}$ to $2.5\,\mbox{fm}^{-1}$ and 
$W_0$ from 0 to $-\,7.0\,\mbox{fm}^{-1}$. 
The width parameter will always be taken to be $b\,=\,0.5\,\mbox{fm}$ and
the reduced mass $\mu\,=\,2.5\,\mbox{fm}^{-1}$.

\subsection{Path integral approach and modified observable}

For the actual evaluation of Eq. (\ref{f3}) we use the Feynman path integral
approach. For this purpose the projector, Eq. (\ref{project}), is rewritten as
\cite{FeyH}
\begin{eqnarray}
   \rho_{\cal{A}} (\vec{r}_{N+1},\vec{r}_1;\beta) &\approx&
   \frac{1}{ {\cal{N}}_{\beta} {{\cal{N}}_{\varepsilon}}^{N} }
   \prod_{n=2}^{N} \; \left[ \int_{-\infty}^{-\infty} d\vec{r}_n \right] \;
   \exp{ \left[-{\cal{A}}[\vec{R}] \right] } \, ,
\label{pathint} \\
   {\cal{A}}[\vec{R}] 
  &=&
   \sum_{n=2}^{N+1} \left[ \,
     \frac{\mu}{2\varepsilon} (\vec{r}_{n}-\vec{r}_{n-1})^{2}  \,+\,
     \frac{\varepsilon}{2} (V_{n} + V_{n-1}) \,\right]\, , 
\label{wirk} \\
   {\cal{N}}_{\varepsilon}
  &=&
   \left[ 2\pi\,\frac{\varepsilon}{\mu} \right]^{\frac{3}{2} } \;\;,\;\;
   \beta = \varepsilon N 
\label{pathintnorm}\mbox{.}
\end{eqnarray}
In the integrand of the path integral ${\cal{A}}[\vec{R}]$ denotes the 
Euclidean action and $\vec{R}=[ \vec{r}_1 ,\vec{r}_2,...,\vec{r}_N ]$
a path consisting of $N$ steps. Each step corresponds to an increment
$\varepsilon$ of the imaginary time, $\beta$.
Parts of the potentials appearing in Eq. (\ref{f3}) will be included into the 
projector by extending the action to
\begin{equation}
  \tilde{{\cal{A}}}[\vec{R}] \equiv
  {\cal{A}}[\vec{R}] \,-\, \ln{v_1} \,-\, \ln{v_{N+1}}\;
  \mbox{,}\quad
  v_n\equiv v(\vec{r}_n)\;
\label{endpunkt}\mbox{,}
\end{equation}
or in our case to
\begin{equation}
  \tilde{{\cal{A}}}[\vec{R}] =
  {\cal{A}}[\vec{R}]+\frac{1}{2b^2}
  \left[\vec{r}_1^{\,2}+\vec{r}_{N+1}^{\,2}\right]
\mbox{.}
\label{endpunktgauss}
\end{equation}

\noindent
Combining different factors into
\begin{equation}
   {\cal{N}}\equiv 
   \frac{ {\cal{N}}_{\beta} \, {{\cal{N}}_{\varepsilon}}^N }{2\pi\mu^2}\;
\mbox{,}
\end{equation}
and with the definitions
\begin{equation}
   P_{\cal{A}}[\vec{R}] = \exp{ \left[-\tilde{\cal{A}}[\vec{R}] \right] }
\end{equation}
and
\begin{equation}
   {\cal{D}}[\vec{R}] = d\vec{r}_1 d\vec{r}_2...d\vec{r}_{N+1}\;
\mbox{,}
\end{equation}
the expression  for the scattering amplitude, Eq. (\ref{f3}), finally becomes
a multi-dimensional integral
\begin{equation}
  f^2_0 \,=\,\frac{V_c^2}{\cal{N}} \int {\cal{D}}[\vec{R}] \,P_{\cal{A}}
  [\vec{R}]\;
\label{feynint}
\mbox{.}
\end{equation}
The integral now also includes the endpoints $\vec{r}_1$ and $\vec{r}_{N+1}$
of the path.

The expression for the scattering amplitude, Eq. (\ref{feynint}), contains 
only known quantities, {\it e.g.} it does not involve the unknown scattering 
wave function. Instead one has to deal with multi-dimensional path integrals. 
In the next chapters we will examine different stochastic methods to evaluate 
these integrals with sufficient accuracy and speed, such that the path integral 
approach can be also be applied to many-body targets. A problem is that the 
potential appearing in the action $\cal{A}$ is complex.
The expression in Eq. (\ref{feynint}) therefore cannot be interpreted as an
integral over paths sampled according to a real and positive distribution,
the starting point for many stochastic methods. Even for real potentials,
which have both attractive and repulsive regions, this interpretation
requires special care \cite{SLen95b}. 

To remedy this difficulty associated with complex potentials, we introduce
a suitable real action, $\tilde{\cal{S}}$, that yields a real positive
distribution, 
\begin{eqnarray}
  P_{\cal{S}}[\vec{R}] &\equiv& \exp{ \left[ -\tilde{{\cal{S}}}[\vec{R}]
   \right] }\mbox{,}\quad
  P_{\cal{S}}[\vec{R}] \geq 0 
\;\mbox{.}
\label{usualact}
\end{eqnarray}
The action $\tilde{\cal{S}}$ should simulate the main features of the actual
problem as closely as possible or, equivalently, $P_{\cal{S}}[\vec{R}]$ provide
a good approximation to 
the kernel $P_{\cal{A}}[\vec{R}]$ in Eq. (\ref{feynint}).
For a many-body target, for example, such a reference problem could be 
described in terms of a simple effective one body potential \cite{SLen95b}:
in regions where the actual interaction is strongly absorptive and the 
wavefunction is small,
a real reference potential $U_{a}(\vec{r})$ may be choosen that is strongly
repulsive.
Using this approximate reference action, Eq. (\ref{feynint}) can be rewritten as
\begin{eqnarray}
   f^2_0
  &=& 
   \frac{V_c^2}{\cal{N}} \int {\cal{D}}[\vec{R}] \; P_{\cal{A}}[\vec{R}]
   = 
   \frac{V_c^2}{\cal{N}} \int {\cal{D}}[\vec{R}] 
   \left[
    \frac{ P_{\cal{A}}[\vec{R}] }{ P_{\cal{S}}[\vec{R}] }
   \right]
    P_{\cal{S}}[\vec{R}]\;
\mbox{.}\label{numobs}\label{numobs1}
\end{eqnarray}
\noindent
The expression has the form of a multi-dimensional integral over a generalized
complex observable, ${\cal{O}}[\vec{R}]$, where the distribution of 
paths is given by 
$P_{\cal{S}}[\vec{R}]$:
\begin{eqnarray} 
&&f_0^2
=
  \frac{V_c^2}{\cal{N}} \int {\cal{D}}[\vec{R}] \;
  {\cal{O}}[\vec{R}]\,\; P_{\cal{S}}[\vec{R}] 
\;\equiv\;
  \frac{V_c^2}{\cal{N}} \; \left\langle{\cal{O}}\right\rangle_{P_{\cal{S}}}
\mbox{,}\label{numobs2}\\
&&{\cal{O}}[\vec{R}] 
=
   \left[ \frac{ P_{\cal{A}}[\vec{R}] }{ P_{\cal{S}}[\vec{R}] } \right]
\mbox{,}
\end{eqnarray}

\noindent
or more explicitely
\begin{eqnarray}
   &&f_0^2
  =
   \frac{V_c^2}{\cal{N}}
   \int \,{\cal{D}}[\vec{R}] \; e^{-( \tilde{\cal{A}}[\vec{R}]- 
   \tilde{\cal{S}}[\vec{R}] ) } e^{-\tilde{\cal{S}}[\vec{R}]}
  =
   \frac{V_c^2}{\cal{N}}
   \left\langle e^{-\Delta {\cal{S}} } \right\rangle_{P_{\cal{S}}}
\mbox{,}\label{numobs5}\\
\end{eqnarray}
where
\begin{eqnarray}
  &&\Delta {\cal{S}}[\vec{R}] =
   \tilde{ {\cal{A}} }[\vec{R}] -\tilde{{\cal{S}}}[\vec{R}]
\label{numobs3}
\;\mbox{.}
\end{eqnarray}
As we will see, in working with complex modified observables one encounters
strong numerical fluctuations which are caused by the presence of a phase that 
varies along the entire path.

Monte Carlo methods used to generate paths distributed according to
$P_{\cal{S}}$ always yield a normalized distribution. Except for few
cases, this normalization is not known. Therefore, instead of 
Eq. (\ref{numobs5}) one calculates the quotient
\begin{eqnarray}
  \frac{f^2_0}{f^2_{0\,a}}
 &=&
  \frac{V_c^2}{U_{a}^{2}} \;
  \frac{ \int {\cal{D}}[R] \;
         e^{ - \left[\tilde{{\cal{A}}}[\vec{R}]
	 -\tilde{{\cal{S}}}[\vec{R}] \right]} \;
         e^{ - \tilde{{\cal{S}}}[\vec{R}] } }
       { \int {\cal{D}}[R] \; e^{ - \tilde{{\cal{S}}}[\vec{R}] } } \\
 &=&
  \frac{V_c^2}{U_{a}^{2}} \;
  \frac{ \int {\cal{D}}[R] \; e^{ - \Delta {\cal{S}}[\vec{R}] } 
         e^{ - \tilde{{\cal{S}}}[\vec{R}] } }
       { \int {\cal{D}}[R] \; e^{ - \tilde{{\cal{S}}}[\vec{R}] } } 
  = 
\frac{V_c^2}{U_{a}^{2}} \;
\frac{\langle {\cal{O}} \rangle}{\langle {\cal{O}}_{a}\rangle}\;
\label{quot3}\mbox{,}
\end{eqnarray}
where $U_{a}$ is the strength of the real reference potential. The advantage of
this method is obvious from Eq. (\ref{quot3}): one does not calculate the whole
problem from the beginning, but uses a solved one as a reference problem. The
computational effort is spent on calculating just the difference between both. 

\section{Random-Walk-Algorithms}
\label{RWA}

The most direct implementation of the path integral expression for the 
scattering amplitude is the method of random walks. We will first discuss this
approach for real potentials and later apply it to complex potentials.
The procedure for generating paths starts with a set of initial 
points ${\vec{r}_1}$.
From each of these points the next point is then reached by means of the 
Euclidean 
propagator $\langle \vec{r}_i \mid e^{-\tilde {\cal{S}} } \mid\vec{r}_1\rangle$. 
The set of paths generated in this way is then distributed according to $P_S$. 
This procedure is continued until a sufficiently large time $\beta$ is reached 
to allow the projector $e^{-\beta\hat{H}}$ to filter out the zero
energy contribution. As the projection time grows, more paths outside
the  potential region contribute, while configurations inside, which sample the
dynamics of the target, only yield a small contribution \cite{SLen94}. 
This causes large fluctuations in the results and we examine methods to improve
the generation of paths through the guided random walk (GRW) algorithms.

\subsection{Simple methods: free diffusion and diffusion with sources and sinks}

\label{SRW}

{\it Free diffusion:} The simplest method to generate paths uses only the
free action ${\cal{S}}_0$ in the Euclidean propagator, 
\begin{equation}
   {\cal{S}}_0[\vec{R}]
   = \sum_{n=1}^{N}\; \frac{\mu}{2\varepsilon}(\vec{r}_{n+1}-\vec{r}_{n})^2
\label{freiprop}\mbox{.}
\end{equation}
\noindent
The matrix element in Eq. (\ref{numobs1}) is then evaluated  by using paths 
distributed according to
\begin{eqnarray}
    P_{\cal{S}}[\vec{R}] &=& 
     \frac{1}{{\cal{N}}_{\varepsilon}^N \,{\cal{N}}_v}
    \; \exp{\left[ -{\cal{S}}_0[\vec{R}] \,+\, \ln{v_1} \right]}\;
\label{vertfrei}\mbox{,} \\
    {\cal{N}}_v &=& \int\,d\vec{r} \;v(\vec{r}) \;<\infty\;
\label{normvertfrei}\mbox{.}
\end{eqnarray} 
The real potential $U$ only enters through the generalized observable 
${\cal{O}}[\vec{R}]$,
\begin{eqnarray}
    {\cal{O}}[\vec{R}] &=&
    2\pi\,\mu^2\;U_0^2\; \frac{ {\cal{N}}_{v} }{ {\cal{N}}_{\beta} }  \;
    \exp{\left[ \ln{v_{N+1}} \,-\, {\cal{S}}_U[\vec{R}] \right]}
\;\mbox{,}\label{freiobs}\\
    {\cal{S}}_U[\vec{R}] &=& \varepsilon \sum_{n=1}^{N+1}\; U_n\;.
\end{eqnarray} 
\noindent
The procedure to generate paths according to the distribution 
$P_{\cal{S}}[\vec{R}]$ consists of the following steps:

\begin{enumerate}
\item
Choose $Z_{in}$ starting coordinates $\vec{r}_1^{\;(\alpha)},\; \alpha 
=1, ... Z_{in}$ according to the initial distribution $\Phi_v(\vec{r}_1)$, 
\begin{equation}
   \Phi_{v}(\vec{r}_1)= \frac{1}{{\cal{N}}_v}\,v(\vec{r}_1)\;\;.
\end{equation}
We only discuss here the case that $v(\vec{r})$ does not change sign, which 
is true for our choice of a Gaussian shape.

\item
Each of the $Z_{in}$ points is successively propagated 
with the transition probablility $p$,
\begin{equation}
 p\,(\vec{r}_n, \vec{r}_{n+1}; \epsilon) = \frac{1}{{\cal{N}}_{\varepsilon}}
\exp{\left[ - \frac{m}{2 \varepsilon} {\left( \vec{r}_n^{\;(\alpha)}
- \vec{r}_{n+1}^{\;(\alpha)}\right)}^2 
\right] }\;,
\end{equation}
for the step from point $n$ to $n+1$ of a given path $\alpha$.
This is numerically implemented by taking
\begin{equation}
   \vec{r}_{n+1}^{\;(\alpha)}
   =\vec{r}_n^{\;(\alpha)}+\sqrt{\frac{\varepsilon}{\mu}}\;\vec{\eta}_n^
   {\;(\alpha)}
   \;\;,\quad\mbox{  }\;\;
   \alpha = 1,2,..Z_{in}
\;\mbox{,}\label{freipropdisk}
\end{equation}
\noindent
where $\vec{\eta}$ is a vector consisting of three independent, uniformly 
distributed random numbers.
\item
Step 2 is repeated $N=\frac{\beta}{\varepsilon}$ times.
\item
Finally, the scattering amplitude is obtained as 
\begin{equation}
  f_0^2 = \frac{1}{Z_{in}} \sum_{\alpha=1}^{Z_{in}}
	  {\cal{O}}[\vec{R}^{(\alpha)}]
\mbox{.}
\end{equation}
\end{enumerate}

{\it Replication of paths:} Properties of the potential can be incorporated
into the generation of paths by the method of sources and sinks for paths: 
each of the segments of a path is assigned a weight $D^{(\alpha)}_n$,
\begin{eqnarray}
 D_n^{\,(\alpha)} &\equiv& 
 \exp{\left[-\,\varepsilon\; U\left(\vec{r}_n^{\,(\alpha)}\right) \;\right]}
\mbox{.}
\label{erzvern}
\end{eqnarray}
After each free diffusion step as described above, the segments of the 
path generated up to this 'time' $n\,\varepsilon$  are replicated or 
deleted, depending on the value of $D^{(\alpha)}$:

\begin{enumerate}
\item[ ]
Let $D_{int}$ be the largest integer which satisfies 
$0< D_{int}\leq D^{(\alpha)}_n$ and draw a random number $w$ in the 
interval $[0,1]$.
For $w\leq(D^{(\alpha)}_n - D_{int})$ one makes $D_{int}$ copies of the path
segment up to this point, otherwise only $(D_{int}-1)$ copies. These
additional paths are then continued independently. 
If however $D_{int}=0$, the segment will be kept with probability
$w$ or deleted with the complementary probability $(1-w)$.
\end{enumerate}
It is easily seen that this leads to the replication of paths
in the region of attractive potentials; for repulsive potentials there
is no replication of paths and the methods ensures that paths with
low weight factor $D^{(\alpha)}$ are not followed too far.
Especially for the case of strongly attractive potentials, not all 
paths one obtains in this fashion are uncorellated, since 
the paths from replication contain common segments.
Compared to the free diffusion, the advantage of this method is that 
the paths are distributed according to 
$P_S$, where $S$ is the full action, {\it i.e.} they are selected in 
a way that reflects the dynamics of the system. 

\begin{figure}[b]
 \epsfysize=9.5cm
 \epsfxsize=14cm
  \centerline{\epsffile{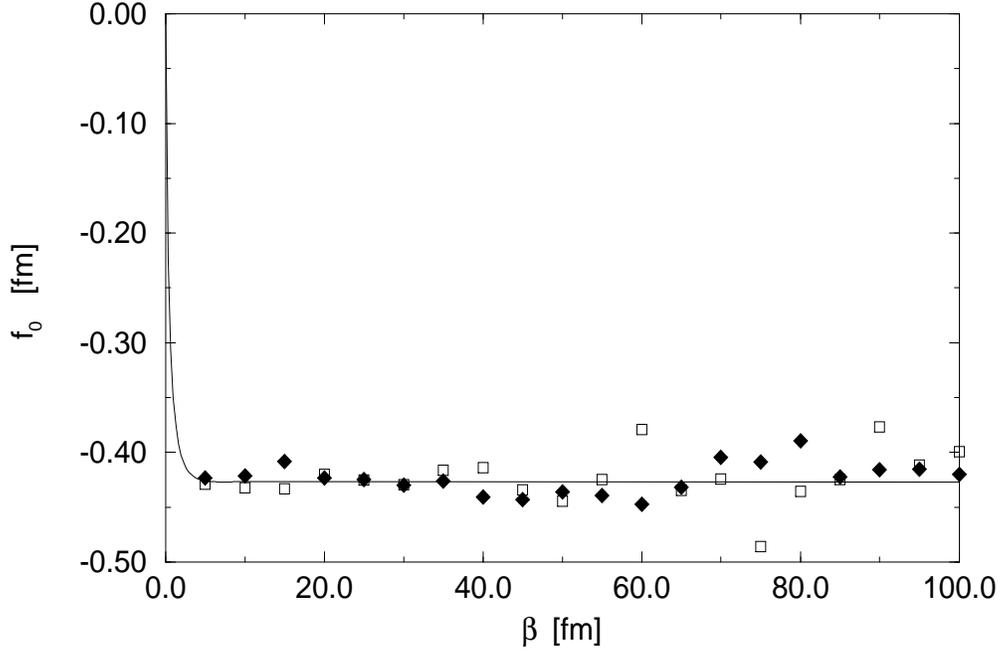}}
  \center{\parbox{11cm}{ 
  \caption{ Scattering length $f_0$ for repulsive potential,
            $U_0 = 1.0\;\mbox{fm}^{-1}$,
            as function of projection time, $\beta$; discretization
	    $\varepsilon = 0.5\;\mbox{fm}$. Open squares: with path 
	    replication; solid squares: free action random walk;
	    solid curve: exact FFT result.
\label{path_fig1}}}}
\end{figure}

\begin{figure}[b]
 \epsfysize=9.5cm
 \epsfxsize=14cm
  \centerline{\epsffile{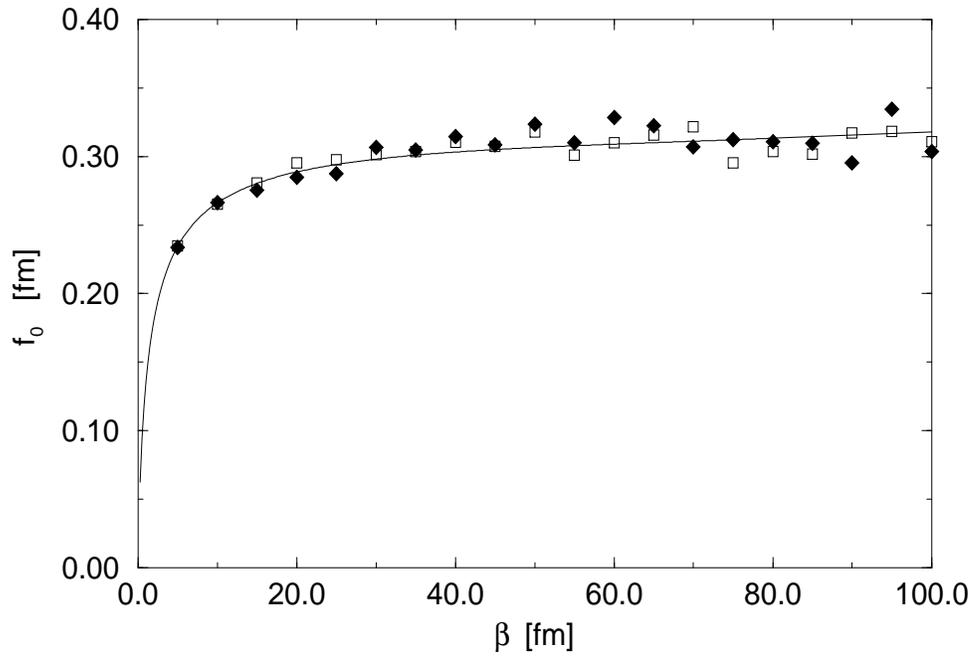}}
  \center{\parbox{11cm}{ 
  \caption{ Scattering length $f_0$ for attractive potential,
            $U_0 = -0.3 \;\mbox{fm}^{-1}$, 
            as function of projection time, $\beta$; discretization 
	    $\varepsilon = 0.5\;\mbox{fm}$. Labelling as in Fig. 1
\label{path_fig3}}}}
\end{figure}

We now illustrate the above by applying these simple random walk methods to a 
repulsive and an attractive real potential. To be able to judge the quality 
of the methods, we calculated the exact expectation value of the observable
as a function of the Euclidean time $\beta$.  This was done by solving 
Eq. (\ref{pathint}) in its phase space representation with the fast Fourier 
transform (FFT) method \cite{Ono88}. Figs. \ref{path_fig1} and \ref{path_fig3} 
show results obtained with the free diffusion and the path replication method.
By comparing to the FFT calculation, we see that for an attractive
potential the convergence is much slower and relatively 
large projection times, typically $\beta = 100 fm$, are needed. 
This long projection time 
requires high stability of the calculations. In the following we will 
therefore often use the attractive potential as a test case. 

The difference in quality between
the two methods, free diffusion and path replication, is small. The main
source of this insensitivity can be traced to the rapid diffusion 
from the potential region,which leads to a relatively small number
of paths that actually contributes to the matrix element of the observable:
starting with 100000 paths, the number grows to 135000 for the attractive 
potential. Of these, only 630 contribute at $\beta = 100 fm$, {\it i.e.} the
generation of new paths does only little to balance the loss of
paths due to the free diffusion from the potential region.
For a repulsive potential, the number of paths drops from the initial
100000 to 50000, of which only 120 contribute. 
For free diffusion, the total number of paths stays constant and 330 
contribute. The number of contributing paths is not sufficiently different 
in the cases considered to lead to a noticable difference in accuracy. 
In view of the achieved accuracy and the long projection time,
none of the methods seems suitable for application to complex obsevables
and many-body targets, which have higher demands on numerical accuracy
and speed of the calculations.

\subsection{Guided random walk algorithms}
\label{GRW}

An improvement of the random walk method can be achieved if an approximate
solution for the wavefunction of the system is known. This solution can be 
used for the generation of the paths one uses to sample a modified observable. 
For the scattering from a many-body target such an approximate solution can
be obtained for example from an effective single particle potential 
\cite{SLen95b}. 
Of course, since we consider the potential scattering of a particle, the 
exact wavefunction in our present studies can be obtained by solving the 
Schr\"{o}dinger equation with standard numerical methods. 
To get an impression of the power of the guided random walk method,
we use here the 'optimal' situation: we work with the exact wavefunction
for the real potential. This provides an upper limit for the performance 
of this method, since with only an approximate wavefunction the accuracy 
can be expected to get worse.

For the zero energy case, we thus use the solution of the Hamiltonian
with a real potential, $ U(\vec{r}) = U_0 v(\vec{r})$, that satisfies
\begin{equation}
  H \; \xi(\vec{r})=0
\mbox{.}
\label{nullew}
\end{equation}
\noindent
The 'time-independent guided random walk method' only makes 
use of this stationary solution.
We have tested the time independent method  to generate paths and evaluate the
appropriate observable. It was found that in our case 
this method also could not prevent the substantial drifting of paths out of 
the potential region. Consequently, there was no noticeable improvement in
accuracy compared to the simple random walk method.
We therefore describe here only the time dependent guided random walk,
which tries to compensate the loss of configurations. An essentially constant 
density of configurations in the potential region can be ensured by 
using a zero energy wavefunction with a norm that depends on the Euclidean 
time, $\tau_e$,
\begin{equation}
  \phi(\vec{r},\tau_e) = \tau_e^{\frac{3}{2}} \,\xi(\vec{r})\;.
\end{equation}
One then considers the evolution of the product 
$\phi(\vec{r},\tau_e) \psi(\vec{r},\tau_e)$, which has the infinitesimal 
evolution operator
\begin{equation}
  \hat{\rho}_{\varepsilon}=
  \phi(\hat{r},\tau_e +\varepsilon)\;e^{-\varepsilon \,\hat H}\;
  \phi^{-1}(\hat{r},\tau_e) \; \mbox{.}
\end{equation}
It can be shown \cite{Neg88} that the matrix element can be written in
terms of integrals over paths distributed according to a modified distribution
\begin{eqnarray}
   P_{\cal{S}}[\vec{R}]
   &=& 
   \frac{\beta^{\frac{3}{2}}}{ {\cal{N}}_{v} }\
   \prod_{n=1}^{N}\left\{\frac{1}{{\cal{N}}_{\varepsilon_n}}\;
   \exp{\left[-\frac{\mu}{2\;\varepsilon_n}
   (\vec{r}_{n+1}-\vec{r}_{n}-\frac{\varepsilon}{\mu}\vec{F}_{n})^2\right]}
   \right\}\,\,v(\vec{r}_0)\;, 
\label{freiproptrial}\label{guidprop}\\
   {\cal{N}}_{\varepsilon_n} &=&
   \left[ \frac{2\pi\varepsilon_n}{\mu} \right]^{\frac{3}{2}}\;,\quad
   \vec{F}_n = \vec{\bigtriangledown}\,\ln{\xi_n}\;,\quad
   \varepsilon_n=\varepsilon\;\left[ 1+\frac{\varepsilon}{\mu}
   \bigtriangleup\;\ln{\phi_n} \right]
\label{normproptrial} 
\mbox{,}
\end{eqnarray}
and a simple modified observable,
\begin{eqnarray}
    {\cal{O}}[\vec{R}]   &=&
    \mu\,(2\,\pi\,\mu)^{-\frac{1}{2}} \,U_0^2 \; {\cal{N}}_{v} \;
    \exp{\left[\;\ln{v_{N+1}}\,-\,\ln{\xi_{N+1}}\,+\,\ln{\xi_1} \; \right]}\;
\label{guidobs} \mbox{.}
\end{eqnarray}
After discretization, the infinitesimal evolution operator reads
\begin{equation}
   \rho_{\varepsilon}=
   \left[\frac{n+1}{n}\right]^{\frac{3}{2}}\;
   \frac{1}{{\cal{N}}_{\varepsilon_n}}\;\exp{\left[-\frac{\mu}{2\;\varepsilon_n}
   (\vec{r}_{n+1}-\vec{r}_{n}-\frac{\varepsilon}{\mu}\vec{F}_{n})^2\right]}
   \mbox{.}
\end{equation}

\begin{figure}[b]
 \epsfysize=9.5cm
 \epsfxsize=14cm
  \centerline{\epsffile{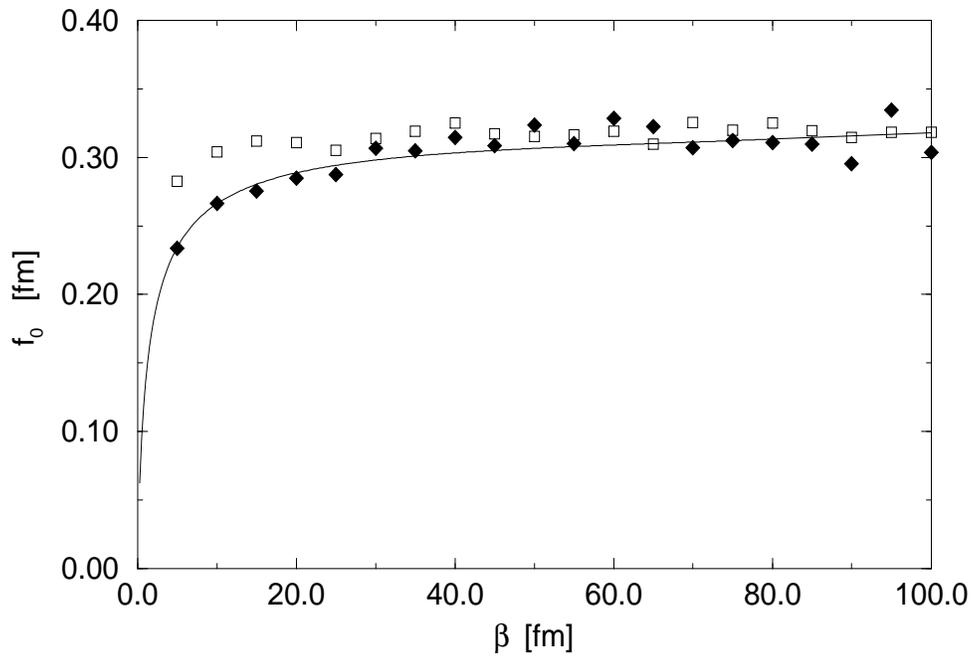}}
  \center{\parbox{11cm}{ 
  \caption{ Scattering length $f_0$ for attractive potential,
            $U_0 = -0.3 \;\mbox{fm}^{-1}$, as function of projection time, 
	    $\beta$; discretization $\varepsilon = 0.5\;\mbox{fm}$.
	    Open squares: time dependent GRW; solid squares: free action 
	    random walk; solid curve: exact FFT result.
\label{path_fig8}}}}
\end{figure}

The numerical implementation of the above is similiar to the diffusion case: 
successive points of a path are obtained according to
\begin{equation}
   \vec{r}_{n+1}^{\;(\alpha)}
   =\vec{r}_n^{\;(\alpha)}-\frac{\varepsilon}{\mu}\vec{F}_n^{\;(\alpha)}
   +\sqrt{\frac{\varepsilon_n}{\mu}}\;\vec{\eta}_n^{\;(\alpha)}
   \;\;,\quad\mbox{where}\;\;
   \alpha = 1,2,..Z_{in}\;
\mbox{.}\label{grwpropdisk}
\end{equation}
After each step, there is a replication of the path segment by a factor
\begin{equation}
 D_n
 \equiv\left[\frac{n+1}{n}\right]^{\frac{3}{2}}\;\;,\quad\;\;
\end{equation}
through a procedure analogous to the one in Section \ref{SRW}. The 
difference is that now the weight factor is not space-, but time-dependent,
reflecting the time-dependent norm of $\phi$. This change of the particle 
number was found to be successful in keeping the number of points in the 
potential region at a reasonably high level. The result is an increase in 
accuracy and stability, as shown in Fig. \ref{path_fig8}.

The price one pays is that the total number 
of points grows rapidly with time. To keep about 2000 points in the potential 
region over a projection time of $\beta = 100 fm$,
an initial number of 100 configurations grows to ca. 200000. 
Since the generated paths all go back to a very small number of initial
configurations, there is a noticeable correlation among the paths for small
$\beta$, which can be seen by the large initial deviations from the exact 
result. However, for large projection times the method is remarkably stable.
This is also due to the presence of the drift-force $\vec{F}$ in Eq.
(\ref{grwpropdisk}), which  together with the random configurations $\eta$ 
has a decorrelating effect. Such effects will be discussed in more detail 
in the next chapter in conection with the Langevin algorithm.

\begin{figure}[b]
 \epsfysize=9cm
 \epsfxsize=14cm
  \centerline{\epsffile{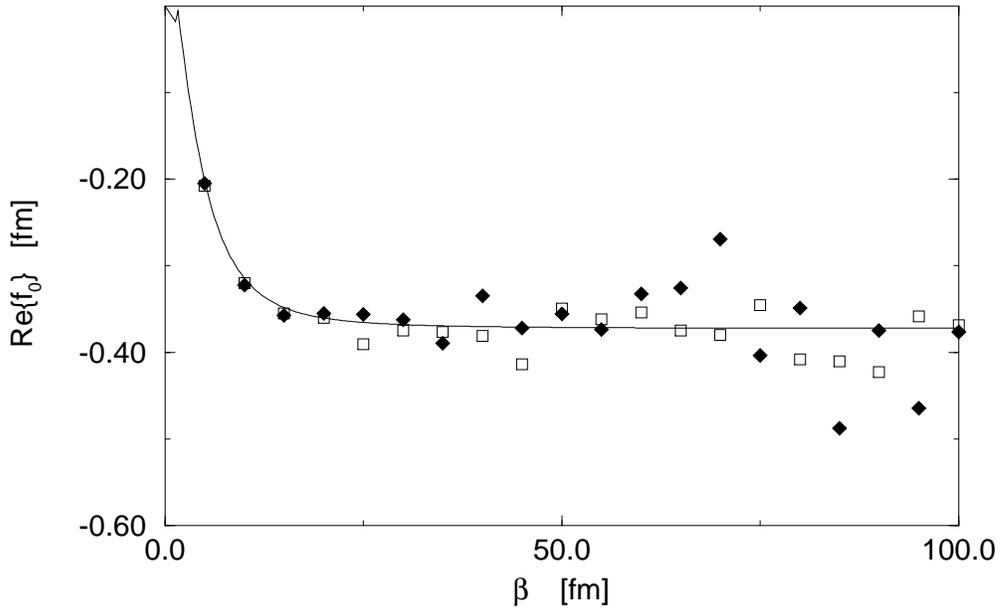}}
 \epsfysize=9cm
 \epsfxsize=14cm
  \centerline{\epsffile{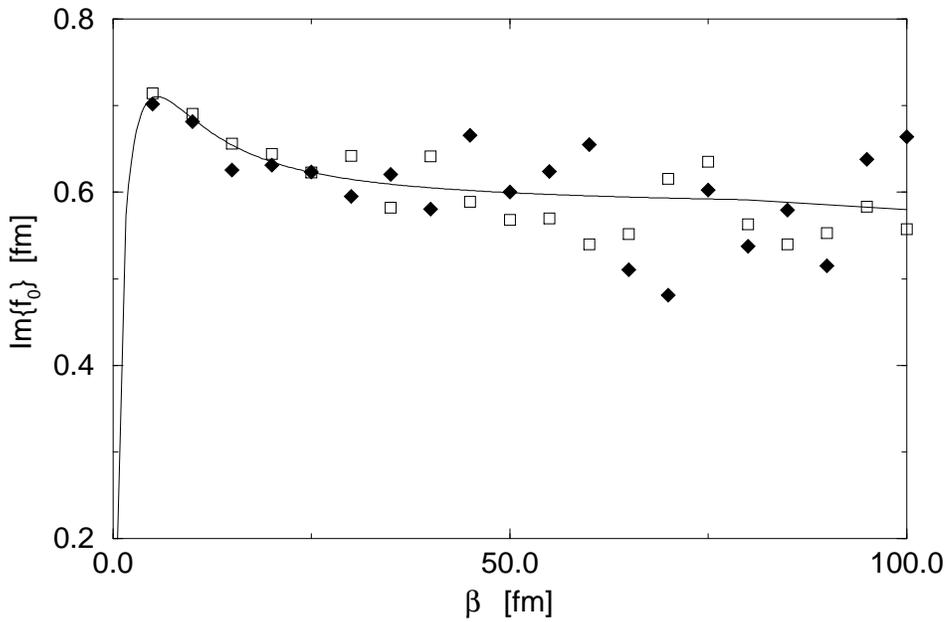}}
  \center{\parbox{11cm}{ 
  \caption{ Scattering length $f_0$  for complex potential 
            $V_c =(-0.3,-1.0) \;\mbox{fm}^{-1}$ 
            as function of projection time $\beta$; discretization 
	    $\varepsilon = 0.5\;\mbox{fm}$. Open squares: time dependent 
	    GRW results; solid squares: free action random walk;
	    solid curve: exact FFT result.
\label{path_fig20}\label{path_fig21}}}}
\end{figure}

Due to the reasonable stability, the time-dependent GRW method looks like a 
promising method for the complex potentials, which require higher numerical
accuracy. The full Hamilton-operator $\hat{H}$ is now
\begin{equation}
 \hat{H}=\hat{K}\,+\,\hat{U}\,+\,i\,\hat{W\;.}
\end{equation}
The wavefunction $\phi(\vec{r},\tau_e)$ is determined through the real 
potential with the time-independent part satisfying
\begin{equation}
  \left[K\,+\,U(\vec{r})\right]\,\xi(\vec{r})\,=\,0
\,\mbox{.}
\end{equation}
The modified observable also includes also the absorptive part of the potential
through the  imaginary action ${\cal{S}}_W[\vec{R}]$,
\begin{eqnarray}
    {\cal{O}}[\vec{R}] &=& 
    \mu\,(2\,\pi\,\mu)^{-\frac{1}{2}} \,U_0^2 \; {\cal{N}}_{v} \;
    \exp{\left[ 
      \ln{v_{N+1}}\,-\,\ln{\xi_{N+1}}\,-\,{\cal{S}}_W[\vec{R}]\,+\,\ln{\xi_1}
    \right]}
\label{guidobscomp}
\mbox{.}
\end{eqnarray}
The resulting scattering length is shown in Fig. \ref{path_fig21}.
The time dependent GRW does somewhat better than the random walk method based
on the free action. The figure shows that there are less fluctuations in the 
results obtained with the time-dependent GRW. However, some fluctuations remain.
They are mainly due to the phase, which sums up the
contributions along each path. Similar results were also found for an
absorptive potential with a repulsive real part and are not shown. 

In summary, we have seen that the random walk algorithm converged reasonably
fast towards the exact scattering length. However, the numerical accuracy was
not satisfactory for any of the different versions we tried. 
A typical calcuation took ca. 25 minutes on a HP UX 9000/720.
Thus while these algorithms are not useful for the calculation of scattering
from a many-body target, they can provide estimates for 
the typical projection times one needs. These projection times
are needed as an input in the algorithms described in the next chapters,
where paths are not built up successively from point to point, but
entire paths of a given length $\beta$ are generated and modified.

\section{Langevin and Hybrid Algorithms}
\label{LAHYB}

\subsection{Standard Langevin Algorithm}
\label{LANG}

The path integral methods presented in the last section describe a diffusion
process in the physical time $\beta$, sequentially generating the elements of 
a path, $R_i$. Another approach is the Langevin 
algorithm, which updates an entire given path $\vec{R}$.
This development of a path proceeds in terms of a new variable $\tau$, the 
'Langevin time', and takes place according to the stochastic 
differential Langevin equation,
\noindent
\begin{equation}
\frac{\partial}{\partial\tau} R_i(\tau)\, =  - \tilde{S}_i\,'(\tau) +
\pi_i(\tau) 
\label{langcont}
\mbox{ ,}
\end{equation}
where $\pi$ is a Gaussian random variable
\begin{equation}
 \langle\pi_i(\tau)\rangle = 0 \;\; \mbox{,}\quad
 \langle\pi_i(\tau),\pi_j(\tau')\rangle\,=\,
 2\, \delta_{ij}\,\delta(\tau-\tau') 
\mbox{ .}
\end{equation}
The variable $\tau$ is different from the physical time, which is
represented here by the index $i$. When discretized, $\tau$ labels the 
elements in a Markov chain.
Again, $\tilde{\cal{S}}$ is a real approximation to the action $\tilde
{\cal{A}}$ and
 \begin{equation}
\tilde{S}_i\,'(\tau) =
 \frac{\partial}{\partial R_i}\,\tilde{{\cal{S}}}[\vec{R}(\tau)]
\mbox{ .}
\end{equation}
\noindent
The stochastic process described by Eq. (\ref{langcont}) is ergodic,
\begin{equation}
  \langle {\cal{O}} \rangle_P \equiv
  \int {\cal{D}}[\vec{R}] \; P_{\cal{S}}[\vec{R}] \; {\cal{O}}[\vec{R}]
  = \lim_{\tau\rightarrow\infty}
    \frac{1}{\tau}\int_0^{\tau} d\tau'  \;{\cal{O}} \; [\vec{R}(\tau')] 
 \label{zeitmittel}
\mbox{,}
\end{equation}
\noindent
where 
\begin{equation}
  P_{\cal{S}}[\vec{R}] =
  \frac{1}{{\cal{N}}_{\cal{S}}} \exp{\left[-\tilde{{\cal{S}}}[\vec{R}] \right]}
  \;.
\label{pfadvertstat}\label{usalact}
\end{equation}
\noindent
Choosing for $\tilde{{\cal{S}}}$ the real part of the action, the
paths generated through Eq. (\ref{langcont})  will be used to evaluate the 
modified observable 
\begin{equation}
  {\cal{O}}[\vec{R}] =  \exp{\left[-{\cal{S}}_W[\vec{R}]\right]} \;.
\label{modobs}
\end{equation}

For numerical evaluation, Eq. (\ref{langcont}) is discretized. We first write 
the contributions to the action $\tilde{\cal{S}}$,
\begin{equation}
  \tilde{{\cal{S}}}[\vec{R}]= \tilde{\cal{S}}_0[\vec{R}]+{\cal{S}}_U[\vec{R}]
  \label{hybvert}
\mbox{,}
\end{equation}
\noindent
in a convenient matrix notation
\begin{eqnarray}
  \tilde{{\cal{S}}}_0[\vec{R}] & = &
  {\cal{S}}_0[\vec{R}] + \frac{1}{2b^2} \left( \vec{r}_1^{\;2}+\vec{r}_{N+1}^{\;2} \right)
  \;\equiv\; \vec{R} \, {\bf{A_0}} \, \vec{R} \;\mbox{,}
  \label{endp}\label{matrixa}\\
  {\cal{S}}_U[\vec{R}] & = &
  \varepsilon\,\sum_{i=1}^{N+1} \; U_{i} \;\equiv\;
  \frac{\varepsilon}{3}\;\mbox{Tr}\;{\bf{D_U}}[\vec{R}] \;\mbox{,} \\
  U_i & \equiv & \;\left\{ \begin{array}{ll} 
     \frac{1}{2}\,U\;v(\vec{r}_i)  &\mbox{for} \,\,i=1,N+1 \\
                  U\;v(\vec{r}_i)  &\mbox{for} \,\,i=2,3,...,N\;\;. \\
\end{array}\right.
\end{eqnarray}
\noindent
The matrix ${\bf{A_0}}$ is symmetric and has $3\,(N+1)$ eigenvalues,
$\lambda_i > 0$, while ${\bf{D_U}}[\vec{R}]$ is diagonal and built from $N$ 
blocks,
\begin{equation}
  {\bf{D_U}}[\vec{R}] = \left[ \;U_i  \;
	       {\bf{E}}_3 \;\right]\;, \;\;i=1,2,...,\,\,(N+1)
\mbox{ ,}
\end{equation}
where ${\bf{E}}_3$ denotes the three-dimensional unit matrix. 
With these definitions we obtain the discretized version of the Langevin 
equation that updates a path $\vec{R}^{(n)}$ and yields $\vec{R}^{(n+1)}$,
\begin{eqnarray}
  \vec{R}^{(n+1)} &=& \left[ {\bf{1}} - \delta\;{\bf{M}}^{(n)}
                     \right]\;\vec{R}^{(n)}\, + h\,\;\vec{\pi}^{(n)}\,,
\label{langdisc4}\\ 
  {\bf{M}}^{(n)} &=& {\bf{A_0}} - \frac{\varepsilon}{b^2} {\bf{D_U}}^{(n)} \;\;.
\end{eqnarray}
The continous variable $\tau$ in the Langevin equation has been replaced 
by the discrete label $n$, 
\begin{equation}
n=\frac{\tau}{\delta}\,,
\label{langdisc2}
\end{equation}
where $\delta$ is the stepsize and we have defined $h$ through
\begin{equation}
\delta \equiv \frac{h^2}{2}\; .
\end{equation}

The Langevin algorithm thus generates from a starting path
$\vec{R}^{(1)}$
a sufficiently large number of path updates through Eq. (\ref{langdisc4}),
which are used to evaluate the observable.
However, not all successive paths produced by the iteration
of Eq. (\ref{langdisc4}) are independent and to assess the efficiency of 
the algorithm we have to consider the autocorrelation time 
$\tau_{cor}$ separating two decorrelated paths. As can be seen from Eq. (\ref{langdisc4}), the evolution in 
$\tau$ is governed by the eigenvalues $\mu_i$ of the matrix $\bf M$. It can be
shown \cite{Sok92} that 
\begin{equation}
\tau_{cor} \sim \frac{1}{\mu_{min}} \;,
\label{cortime}
\label{dectime}
\end{equation}
\noindent
where $\mu_{min}$ is the smallest eigenvalue of $\bf M$. 
On the other hand, for the algorithm to be numerically stable, the step size
$\delta$ should be small enough to resolve also the evolution of the
quickly changing modes, {\it i.e.}
\begin{equation}
  \delta \; << \; \frac{2}{\mu_{max}}
\mbox{ .}
\label{stabil}
\end{equation}
\noindent
Consequently, the number of intermediate steps $n_{cor}$, required such that
a path $\vec{R}^{(m)}$ is decorrelated from its predecessor $\vec{R}^{(n)}$,
satisfies
\begin{equation}
  n_{cor} = \frac{\tau_{cor}}{\delta} >>
            \frac{\mu_{max}}{\mu_{min}}\;,\;\;\;m=n+n_{cor} 
\mbox{.}
\label{vglcorla}
\end{equation}

In general, the spectrum of the full matrix $\bf M$ cannot be calculated. 
Since our scattering potential has a short range, the potential matrix
${\bf{D_U}}^{(n)}$ vanishes in most regions of the space of paths. As an
approximation, we therefore replace
the full dynamical matrix, $\bf M$, by the matrix containing the free action,
$\bf A_0$, and get a lower limit for the autocorrelation time in terms 
of the eigenvalues $\lambda_i$,
\begin{equation}
  n_{cor}  >>
            \frac{\lambda_{max}}{\lambda_{min}} \;.
\label{freecon}
\end{equation}

The ratio of the largest and smallest eigenvalues is shown in Fig. 5. As can 
be seen, a large number of intermediate steps must be taken
before an uncorrelated path is obtained. This is due to the presence of 
two very different length scales in scattering: The average the distance 
of the middle of the path from the origin is of order 
$O[\frac{\beta}{\mu}]$ and grows with the projection time. On the other hand,
since the endpoints of the paths are required to lie in the potential,
a small scale $b$ enters. Of course, the actual spectrum of eigenvalues 
in scattering is continous and the presence of a lowest eigenvalue 
$\lambda_{min} > 0$ is due to discretization.
The finite projection time, the existence of a condition for the endpoints
and the discretization of the action are thus crucial for the
convergence of the Langevin algorithm in scattering problems \cite{SLen95b}.

The change of the stability criteria due to the potential can be 
roughly estimated in the region where $\bf D_U $ is approximately constant.
For a path entirely in this region one has 
\begin{equation}
  \delta \; << \; \frac{2}{\lambda_{i}-\frac{\varepsilon U_0}{b^2}}
\mbox{,}
\label{genstab}
\end{equation}
\noindent
For an attractive  potential, numerical stability may therefore 
require a stepsize $\delta$ which is considerably smaller than the 
estimate based on the eigenvalues of $\bf A_0$; 
for repulsive potentials the stepsize may be taken larger, at least 
as long as the values of $U_0$ are such that the r.h.s. of 
Eq. (\ref{genstab}) stays positive and finite and the condition
doesn't become meaningless. 
The presence of the potential thus may require an increase in the number 
of intermediate steps one needs before another uncorrelated path is obtained 
or, even worse, it can lead to to numerical instabilities.
The condition Eq. (\ref{genstab}) does not apply in the region where the
potential varies significantly. 
Lowering the stepsize in the physical time, $\varepsilon$, to reduce the 
influence of the potential on the stability it is no remedy, since the
eigenvalues $\lambda_i$ also dependend on $\varepsilon$. However, since the 
range of the potential is relatively small, most steps take
place in the outer region and Eq. (\ref{dectime}) still provides 
a reasonable indicator for the overall autocorrelation time.

Finally, to avoid local instabilities and errors due to the
discretization of $\tau$, the updated path obtained from the
discretized Langevin equations are treated as
the proposal step in a Metropolis procedure \cite{Cre89}. The
resulting algorithm is often referred as the 'exact' Langevin algorithm.

The above considerations are illustrated by the examples in Figs. \ref{la_fig1} 
and \ref{la_fig2}. For a potential with a strong repulsive real part and 
strong absorption as in Fig. \ref{la_fig1}, a projection time
of $\beta\,=\,50\,\mbox{fm}$ and $\varepsilon\,=\,0.5\,\mbox{fm}$ is
sufficient; as in Section \ref{POTSCAT} this was checked by
comparing to the results of deterministic FFT 
methods \cite{Ono88}. Inspection of Fig. \ref{corldi} shows that one needs 
according to Eq. (\ref{freecon}) $n_{cor} >> 5000$
intermediate Langevin steps in $\tau$ to generate an independent path. The
numerical values for $\lambda_{max}$ and $\lambda_{min}$ are $20$ and
$4.7\,\cdot\,10^{-3}$. The stepsize $\delta$ is chosen to be 
$2.5\,\cdot\,10^{-2}$ in order to meet condition Eq. (\ref{stabil}) at
least in the region where the potential vanishes.
Therefore the result shown in Fig. \ref{la_fig1} contains less then $80$ 
decorrelated paths. After an initial equilibration phase of about 5000 Langevin
steps, the the fluctuations get smaller and the result for the real part of 
the scattering length gets very close to the exact value. However, the 
imaginary part is still quite far off and shows little sign of further 
convergence to the exact answer with increasing $\tau$. 

For attractive potentials the situation is worse and the necessary projection
time increases. For the strength parameters in Fig. \ref{la_fig2}, 
$U_0\,=\,-0.3\, \mbox{fm}^{-1}$ and $W_0\,=\,-1.0\,\mbox{fm}^{-1}$, the 
FFT calculations show that a projection time $\beta = 80\,\mbox{fm}$ should 
be chosen. This introduces a larger length scale and the number of intermediate
steps $n_{cor}$ increases quickly. For $\varepsilon\,=\,0.5\,\mbox{fm}$ the 
number $n_{cor}$ is already of the order of 10000 and less than $40$ 
independent paths contribute to the result in Fig. \ref{la_fig2}. 
Correspondingly, one has $\lambda_{max}=20$ and $\lambda_{min}=1.9\,\cdot\,
10^{-3}$; $\delta$ is chosen to be $1.0\,\cdot\,10^{-2}$.
The equilibration time is now larger, ca. 10000 steps, and fluctuations in 
the result remain as $\tau$ increases. One improvement would be to reduce 
discretization errors in $\beta$ for the longer projection time by reducing
the stepsize $\varepsilon$. This would lead to an even larger $n_{cor}$ 
as illustrated by Fig. \ref{corldi}. However, the calculation  in 
Fig. \ref{la_fig2} already took approximately 2.5 hours on one processor 
of a CONVEX EXEMPLAR 1000 using standard tools for algebraic manipulations. 
Therefore this possibility of improving the performance seems not practical.

The rather poor performance of the Langevin algorithm seen in the two
examples considered here is very different from
the successful application to real potentials in Ref. \cite{SLen95b},
which could even be extended to many-body targets.
The major cause of this difference is the phase in the modified observable,
Eq. (\ref{modobs}), which is responsible for strong fluctuations. 
Clearly the few independent paths in the examples above are not sufficient
to probe the rapidly varying observable and their number has to be increased to 
achieve a satisfactory result. Straightforward extensions of the above method 
by increasing the number of Langevin steps would lead to prohibitively long 
computation times. We therefore now consider two possibilities which can help to
accelerate the algorithm: The first one modifies the dynamical
properties of the Langevin process by introducing a kernel into the
stochastic differential equation, Eq. (\ref{langcont}), while the second 
method, the hybrid algorithm, combines the Langevin algorithm with a 
microcanonical approach.

\subsection{Langevin Algorithm with Kernel}
\label{KERN}

To speed up the convergence of the Langevin algorithm, an appropriate matrix 
kernel, ${\bf K}$, can be inserted into the discretized Langevin 
equation, Eq. (\ref{langdisc4}). This does not change the distribution
$P_{\cal{S}}[\vec{R}]$ according to which the paths are sampled. The conditions 
under which the use of a kernel is valid are discussed in {\it e.g.} Refs. 
\cite{Nam93} - \cite{Tan92}. For position-independent kernels the equation 
becomes
\begin{equation}
   \vec{R}^{(n+1)} = \left[
   {\bf{1}} - \frac{h^2}{2} {\bf{K}} \;
   \left[{\bf{A_0}}-\frac{\varepsilon}{b^2}\;{\bf{D_U}}^{(n)}\right]\;
   \right] \vec{R}^{(n)} +  h\;\sqrt{\bf{K}} \;\vec{\pi}^{(n)}
\label{kela}
\;\mbox{.}
\end{equation}
Note that the square root of the matrix ${\bf K}$ re-scales $h$, the stepsize 
in the 'Langvin time' $\tau$. The dynamical matrix, ${\bf{M}}$, which 
now governs the evolution of the algorithm in $\tau$, is
\begin{equation}
   {\bf M}^{(n)} \,=\, {\bf K} \;
   \left[{\bf{A_0}}-\frac{\varepsilon}{b^2}\;{\bf{D_U}}^{(n)}\right]
\mbox{,}
\end{equation}
with $n=\frac{\tau}{\delta}$. 

Again the region outside the potential dominates the dynamical properties of
the algorithm. To increase the efficiency, it is desirable to decouple the slow,
long ranged modes of the free action $\tilde{\cal{S}}_0$, corresponding to
the small eigenvalues of $\bf{A_0}$, from the fast modes. 
A natural choice for the kernel is ${\bf{K}}={\bf{A_0}}^{-1}$, which
leads to 
\begin{equation}
 {\bf{M}}^{(n)} = {\bf{1}}-\frac{\varepsilon}{b^2}
             \left[{\bf{A_0}}^{-1}\,{\bf{D_U}}^{(n)} \right]
\mbox{,}
\end{equation}
reducing to the unity operator outside the potential region and thus has
eigenvalues $\lambda_i = 1$. This lowers the correlation time,
Eq. (\ref{cortime}), quite dramatically to $ \tau_{cor} \sim 1$!

As discussed in the previous section, numerical instabilities can occur in 
the region of the potential. With the choice 
${\bf{K}}={\bf{A_0}}^{-1}$
the potential does not enter into the kernel and the scaling of the 
stepsize in the inner and the outer region of the potential is the same.
Nevertheless, since the correlation time is dominated by the behavior
in the outer region, where the different eigenmodes are decoupled by the use of
the above kernel, far less intermediate steps are required to generate the 
necessary independent paths.

The improvement of the algorithm through the kernel does not permit an increase
in the stepsize $\delta$. The coupling of the modes in the interior 
remains and the numerical stability
requires again to follow each step proposed by the Langevin
equation with a kernel by a Metropolis acceptance-step. Since the algorithm
is dominated by the
long ranged modes of the free action, a large rejection rate is obtained.
This  has to be remedied by a decrease in the step-size $\delta$ 
compared to the case without kernel in order to have the Metropolis step 
accept $50\%$ to $80\%$ of the proposed paths. Another factor that 
leads to longer computation times is the matrix structure of the kernel and
thus the increased algebraic complexity of the algorithm. The number of the 
required numerical operations scales at least quadratically with the dimension 
of the matrix $\bf{A_0}$. 

Fig. \ref{lak_fig3} shows the result for a complex potential with a repulsive
real part. The calculation with the projection time $\beta=80 \;\mbox{fm}$,
consumed approximately 2 hours, the runs with $\beta=50 \;\mbox{fm}$ and 
different discretization $\varepsilon$ took 1 hour and 3 hours, respectively. 
The computational effort and the accuracy is similar as in the calculation 
without a kernel in Fig. \ref{la_fig1}. But the increased number of about 
$5000$ independent paths leads to a very stable behaviour; furthermore the
number of steps for the initial equilibration is now reduced to about 1000. 
As can be  concluded from Fig. \ref{lak_fig3}, a smaller discretization in 
the projection time $\beta$ is needed to improve the accuracy of the result. 
This trend was veryfied by a run with $\varepsilon = 0.1 \;\mbox{fm}$, 
$\beta=50 \;\mbox{fm}$ and $n = 5000$ Langevin steps, which however took more 
than 5 hours (not shown).

Fig. \ref{lak_fig4} shows results for a potential with an attractive real part
and $\beta = 80\;\mbox{fm}$, which took about 2 hours. Although the results 
are again remarkable stable after relatively few Langevin steps due to the use 
of a kernel, they deviate from the exact results.
The calculations were repeated (see Fig. \ref{lak_fig5})
with a longer projection time, $\beta = 200\;\mbox{fm}$, and fewer Langevin 
steps, $n = 5000$, which required 5 hours of CPU time. The results are somewhat 
improved, but are less stable. The fluctuations are caused by the decreased 
number of independent paths. Calculations for more attractive potentials 
were also carried out and it was found that no reasonable convergence could be
achieved within the chosen time limits. 

The usefulness of introducing a kernel into the Langevin algorithm is
therefore limited to a small range of potential parameters, where it 
can produce very stable results.
However, as this improved algorithm is rather time-consuming,
its application to multi-particle systems looks not promising. 

\subsection{Hybrid-Algorithm}

The Langevin algorithm can be combined with deterministic molecular-dynamical
methods to a 'hybrid algorithm' with the advantages of both
methods \cite{Mak88,Dua85, Dua86, Dua87, Kro93}.
The deterministic updating of entire paths in a new 'hybrid time', $t$, is governed 
by a 'CPU Hamiltonian', $\cal H$, which is different from the 
physical Hamiltonian, $H$,
\begin{eqnarray}
  {\cal{H}} &=& \frac{1}{2}\vec{\pi}^{\,2}+\tilde{{\cal{S}}}[\vec{R}] \\
  \frac{\partial}{\partial t}{R_i} &=& 
  \frac{\partial{\cal{H}}}{\partial\pi_i} \\
  \frac{\partial}{\partial t}{\pi_i} &=& 
  -\, \frac{\partial{\cal{H}}}{\partial R_i}  =
			      - \, \tilde{S}\,'_i[\vec{R}]\;
\mbox{.}\
\end{eqnarray}
In this approach $\vec{\pi}$ is the CPU momentum of an object with 
position vector $\vec{R}$, moving in a potential $\tilde{{\cal{S}}}[\vec{R}]$. 
The discretized version of these equation for the transition from
$n$ to $n+1$ reads to order $0[h^2]$
\begin{equation}
  T_2(h) \sim \left\{ \begin{array}{ll}
     \pi_i^{(n+\frac{1}{2})} &=\;
     \pi_i^{(n)} \; - \; \frac{1}{2}\; h \; \tilde{S}\,'^{\;(n)}_i \;,\\
      R_i^{(n+1)} &=R_i^{(n)}\;+\;h\,\pi_i^{(n+\frac{1}{2})} \;, \\
     \pi_i^{(n+1)} &=\pi_i^{(n+\frac{1}{2})}-\,
     \frac{1}{2}\,h\,\tilde{S}\,'^{\;(n+1)}_i\;. \\
    \end{array} \right.
\label{hybdisc}
\end{equation}
\noindent
After several of deterministic steps according to the above equations of
motion, the momentum $\vec{\pi}$ is updated by a Gaussian-distributed random 
number. Therefore, all momenta and position vectors, if they they belong 
to a deterministic step or not, are random variables.

The trajectory of a path $\vec{R}$ is split into sub-trajectories, 
${\cal{T}}_{\alpha}$, each starting with a Gaussian random momentum, 
$\vec{\pi}^{(n_{I\,\alpha})}$, and ending with $\vec{\pi}^{(n_{F\,\alpha})}$,
containing  
\begin{equation}
  \bigtriangleup n_{\alpha}\,=\,n_{F\,\alpha} \,-\, n_{I\,\alpha}
\mbox{.}
\end{equation}
sucessive deterministic steps $T_2(h)$.
The next sub-trajectory ${\cal{T}}_{\alpha+1}$ then starts again with a 
Gaussian random momentum. Each move along a sub-trajectory is completed by a 
Metropolis acceptance step: The entire sub-trajectory is considered as a 
proposed move, which is accepted with probability
\begin{eqnarray}
  P_{acc} &=& \mbox{min} \left[1,e^{-\bigtriangleup {\cal{H}}}\right] \;, \\
  \bigtriangleup {\cal{H}} &=&
  {\cal{H}}[\vec{R}^{\,(n_{F\,\alpha})},\vec{\pi}^{\,(n_{F\,\alpha})}]
 -{\cal{H}}[\vec{R}^{\,(n_{I\,\alpha})},\vec{\pi}^{\,(n_{I\,\alpha})}]\;
\label{methyb} \mbox{.}
\end{eqnarray}
This combination with the Metropolis step, which we have used in all
examples below, is referred to as 'exact hybrid algorithm'.

\begin{figure}[p]
  \epsfysize=9.5cm
  \epsfxsize=11.5cm
  \centerline{\epsffile{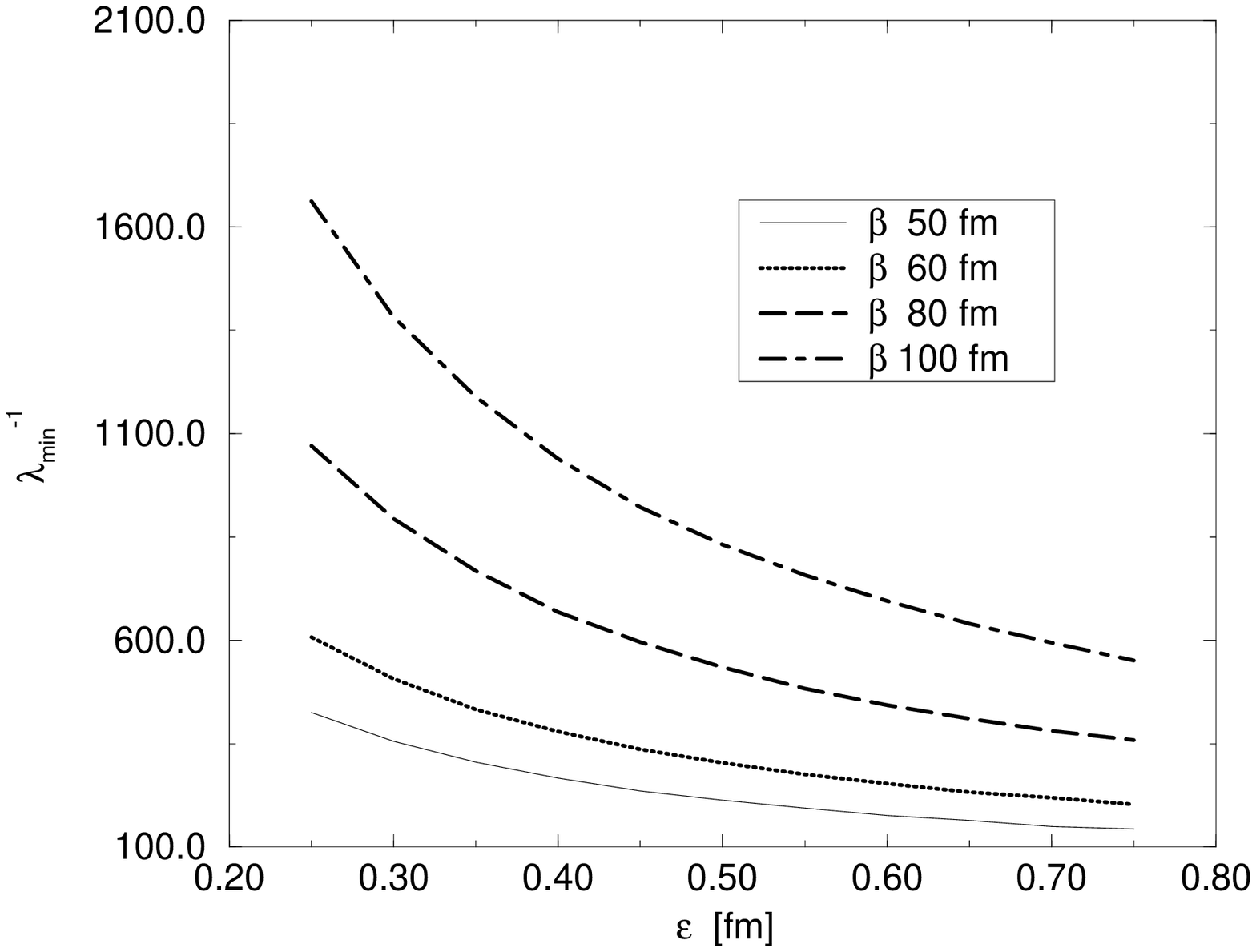}}
  \epsfysize=9.5cm
  \epsfxsize=11.5cm
  \centerline{\epsffile{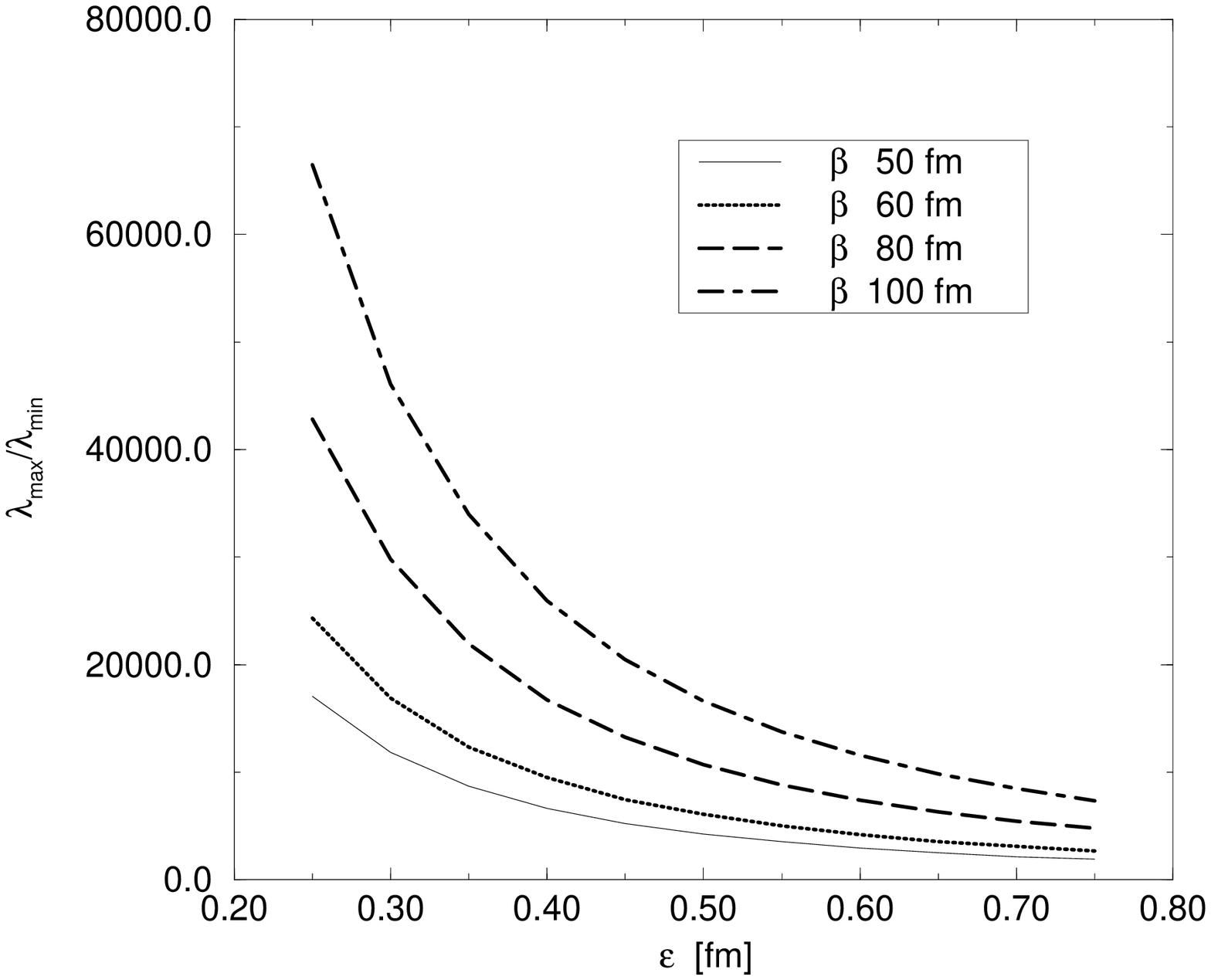}}
  \center{\parbox{11cm}{ 
  \caption{Upper graph: smallest eigenvalues of free action matrix $ \bf A_0$ 
	   as function of discretization $\varepsilon$ for different projection 
	   times $\beta$. 
    	   Lower graph: ratio of largest and smallest eigenvalue of $\bf A_0$. 
	   \label{corlt}\label{corldi} }}}
\end{figure}

\begin{figure}[p]
  \epsfysize=9.5cm
  \epsfxsize=11.5cm
  \centerline{\epsffile{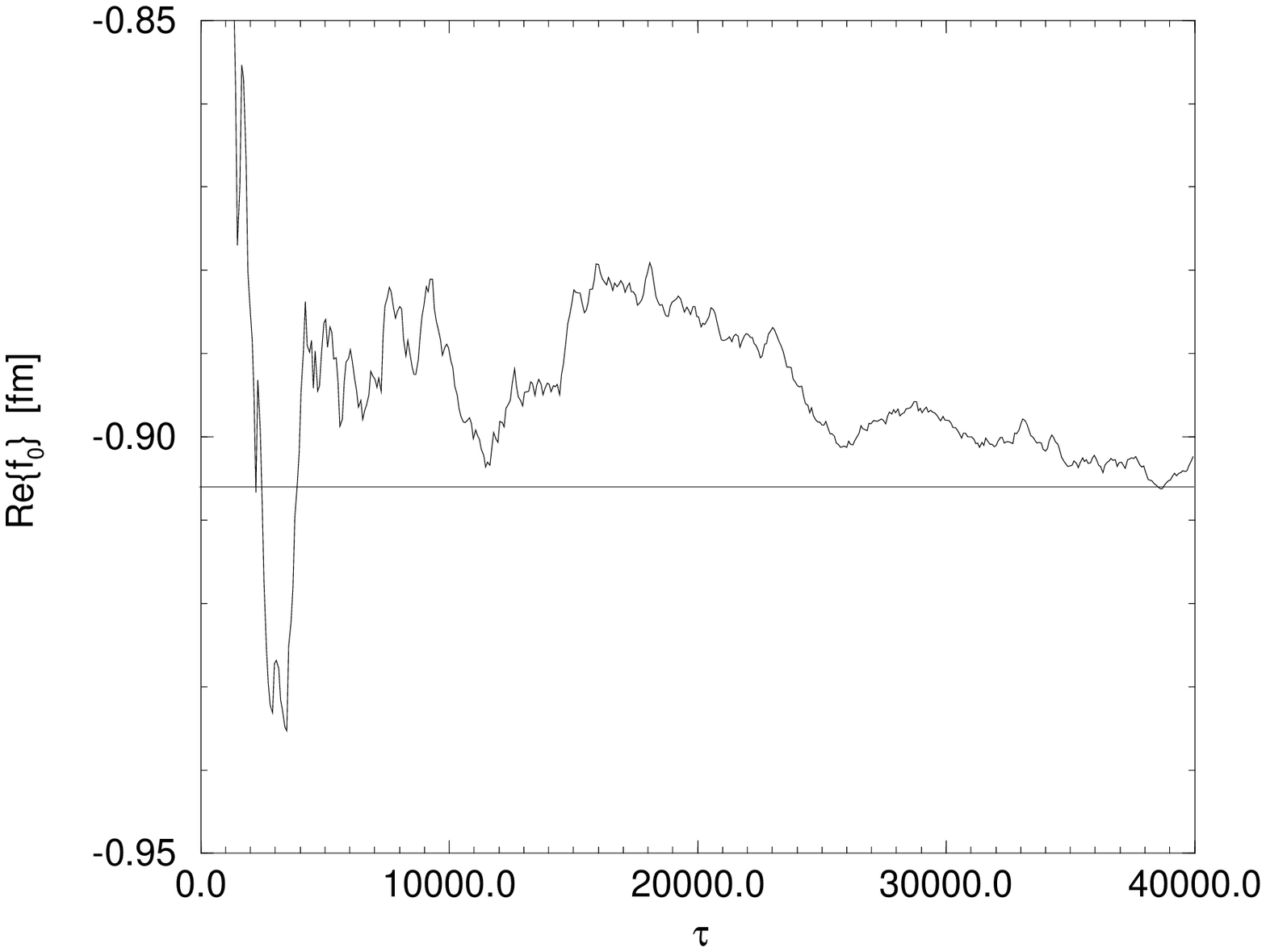}}
  \epsfysize=9.5cm
  \epsfxsize=11.5cm
  \centerline{\epsffile{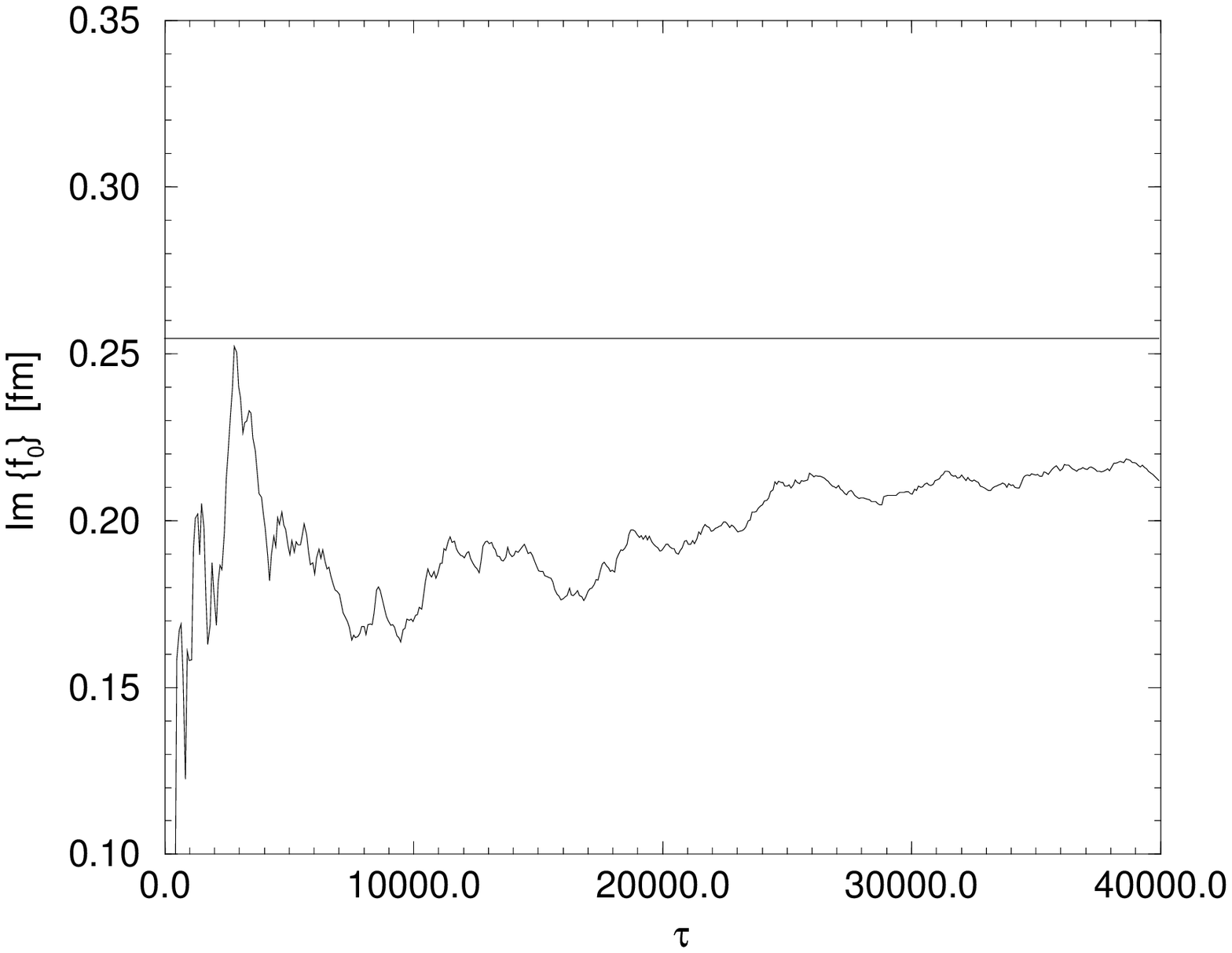}}
  \center{\parbox{11cm}{ 
  \caption{ Scattering length for complex potential 
            $V_c =(2.3,-7.0) \;\mbox{fm}^{-1}$, 
            obtained from the 'exact' Langevin algorithm; 
	    $\beta = 50\;\mbox{fm},
	    \varepsilon = 0.33\;\mbox{fm}$. 
	    Horizontal line: exact result.
	    \label{la_fig1} }}}
\end{figure}

\begin{figure}[p]
  \epsfysize=9.5cm
  \epsfxsize=11.5cm
  \centerline{\epsffile{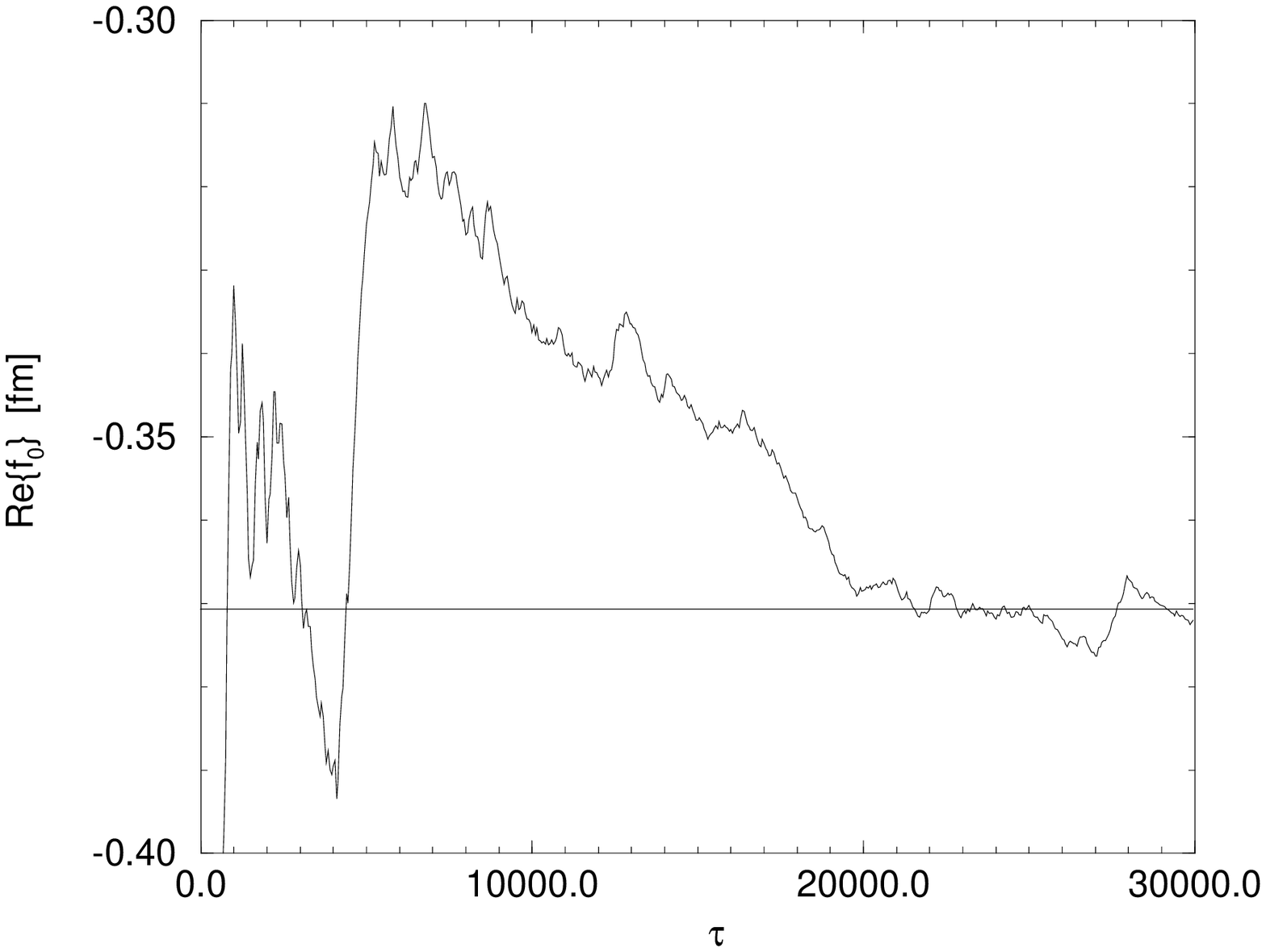}}
  \epsfysize=9.5cm
  \epsfxsize=11.5cm
  \centerline{\epsffile{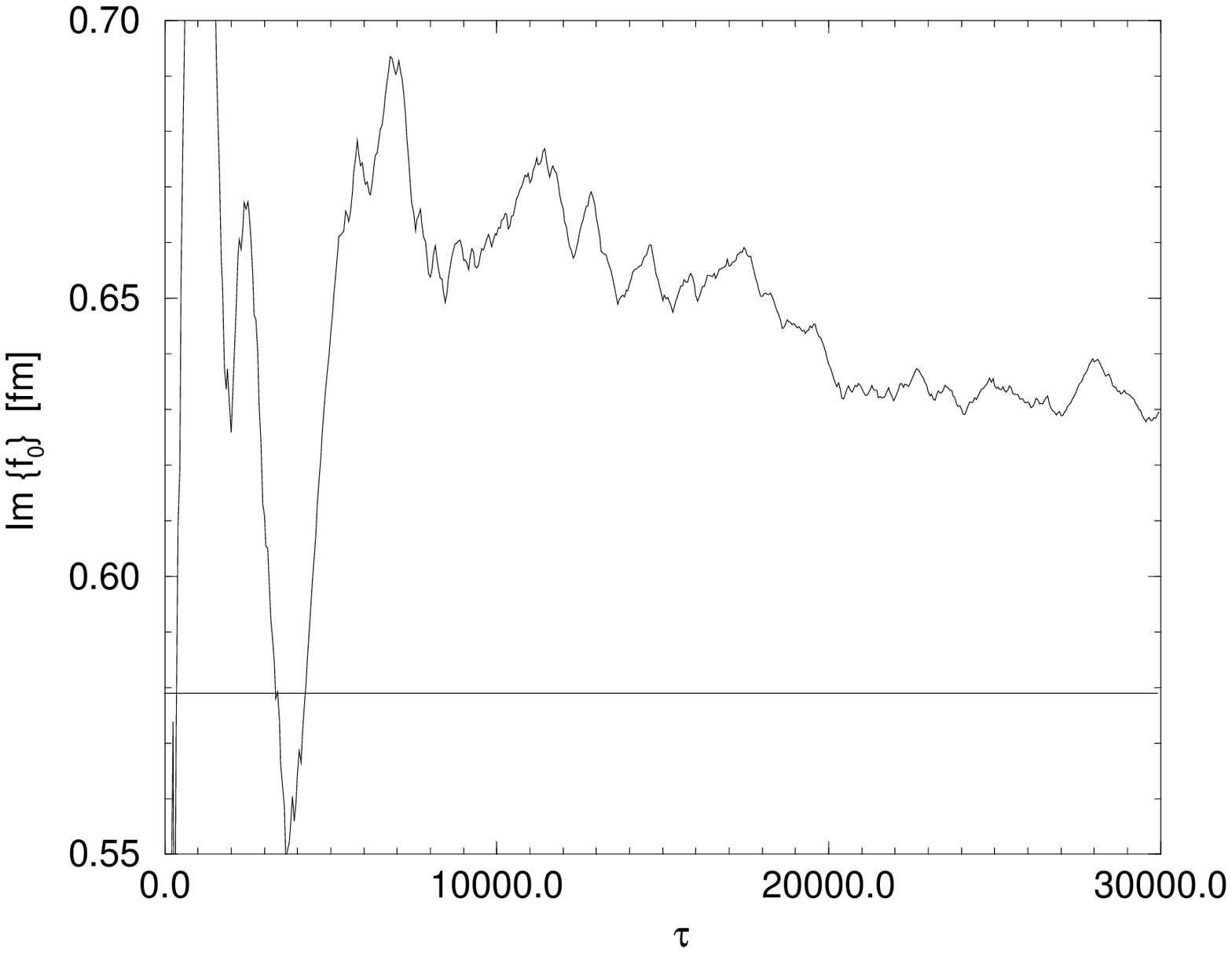}}
  \center{\parbox{11cm}{ 
  \caption{ Scattering length for complex potential 
            $V_c =(-0.3,-1.0) \;\mbox{fm}^{-1}$, 
            obtained from the 'exact' Langevin algorithm; 
	    $\beta = 80\;\mbox{fm},\varepsilon = 0.5\;\mbox{fm}$. 
	    Horizontal line: exact result.
	    \label{la_fig2} }}}
\end{figure}

\begin{figure}[p]
  \epsfysize=9.5cm
  \epsfxsize=11.5cm
  \centerline{\epsffile{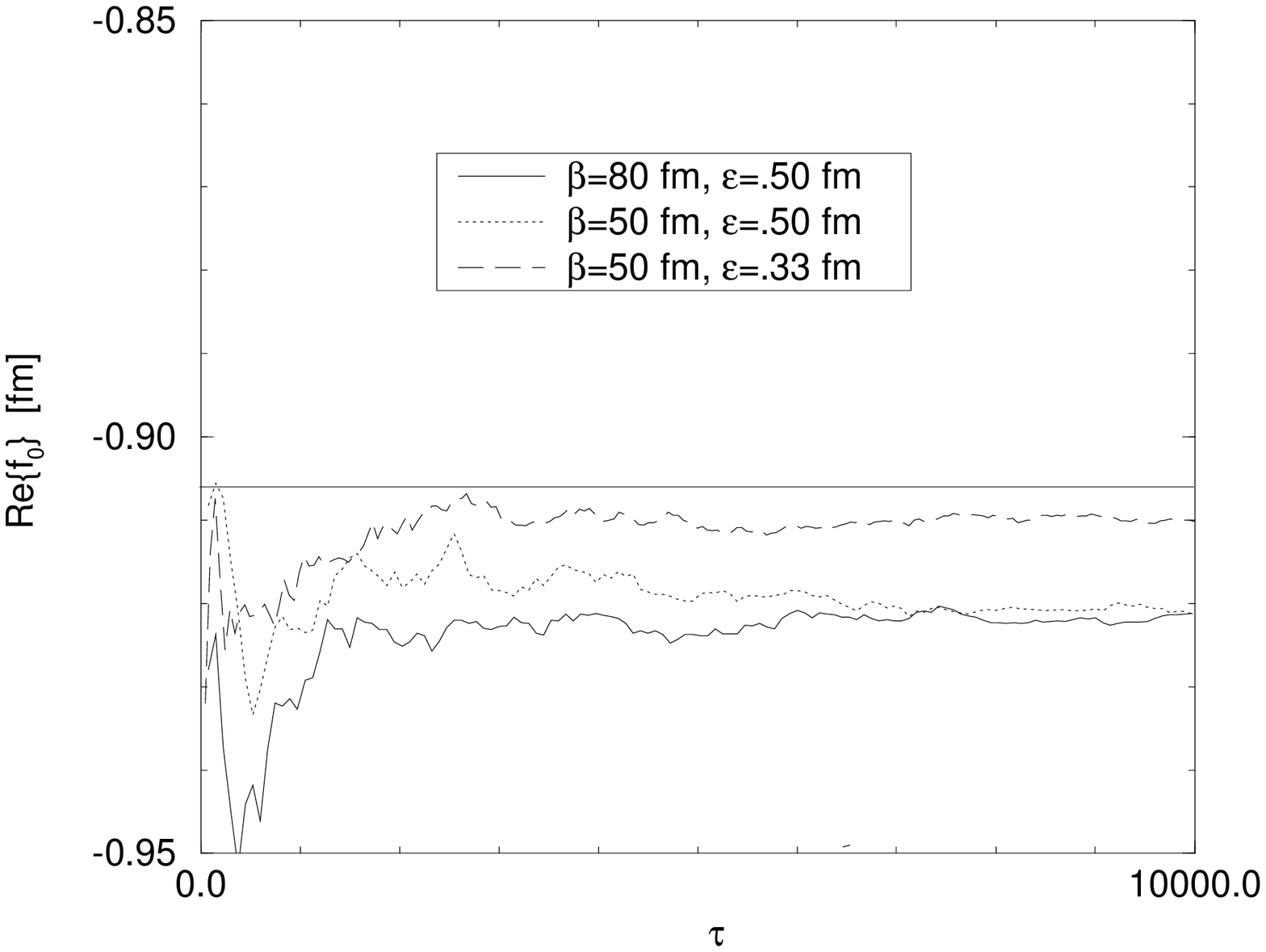}}
  \epsfysize=9.5cm
  \epsfxsize=11.5cm
  \centerline{\epsffile{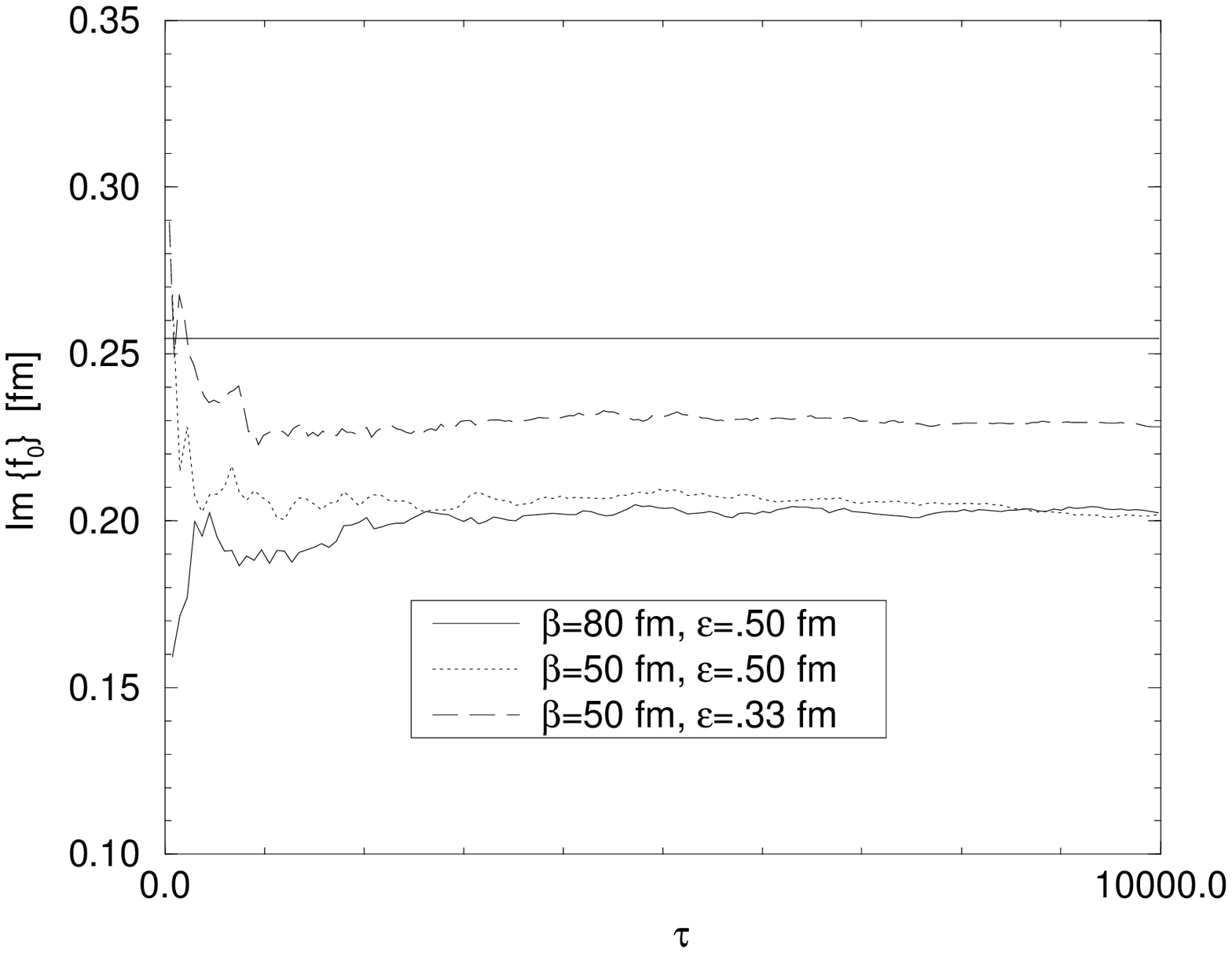}}
  \center{\parbox{11cm}{ 
  \caption{ Scattering length for complex potential 
            $V_c =(2.3,-7.0) \;\mbox{fm}^{-1}$, 
            obtained from the 'exact' Langevin algorithm with kernel. 
	    Horizontal line: exact result.
	    \label{lak_fig3} }}}
\end{figure}

\begin{figure}[p]
  \epsfysize=9.5cm
  \epsfxsize=11.5cm
  \centerline{\epsffile{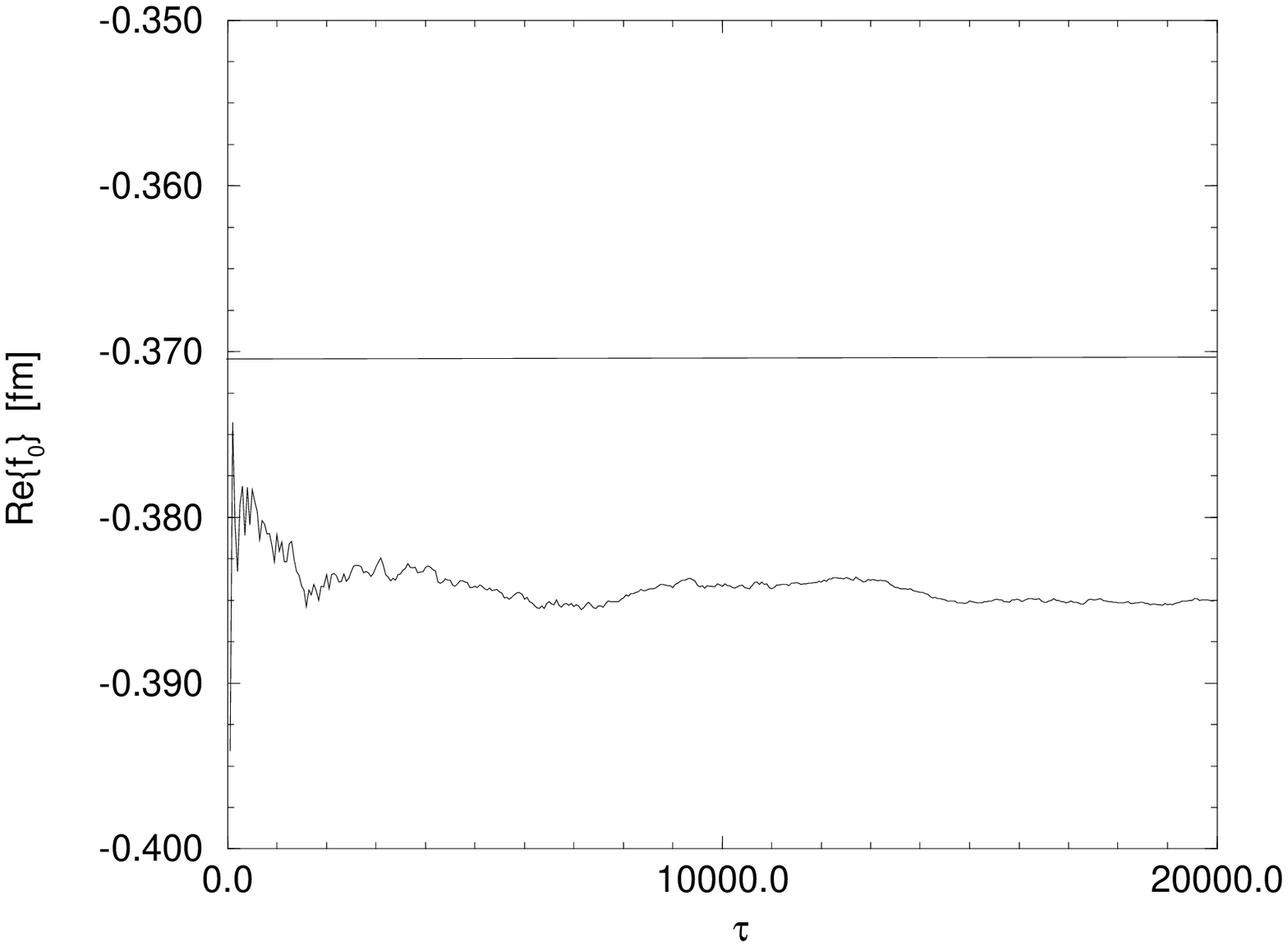}}
  \epsfysize=9.5cm
  \epsfxsize=11.5cm
  \centerline{\epsffile{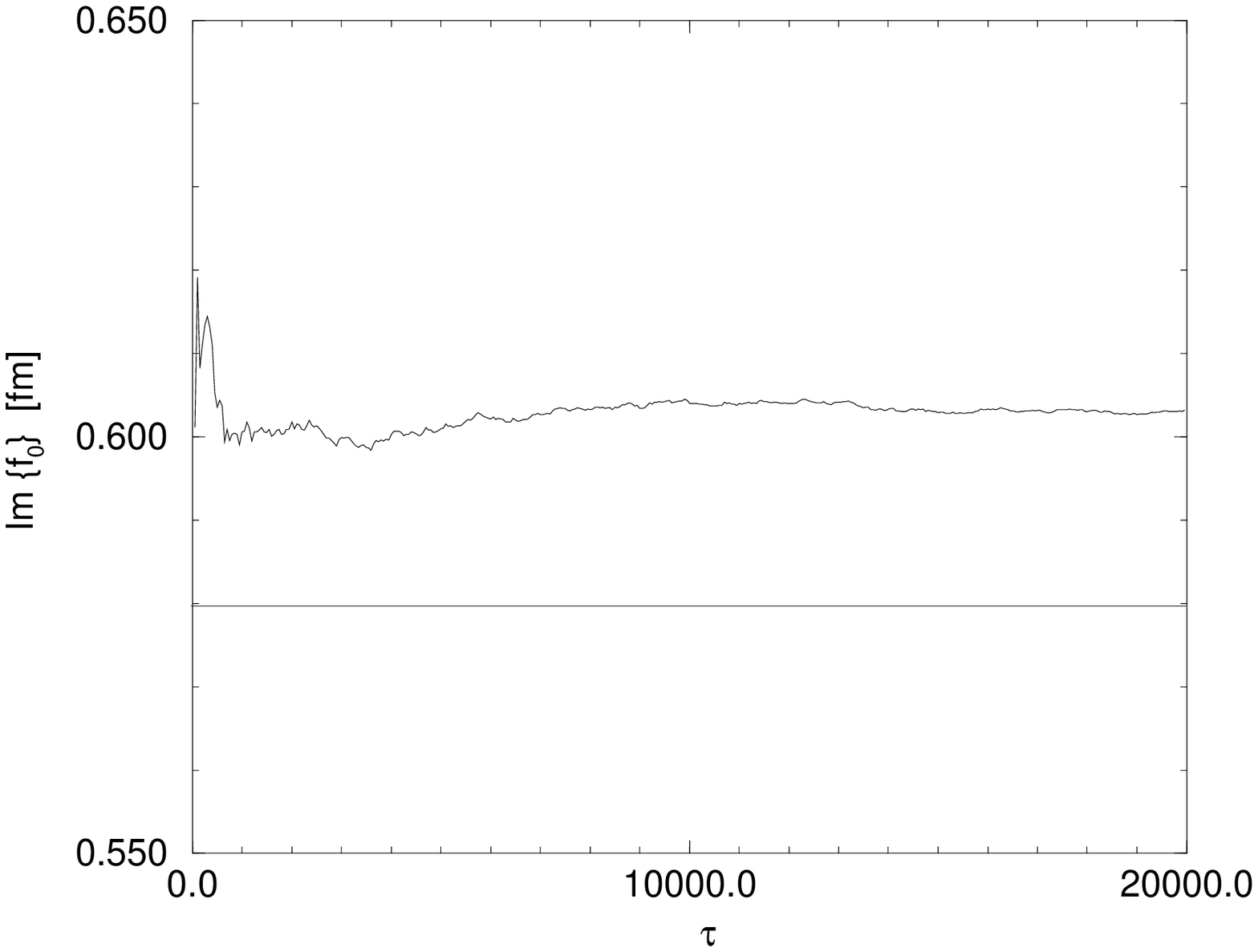}}
  \center{\parbox{11cm}{ 
  \caption{ Scattering length for complex potential 
            $V_c =(-0.3,-1.0) \;\mbox{fm}^{-1}$, 
            obtained from the 'exact' Langevin algorithm with kernel;
	    $\beta = 80\;\mbox{fm}, \varepsilon = 0.5\;\mbox{fm}$. 
	    Horizontal line: exact result.
	    \label{lak_fig4} }}}
\end{figure}

\begin{figure}[p]
  \epsfysize=9.5cm
  \epsfxsize=11.5cm
  \centerline{\epsffile{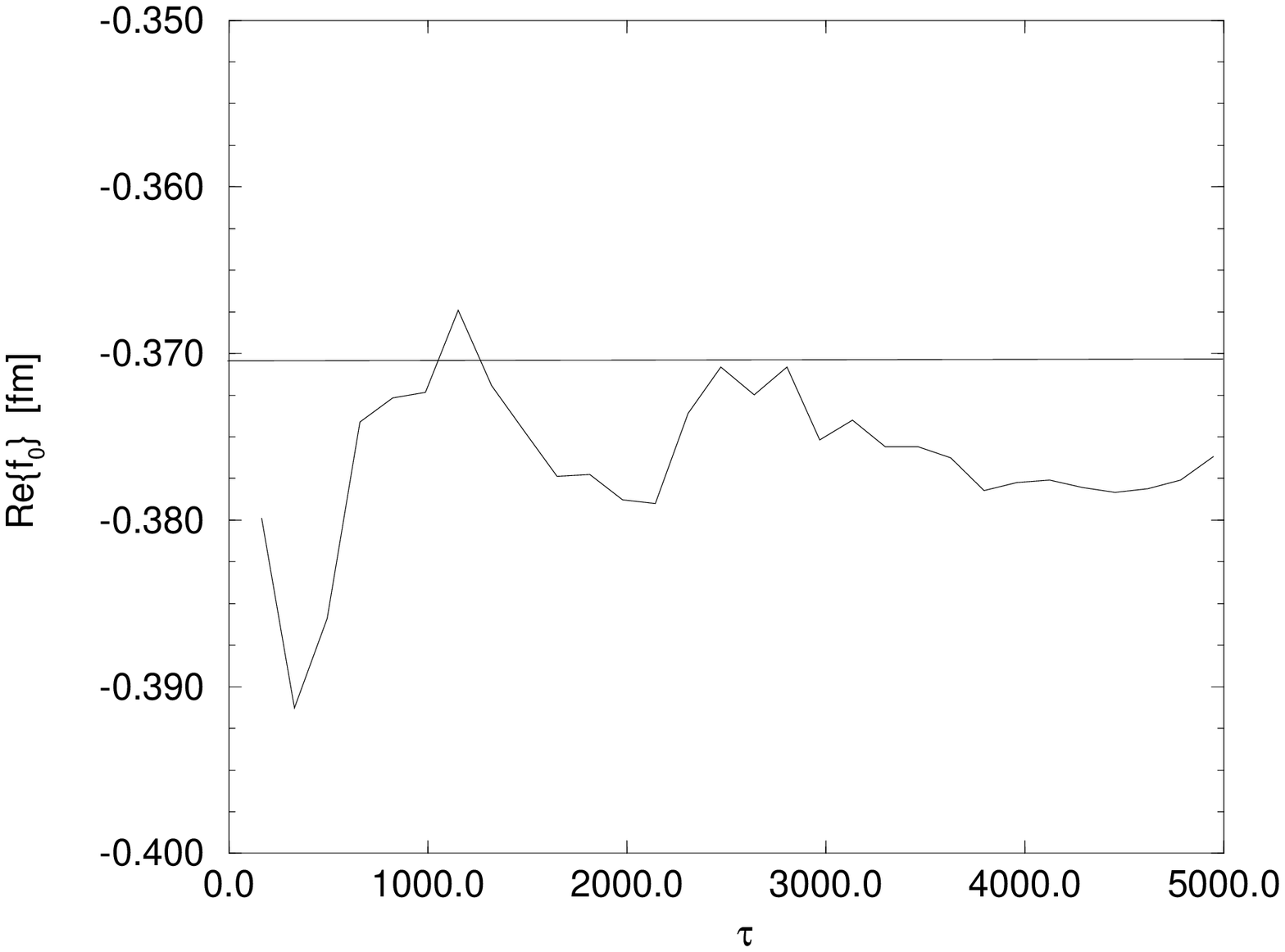}}
  \epsfysize=9.5cm
  \epsfxsize=11.5cm
  \centerline{\epsffile{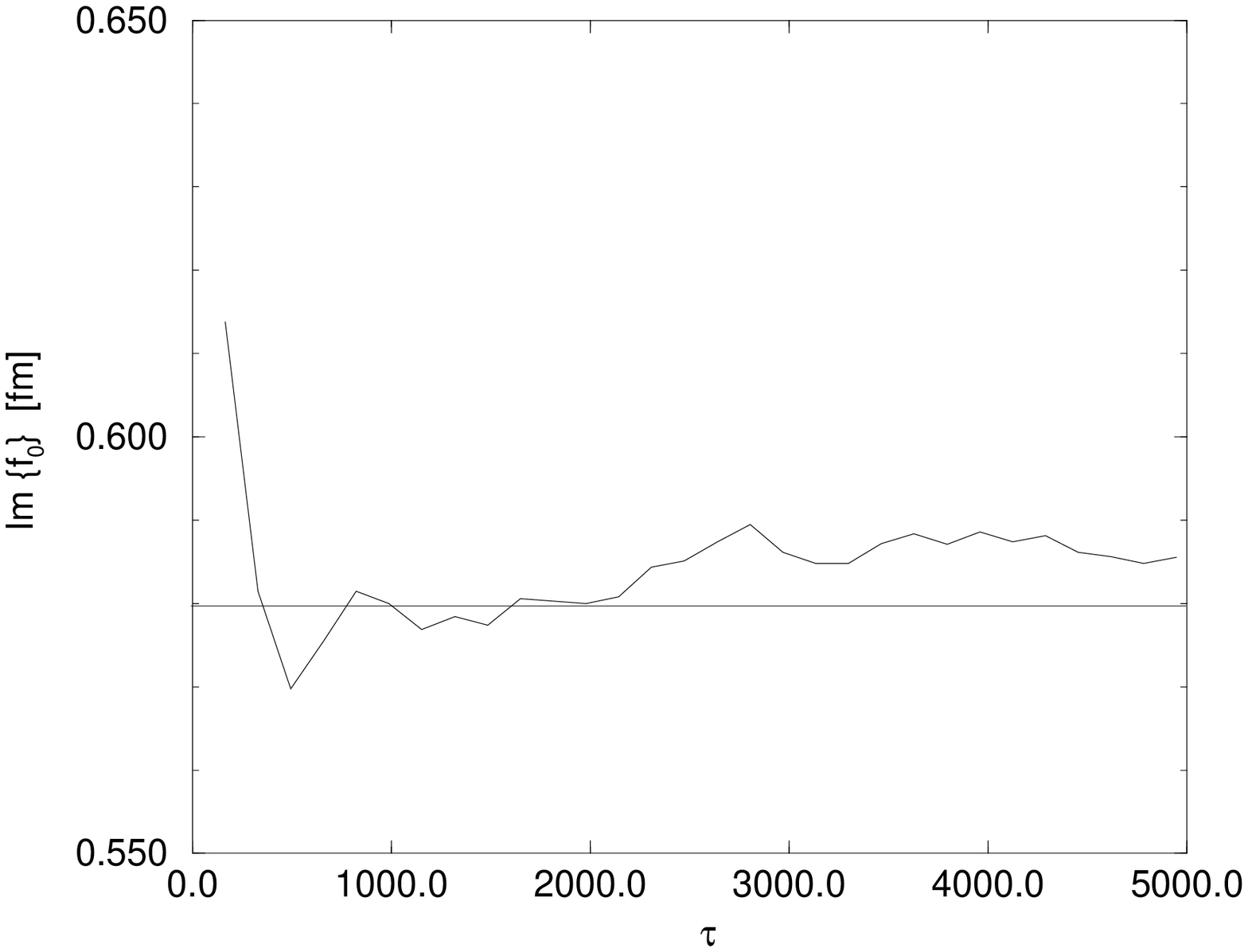}}
  \center{\parbox{11cm}{ 
  \caption{ Scattering length for complex potential 
            $V_c =(-0.3,-1.0) \;\mbox{fm}^{-1}$, 
            obtained from the 'exact' Langevin algorithm with kernel;
	    $\beta = 200\;\mbox{fm}, \varepsilon = 0.5\;\mbox{fm}$. 
	    Horizontal line: exact result.
	    \label{lak_fig5} }}}
\end{figure}

The length of ${\cal{T}}_{\alpha}$, {\it i.e.} the number of deterministic steps
$\bigtriangleup n_{\alpha}$, should be such that the paths
$\vec{R}^{(n_{I\,\alpha})}$ and $ \vec{R}^{(n_{F\,\alpha})}$ are decorrelated.
Note that the first two steps of Eqs. (\ref{hybdisc}) can be recombined to yield
the ordinary Langevin equation, Eq. (\ref{langdisc4}). In the Langevin case, a
sub-trajectory $\cal T$ then just consists of one step and the Metropolis step 
in both methods is analogous.
The Hamiltonian ${\cal{H}}$ describes a classical motion confined in a 
high-dimensional harmonic oscillator potential. In contrast to the previous 
approaches, the correlation between two paths both belonging to the same 
sub-trajectory has not an exponential but oscillatory dependence on
the time which labels successive paths. The period of this dependence is
$t_i=\frac{1}{\sqrt{\lambda_i}}$, where $\lambda_i$ again denotes the
eigenvalues of the free action matrix $\bf A_0$. To assure decorrelation of the 
smallest mode within one sub-trajectory, its length should satisfy
\begin{equation}
  t_{tra} \sim \frac{1}{\sqrt{\lambda_{min}}}
\mbox{ .}
\end{equation}
However, this does not prevent strong correlations between some other modes 
due to their different periodic time dependence of their correlation. 
Therefore in practice the length of each sub-trajectory is varied randomly 
in the vicinity of $t_{tra}$. Stability requires a number, $n_{tra}$, of steps 
in a subtrajectory that satisfies
\begin{equation}
  n_{tra} = \frac{t_{tra}}{h} \sim \sqrt{\frac{\lambda_{max}}{\lambda_{min}}} \;.
\label{vglcorhyb}
\end{equation}
Comparison of Eqs. (\ref{vglcorhyb}) and (\ref{vglcorla}) shows that the 
decorrelation in the hybrid algorithm is faster by a factor 
$\sqrt{\frac{\lambda_{max}}{\lambda_{min}}}$ compared to the standard Langevin 
algorithms; the substantial size of this factor can be seen in Fig. 
\ref{corldi}. In addition to the decrease in the required number of 
intermediate steps, a further advantage of the hybrid algorithm is that it 
needs fewer Gaussian random numbers, which are very time consuming to generate.

Eqs. (\ref{hybdisc}) represent only the simplest discretization in the time $t$.
With increasing sub-trajectory length the discretization error increases also 
the drift from the CPU energy shell in the deterministic steps of a 
sub-trajectory. This drift is measured by $\bigtriangleup {\cal{H}}$, 
Eq. (\ref{methyb}). Therefore the acceptance rate is lowered considerably and an
improvement of the discretization in Eqs. (\ref{hybdisc}) is necessary. We have 
studied examples with projection times in the range $\beta = 50$ -- $200\,
\mbox{fm}$ and stepsizes $\varepsilon = 0.1$ -- $1.0\,\mbox{fm}$. For 
sub-trajectories containing approximately $n_{tra}=50$ steps, the $O[h^2]$ 
discretization was found adequate. For sub-trajectories of about $100$ steps
a discretization procedure up to terms of order $O[h^4]$ and for 200 steps
even $O[h^6]$  was needed to lower $\bigtriangleup {\cal{H}}$, requiring 
three or nine times more steps, respectively, as with the $O[h^2]$ 
discretization. We used the higher order scheme of Ref. \cite{Cre89}, where 
Eq. (\ref{hybdisc}) is extended to $O[h^{k+2}]$ by the recursive relation
\begin{eqnarray}
  T_{k+2}(h) &=& T_k(\tilde h)\;T_k(-s\tilde h)\;T_k(\tilde h)\\
    \tilde h &=& \frac{1}{2-s}\;h\;,\quad s=2^{\frac{1}{k+1}}
\mbox{.}
\end{eqnarray}
We found that in the range of parameters shown in Fig. \ref{corldi} the hybrid
algorithm, while technically more involved, works efficiently; 
discretization to order $O[h^4]$ was sufficient.

The calculations shown in Figs. \ref{hyb_fig1} and \ref{hyb_fig2} are
obtained with the exact hybrid algorithm, without a kernel. The potential is 
$V_c\,=\,(2.3,\,-7.0)\;\mbox{fm}^{-1}$, the projection time $\beta$ and the
discretization $\varepsilon$ are varied.
As before, we found for this repulsive potential a projection time of 
$\beta = 50\,\mbox{fm}$ sufficient. In this case discretization in the 
stepsize, $h$, to order $O[h^4]$ was needed for $\varepsilon =0.33\,\mbox{fm}$,
while for $\varepsilon =0.1\,\mbox{fm}$ also terms of order $O[h^6]$ were 
necessary. The calculations of discretization, $\varepsilon=0.33\,\mbox{fm}$,
and order, $O[h^4]$, consumed about $30$ minutes of computing time, while the 
$\varepsilon=0.10\,\mbox{fm}$,  $O[h^6]$ example needed more than $5$ hours!
Note that compared to $O[h^4]$ calculations not only the number of algebraic
manipulations is increased by a factor of three, but according to 
Eq. (\ref{vglcorhyb}) and Fig. \ref{corlt}, longer projection times $\beta$ 
and smaller values of $\varepsilon$ also lead to an significant increase in
$n_{tra}$. The length of one trajectory can be estimated to range from
$t_{tra}\sim 15$ to $t_{tra}\sim 30$ (see Fig. \ref{corlt}). This means, that
Fig. \ref{hyb_fig2} contains approximately $6000$ independent samples for the
discretization $\varepsilon = 0.33\,\mbox{fm}$ and $2500$ samples for 
$\varepsilon = 0.10\,\mbox{fm}$. 
The algorithm shows a good and stable performance. Reasonable results
can be obtained for a smaller number of steps in the 'hybrid time' $t$ 
which required about $10$ minutes of computational time. Figs. \ref{hyb_fig1}
and \ref{hyb_fig2} show that the discretization is important, but a slight
discrepancy in the imaginary part still persists.

The limitations of the method show up for attractive potentials. While the 
hybrid algorithm was found to work equally well for potentials with weak 
attractive parts down to $-0.3\,\mbox{fm}^{-1}$, it starts to fail for 
more attractive values of $U_0$. 
This is due to the fact that with our standard parametrization of 
$b=0.5\,\mbox{fm}$ and $\mu=2.5\,\mbox{fm}^{-1}$, a real potential develops 
a bound state at $U_0 = -1.2\,\mbox{fm}^{-1}$. 
Fig. \ref{hyb_fig3} shows the the results for
$V_c=(-0.5,-1.5)\,\mbox{fm}^{-1}$.
The time needed for the calculations was $90$ minutes for a projection time 
$\beta = 100\,\mbox{fm}$. Attempts with 
$\beta = 200\,\mbox{fm}$ (not shown) were also found insufficient
to filter the ground state well enough and therefore the algorithm cannot be
used for attractive potentials of this strength.

Both extensions of the Langevin algorithm --- use of a kernel and hybrid 
method --- show a comparable behaviour. In our present applications of the
algorithm, all versions comprise a sequence of steps given by Eq. 
(\ref{hybdisc});
if a kernel used, $h$ is scaled by the square root of $\bf K$. For stability
reasons $h$ was chosen to be approximately five times smaller then 
minimum required.
The hybrid algorithm performs slightly faster because it needs far less
Gaussian random numbers.
For refined discretizations and long projection times, the two improved 
versions become equally time-consuming. 
It is hard to compare both extended schemes in general, since the calculation
time depends on the algebraic structure of the chosen kernel, the 
implementation of the stochastic and deterministic differential equations and 
the discretization in $h$ for the hybrid algorithm, which will vary from case 
to case.
\noindent
A kernel can also be introduced into the hybrid algorithm. We did not examine 
this possibility in detail, but tests showed that long trajectories cause 
stability problems. Therefore, one has to decrease the step size further and no 
gain in velocity is obtained. 

\section{Cluster-Algorithms}
\label{CLUST}

The hybrid and Langevin Algorithms are hindered by the fact that the
coupling of two points of a path is restricted to nearest neighbor
interactions. The physical potential is purely local and acts only within a 
small region with typical scale $b$ of the potential width. Outside this region 
one is left with the free action. Hence, strongly oscillating paths are 
suppressed and long range correlations of order $\beta$ dominate. 
Only the end point contributions containing the logarithm of the potential 
prevent the paths from drifting away to infinity.  Hence, the
displacement of a path on a large scale proceeds very slowly. 

To improve the computational efficiency it is desirable to speed up these slow 
long-ranged modes without neglecting small fluctuations introduced by the 
short range of the potential.
One would like a non-local or collective-mode updating based on information 
over a larger range of coordinates. We considered already one example of a such 
an algorithm in Section \ref{LAHYB}, where we introduced a kernel into the Langevin 
algorithm in order to deal with modes of the path motion and not just 
individual points when generating new paths. This approach suffered from the 
fact that we were not able to diagonalize the complete action. A coupling of the
different modes through the potential remained.

A similiar situation often occurs for spin systems with nearest neighbor
interactions. It is well known that large-scale modes of spin waves lead 
to sizeable slowing down. To remedy this, Swendsen and Young introduced a new 
type of collective algorithm for Potts models \cite{Swe87}. Additional auxiliary
degrees of freedom are introduced into the original model: groups of correlated 
spins are treated as new degrees of freedom which are called 'clusters'. 
Changes are applied to these collective objects and not to single physical 
degrees of freedom, the indivual spins, which leads to a much more efficient and 
faster way to deal with such systems. Subsequently many related applications 
have been developed for Ising models, O(N) spin models \cite{Wol89} and lattice 
gauge theories \cite{Eve91}, commonly referred to as 'cluster algorithms'.
There is considerable freedom in how to define a cluster. Some special aspect 
of the physical problem has to be used in each case to define the appropriate 
collective variables.

In this section, we extend the idea of clusters from 
bound systems to scattering problems.
Adjacent coordinate points of a path having similiar positions will under 
certain 
conditions be considered as clusters and dealt with collectively. The size of 
these clusters must reflect the two different length scales of the physical 
system under consideration, the short range of the potential and the large 
exterior region. There are two basic steps of the cluster algorithm which
are carried out to a given path $\vec R$ to generate a new path, $\vec R'$.
They are repeated until satisfactory convergence of the matrix element of
the chosen obserevable is achieved:

\begin{enumerate}
\item
The first step consists of grouping the points of a path into clusters. 
As will be discussed in detail below, clusters are designated in a way that 
will favor a decrease in kinetic energy when updating the path by
moving these clusters in step 2.

\item
Once the clusters have been defined, a number of them are reflected on a 
randomly oriented plane ${\cal{E}}$, yielding a new path $\vec{R}'$. This new 
path consists thus of some clusters which were reflected and some which did
not change. The decision wether a cluster is reflected or not is chosen
in way way which favors the decrease of the potential energy. 
To preserve ergodicity, the position of ${\cal{E}}$ is changed each time this
step is carried out.
\end{enumerate}

\noindent
Both steps together lead to a splitting into smaller clusters in regions
where the scale of the potential becomes important and to larger clusters
where the influence of the kinetic part of the action is dominant.
This is a great advantage compared to {\it e.g.} the methods in Section 
\ref{RWA}. There are two differences between our method and the application 
of cluster methods to spins of fixed length. First, it is in our case
not sufficient to reflect the coordinate points through a plane which 
contains the origin; instead
the position of the plane is varied throuhgout the entire space.
Second, the scattering potential influences the reflection probability of
clusters and therefore scales the effective size of reflected clusters. 

\subsection{Introduction of auxiliary variables and collective
updating of paths}

To group points of a given path $\vec{R}$ into clusters, first auxiliary 'bond'
variables $b_i$ are defined for each pair of adjacent points 
$\vec{r}_i$ and $\vec{r}_{i-1}$. The $b_i$ can take on the values 
$0$ or $1$. The whole set of these variables for a path is denoted by $B$.
A cluster $\cal{C}$ is a set of neigbouring coordinates $\vec{r}_i$ which are
connected by non-zero bonds $b_i$
\begin{equation}
  {\cal{C}}^s \equiv [\vec{r}_i,\vec{r}_{i+1},......,\vec{r}_{j}]
\mbox{,}
\end{equation}
where  $s$ stands for the set of indices $i,i+1,...,j$ of coordinate  points
contained in ${\cal{C}}^s$.

To decide which value should be assigned to a 'bond' variable $b_i$, we 
use a plane ${\cal{E}}$, characterized by a normal vector $\hat{n}$ and its 
distance $\xi$ to the origin. We then consider the point 
$\vec{r}_{i}^{\,\star}$, which is obtained by  reflecting $\vec{r}_{i}$ 
through the plane ${\cal{E}}$, {\it i.e.}
$\vec{r}_{i}^{\,\star}\equiv\vec{r}_{i}^{\,\star}[\vec{r}_{i},{\cal{E}}]$.
The probablitiy of setting the bond variable $b_i$ to zero is then given by
\begin{equation}
    P_{nb,i}  =  \min
    \left[ 1,\exp{\left(-\Delta {\cal{S}}_{0}[\vec{\Delta}_i]\right)}\;\right] ,
    \label{nobond}\\
\end{equation}
where $\Delta{\cal{S}}_{0}$ is determined by the free part of the action,
\begin{eqnarray}
    \Delta{\cal{S}}_{0}[\vec{\Delta}_{i}] &=& \frac{\mu}{2\varepsilon} 
    \left[ \vec{\Delta}_{i}^{\star\,2}-\vec{\Delta}_i^2 \right] ,\\
    \vec{\Delta}_{i}          &=&   \vec{r}_{i} - \vec{r}_{i-1} ,\\
    \vec{\Delta}_{i}^{\star}  &=&   \vec{r}_{i}^{\star} - \vec{r}_{i-1}
\label{implic}
\;\mbox{.}
\end{eqnarray}
$P[B\mid\vec{R},{\cal{E}}]$ is then conditional probability to obtain a
set of bonds $B$ for given plane ${\cal{E}}$ \cite{Wol89}, \cite{Wie95},
\begin{eqnarray}
    P[B\mid\vec{R},{\cal{E}}]   &=&
    \prod_{i=2}^{N+1} \left[
       P_{nb,i} \; \delta_{b_i,0}+ (1-P_{nb,i}) \;\delta_{b_i,1}
    \right]\;,
 \label{condit}\\
    B &=& [b_2,b_3,....,b_{N+1}]\;\quad
    \vec{R}=[\vec{r}_1,\vec{r}_2,....,\vec{r}_{N+1}]
\;\mbox{.}
\end{eqnarray}

Taking into account these additional variables, the probability
distribution $P_{\cal{S}}[\vec{R}]$ can be extended to the combined 
distribution $\cal{P}$
\begin{equation}
    {\cal{P}}[\vec{R},{\cal{E}},B]
    =
    P[B\mid\vec{R},{\cal{E}}]  
    P_{\cal{S}}[\vec{R}] \; \, P_{{\cal{E}}}[\hat{n},\xi] \;\,
\mbox{.}
\label{combdist}
\end{equation}
Here $P_{\cal{S}}[\vec{R}]$ is as defined before in Eq. (\ref{usalact}) and
$P_{\cal{E}}[\hat{n},\xi]$ describes the probability of finding a plane
$\cal{E}$ with normal vector $\hat{n}$ and at a distance $\xi$ to the origin. Its form
will be specified below; in general it does not depend on the path $\vec{R}$.
One has by construction
\begin{equation}
  P_{\cal{S}}[\vec{R}] = \int d\,\xi \int d\Omega \;
  \sum_{\{b\}}\;\, {\cal{P}}[\vec{R},{\cal{E}},B]  
\;\mbox{,} 
\end{equation}
where integration over $d\,\Omega$  extends over all possible orientations 
of $\hat{n}$. In terms of the extended probality distribution, 
${\cal{P}}[\vec{R},{\cal{E}},B]$, we can re-write the expectation value of 
the modified observable ${\cal{O}}[\vec{R}]$ as
\begin{equation}
\langle {\cal{O}} \rangle =
\int {\cal{D}}[\vec{R}] \;{\cal{O}}[\vec{R}] \;P_{\cal{S}}[\vec{R}] =
\int{\cal{D}}[\vec{R}]\int d\xi \int d\Omega \, \sum_{\{b\}} \;
{\cal{O}}[\vec{R}] \; {\cal{P}}[\vec{R},{\cal{E}},B]
\;\mbox{.}
\end{equation} 

We now have to specify how we update the path $\vec{R}$ by
reflecting entire clusters on the plane $\cal{E}$. We denote
by ${\cal{C}}^{s\star}$ the cluster one obtains by reflecting ${\cal{C}}^s$
on $\cal{E}$. The probability for a cluster to be reflected is given by 
\begin{eqnarray}
   T[{\cal{C}}^{s\star}\mid{\cal{C}}^{s}] &=& \alpha\;\mbox{min} 
   \left[ 1, \exp{\left(-\Delta{\cal{S}}_{U}[{\cal{C}}^{c}]\right)} \;\right] 
   \;,
\label{transprob} \\
   \Delta{\cal{S}}_{U}[{\cal{C}}^{c} ] &=&
   {\cal{S}}_{U}[{\cal{C}}^{c \star}] -{\cal{S}}_{U}[{\cal{C}}^{c}]
\mbox{.}
\end{eqnarray}
Here $0<\alpha<1$ is a real parameter; its choice will be discussed below.
The quantity ${\cal{S}}_{U}[{\cal{C}}^s]$ is the contribution of the potential
to the action for the points within the cluster ${\cal{C}}^s$. Clusters which
contain the endpoints also contribute $\ln{v}$ according to 
Eq. (\ref{endpunkt}).
As can be seen from the above prescription, the probablity for reflecting
a cluster is chosen such that a transition to clusters with a lower potential
energy is favored.
Since the new path  $\vec{R}'$ in general contains reflected 
clusters, ${\cal{C}}^{s \star}$, as well as unreflected ones, ${\cal{C}}^{s}$,
the total transition probability for given $\cal{E}$ and $B$ reads
\begin{equation} 
  {\cal{T}}_{{\cal{E}}B}[\vec{R}'\mid\vec{R}] =
  \prod_{s=1}^{N_{r}}\, T[{\cal{C}}^{s\star}\mid {\cal{C}}^{s}] \; 
  \prod_{n=1}^{N_{nr}}
  \left\{1- T[{\cal{C}}^{n\star}\mid {\cal{C}}^{n}] \;  \right\}
 \mbox{.}
\end{equation}
The first factor in the above equation contains only the $N_{r}$ clusters
which are reflected to transform $\vec{R}$ into $\vec{R}'$, while the second
is due to the $N_{nr}$ clusters which retain their position.  
After having updated a path $\vec{R}$ to $\vec{R}'$, the procedure is then repeated
with now $\vec{R}'$ as starting path to obtain a new $\vec{R}'$ {\it etc.},
until a sufficient number of paths is obtained for the evaluation of the 
observable.  

It remains to be shown that the transition from $\vec{R}$ to $\vec{R}'$
satisfies the criterion of detailed balance,
\begin{eqnarray}
  \frac{ {\cal{P}}[\vec{R}',{\cal{E}},B] }
   { {\cal{P}}[\vec{R},{\cal{E}},B] } &=&
   \frac{{\cal{T}}_{{\cal{E}}B}[\vec{R}'\mid\vec{R}]} 
   {{\cal{T}}_{{\cal{E}}B}[\vec{R}\mid\vec{R}']}
\label{detbal}
\mbox{.}
\end{eqnarray}

\noindent
Making use of the general relation
\begin{equation}
  \frac{\mbox{min}[1,e^{-\kappa}]}{\mbox{min}[1,e^{\,\kappa}]}
   =
   e^{-\kappa} 
\mbox{,}
\label{minquot}
\end{equation}
one finds for the r.h.s. of Eq. (\ref{detbal})
\begin{equation}
  \frac{ {\cal{T}}_{{\cal{E}}B}[\vec{R}'\mid\vec{R} ] }
       { {\cal{T}}_{{\cal{E}}B}[\vec{R} \mid\vec{R}'] } =
	  \exp{\left[ -\sum_{s=1}^{N_{r}}
	  \Delta {\cal{S}}_{U} [{\cal{C}}^{s}] \right]}
\label{potwolff}
\mbox{.}
\end{equation}
The sum runs only over those $N_{r}$ clusters which are reflected to make
the transition from $\vec{R}$ to $\vec{R}'$. The l.h.s. of Eq. (\ref{detbal}) 
can be re-written by making use of Eq. (\ref{minquot}) to yield 
\begin{eqnarray}
     \frac{ {\cal{P}}[\vec{R}',{\cal{E}},B] }
     { {\cal{P}}[\vec{R},{\cal{E}},B] } &=&
     \exp{\left[-\Delta{\cal{S}}[\vec{R}',\vec{R}] \right] } \;\;
     \frac{P[B\mid\vec{R}',{\cal{E}}]}
          {P[B\mid\vec{R}, {\cal{E}}]} ,
    \label{kombquot}
\end{eqnarray}
where the exponent is given by
\begin{eqnarray}
     \Delta{\cal{S}}[\vec{R}',\vec{R}] &=&
     \tilde{\cal{S}}[\vec{R}']-\tilde{\cal{S}}[\vec{R}]=
     \sum_{t=1}^{N_{tr}} \Delta{\cal{S}}_{0}[\vec{\Delta}_{t}]+	  
     \sum_{s=1}^{N_{r}} \Delta{\cal{S}}_{U}[C^{s}]\;
     \label{quotwolff}
\mbox{.}
\end{eqnarray}
The index $t$ in Eq. (\ref{quotwolff}) runs only over
those $N_{tr}$ junctions $\vec{\Delta}_{t}$ which are changed into
$\vec{\Delta}_{t}^{\star}$; junctions which are contained in one cluster
and junctions between two clusters which are reflected simultaneously do not 
contribute. 
The remaining term in Eq. (\ref{kombquot}) is obtained from Eqs.
(\ref{nobond}), (\ref{condit}), which yield
\begin{equation}
   \frac{ P[B\mid\vec{R},{\cal{E}}] }
        { P[B\mid\vec{R}',{\cal{E}}] } =
   \exp{\left[ -\sum_{t=1}^{N_{tr}} \; \Delta{\cal{S}}_{0}[\vec{\Delta}_{t}] 
   \right]}
  \label{difwolff}
\;\mbox{.}
\end{equation}
Combining Eqs.  (\ref{kombquot}), (\ref{quotwolff}) and (\ref{difwolff}) and 
inserting the result into the l.h.s. in Eq. (\ref{detbal}),
it is readily seen that the detailed balance condition is satisfied.

\subsection{Scaling of the Cluster Size and Choice of Parameters}

Important for the convergence and accuracy of the algorithm are the 
size and relative distribution of the clusters. Too many large
clusters leads to inefficient probing of the potential region,
while too many small clusters slow down the algorithm. The cluster distribution
depends on the function $P_{\cal{E}}$ and the coefficient 
$\alpha$ in Eq. (\ref{transprob}).  We will discuss these two aspects in this 
section. 

Geometrical considerations show that two 
neighbouring points belong to different clusters if the line connecting
these points is intersected by the plane $\cal{E}$. Furthermore, it is
readily seen that the size of the path segment contained 
in a cluster and the relative distribution of clusters with different sizes
depend on the average distance of the plane $\cal{E}$ to 
the region of classical paths where there are many points of a path.
A distribution $P_{\cal{E}}$, which mainly places the plane into
these physically preferred areas, results in many intersections and thus a 
large number of small clusters. If on the other hand the plane $\cal{E}$ is 
placed at large distances from these regions, it is likely to intersect the 
path only at a few points, resulting in relatively few, large clusters. 
Thus, while  at first sight the setting of bonds and grouping of points into 
clusters seems to entirely depend on the kinetic energy
through the condition in Eq. (\ref{nobond}), the above considerations show that
also the potential enters into the eventual distribution of clusters through
the action $\tilde{\cal{S}}$: If the 
algorithm is equilibrated and starts to converge, the paths will be 
denser in regions of local minima of the real action $\tilde{\cal{S}}$;
reflection planes placed into these regions will lead to a splitting into 
many small clusters that allow for an accurate probing of the details of the 
potential.

The location of the plane $\cal{E}$ is important for another reason.
If  $\cal{E}$ is placed between two classically allowed regions,
which are separated by a repulsive potential well, the reflection
makes the fast interchange of path segments between both minima possible. 
A very large number of intermediate steps would be needed if, instead of 
clusters, we dealt with individual points as in the other algorithms.
With the Langevin method, for example, the path first 
has to get through the repulsive potential barrier step by step before it 
reaches the neighboring minimum. This advantage of the cluster method is 
thus especially advantageous for scattering problems with a more complicated
target structure such as many-body targets. Projectile paths will not remain 
trapped in regions separated by potential barriers, but move quickly and 
probe efficiently the entire target region. 

We now discuss the choice of the parameters defining the 
distribution $P_{\cal{E}}[\hat{n},\xi]$ of the reflection planes.
The orientation of the plane ${\cal{E}}$ is defined by the normal vector 
$\hat{n}$; in spherical coordinates $\hat{n}=(\cos{\psi}\;\sin{\theta},
  \sin{\psi}\;\sin{\theta},\cos{\theta})$.
Since we are considering scattering problems for spherically symmetric 
potentials no direction in space should be preferred and therefore the
variables $\cos{\theta}$ and $\psi$ are randomly distributed over the
intervals 
\begin{equation}
 \; \cos{\theta} \in [-1,1]
  \;\mbox{,}\quad \psi \in [0,2\pi] .
\end{equation}
\noindent
The distance of the plane to the origin, $\xi$, is taken to be of Gaussian type,
\begin{equation}
  P[\xi]=\left[\frac{1}{2\pi\,w^2}\right]^{\frac{1}{2}} 
      \exp{\left[ -\frac{1}{2\,w^2} \xi^2 \right]} 
\label{breitew}\mbox{.}
\end{equation}
If the projection time is large enough such that it satisfies 
$b^2 << \frac{\beta}{\mu}$, it can be shown that one has for the mean square 
distance of the midpoint of the paths to the origin, 
\begin{equation}
  \langle d^2_M\rangle  \approx 3\frac{\beta}{4\mu}
\mbox{,}
\end{equation}
while the mean square distance of the endpoints is
\begin{equation}
  \langle d^2_{1,N+1}\rangle \approx 3\,b^2
\mbox{.}
\end{equation}

For repulsive potentials the choice $w\sim b$ for the width of the
distribution leads to the best results. By concentrating the position 
of the reflecting plane in this region around the origin, one avoids
that large segments in the exterior region, typically located at a
distance $\sqrt{\langle d^2_M\rangle} >> b$ are reflected into the 
potential region.
Avoiding these paths, which are practically all rejected, speeds up the
algorithm and reduces fluctuations. At the same time, 
this positioning of the reflecting planes takes into account the detailed
shape of the potential in updating a path. On the other hand, taking $w$ 
too small would lead to too many almost symmetrical reflections, as the
planes are concentrated close to the symmetry center of the potential. 
The updated paths then only differ in the kinetic energy, while the
contribution from the potential remains essentially unchanged. As the
observable we evaluate only depends on the absolute value of the distance
of the points of a path to the origin, such updates would only increase
the computational time for the algorithm.
In the examples below with a repulsive real part of the potential, 
which were all
carried out with a projection time of $\beta=80\,\mbox{fm}$, we chose a
width of $w= 0.5 \,\mbox{fm}$.

Attractive potentials require a much longer projection time. 
As a result, the midpoint of the paths moves further away from the 
origin and a larger distribution parameter $w\sim O[d_M / 2]$ should be chosen,
yielding planes ${\cal{E}}$ located about halfway between
the midpoint of a path and the origin. Exterior segments are then 
often reflected into the energetically favored potential region;
the average time a cluster remains in the interior region is short, 
since the next path-update will tend to move it out of that region again.
Symmetric reflections through the origin are suppressed with this
choice for $w$. A disadvantage of this choice results
for the endpoints of the path, which should remain in the potential
region: a large $w$ leads to reflections out of this region and
thus to configurations with a low acceptance. We used {\it{e.g.}} for a
projection time of $\beta=300\,\mbox{fm}$ a width of $w=2.0\,\mbox{fm}$,
while for $\beta=500\,\mbox{fm}$ a choice of $w\approx\,4.0\,\mbox{fm}$ yielded 
the best results.

The above discussion shows that one has to balance competing effects
when choosing the distribution of the planes: If $w$ is chosen too small
the tendency exists that many clusters are reflected to a position 
at the same distance to the origin; a too large $w$ leads to motions
of the endpoints with a low acceptance. Both extremes significantly lower
the convergence speed of the algorithm. Clearly a different shape of the
surface on which one reflects is also possible to avoid these problems and to
further improve the efficiency of the algorithm; we do not pursue this 
here.

The probability for the reflection of a cluster depends on the
potential and on the choice of the parameter $\alpha$ in Eq. (\ref{transprob}),
which also has an important influence on the relative distribution of 
clusters of different sizes. The decision if a cluster should be reflected is 
made independently for each cluster. 
If we take $\alpha\approx 1$, Eq. (\ref{transprob}) shows that a single cluster
is then practically always reflected if the potential energy is not 
increased by this operation. Therefore in regions with no structure 
due to the potential, it 
is then likely that many adjacent clusters are reflected simultaneously and 
'effective clusters' of large size are formed. Within the range of the 
potential small clusters are formed since the reflection probability, 
Eq. (\ref{transprob}), now changes over 
short distances due to changes of the potential.
The algorithm then provides a fast collective movement in regions where 
the potential vanishes and leads at the same time to a fine scaling when
the path crosses the relatively small interaction region. Therefore the choice 
of $\alpha\approx 1$ is especially suited for scattering problems.
On the other hand for $\alpha = 1$ a cluster would always be reflected and
effective clusters would tend to stay together in the exterior. This leads
to a slowing down due to insufficient decorrelation of modes with 
a longer range. We obtained the best results with $\alpha \approx 0.75$. 
In choosing the parameters to optimize the cluster motions in the exterior
we do not have to worry about how this affects the modified observable. 
In our case the observable depends on the imaginary part of the potential and
has no structure outside of the range of the potential. The choice of $\alpha$ 
can thus entirely be used to speed up the algorithm in the 
large exterior region.

Fig. \ref{eval} demonstrates for a repulsive potential how the algorithm 
selects automatically the appropriate scale for the clusters. Small 
clusters, necessary to probe the potential region in detail, can be 
found at short distances from the origin, while larger clusters can be 
found at larger distances. That the cluster size does
not grow stronger at very large distances from the origin is due to the
fact that the endpoints of the path are restricted to lie in the 
potential region. Too large clusters lead to too few acceptable paths.

\subsection{Examples}

\begin{figure}[p]
  \epsfysize=9.5cm
  \epsfxsize=11.5cm
  \centerline{\epsffile{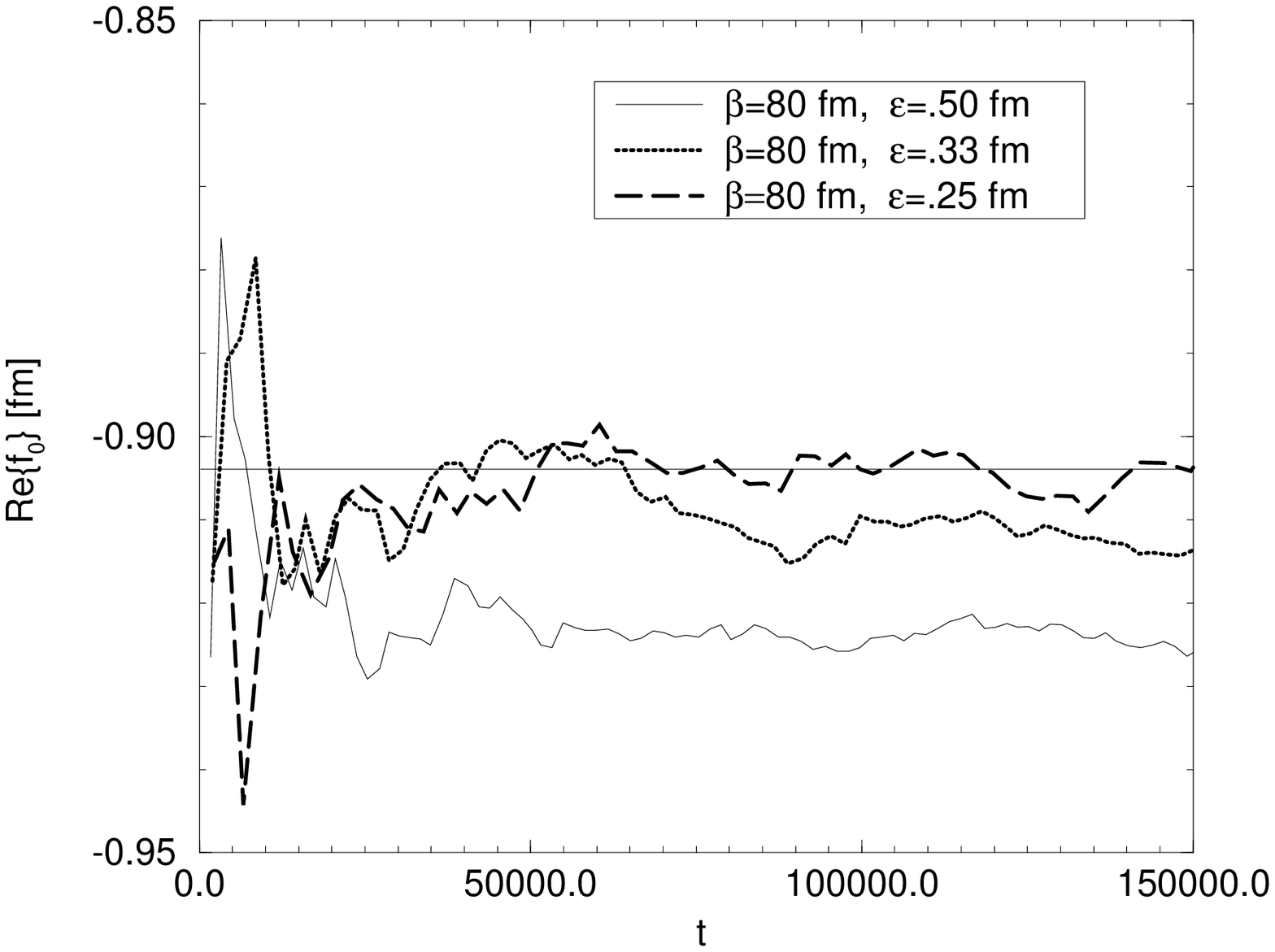}}
  \epsfysize=9.5cm
  \epsfxsize=11.5cm
  \centerline{\epsffile{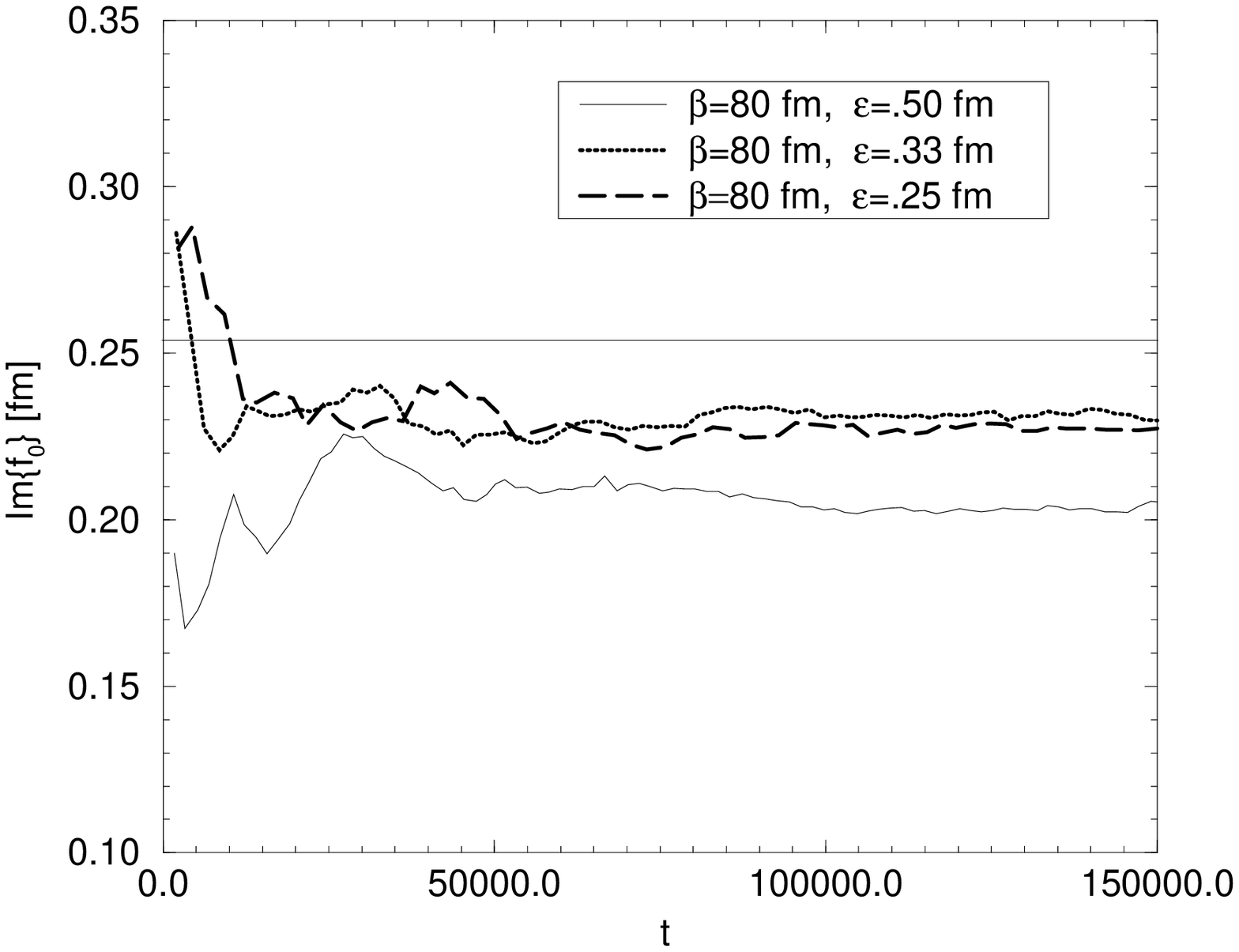}}
  \center{\parbox{11cm}{ 
  \caption{ Scattering length for complex potential 
            $V_c =(2.3,-7.0) \;\mbox{fm}^{-1}$, 
            obtained from the 'exact' hybrid algorithm. 
	    Horizontal line: exact result.
	  \label{hyb_fig1} }}}
\end{figure}

\begin{figure}[p]
  \epsfysize=9.5cm
  \epsfxsize=11.5cm
  \centerline{\epsffile{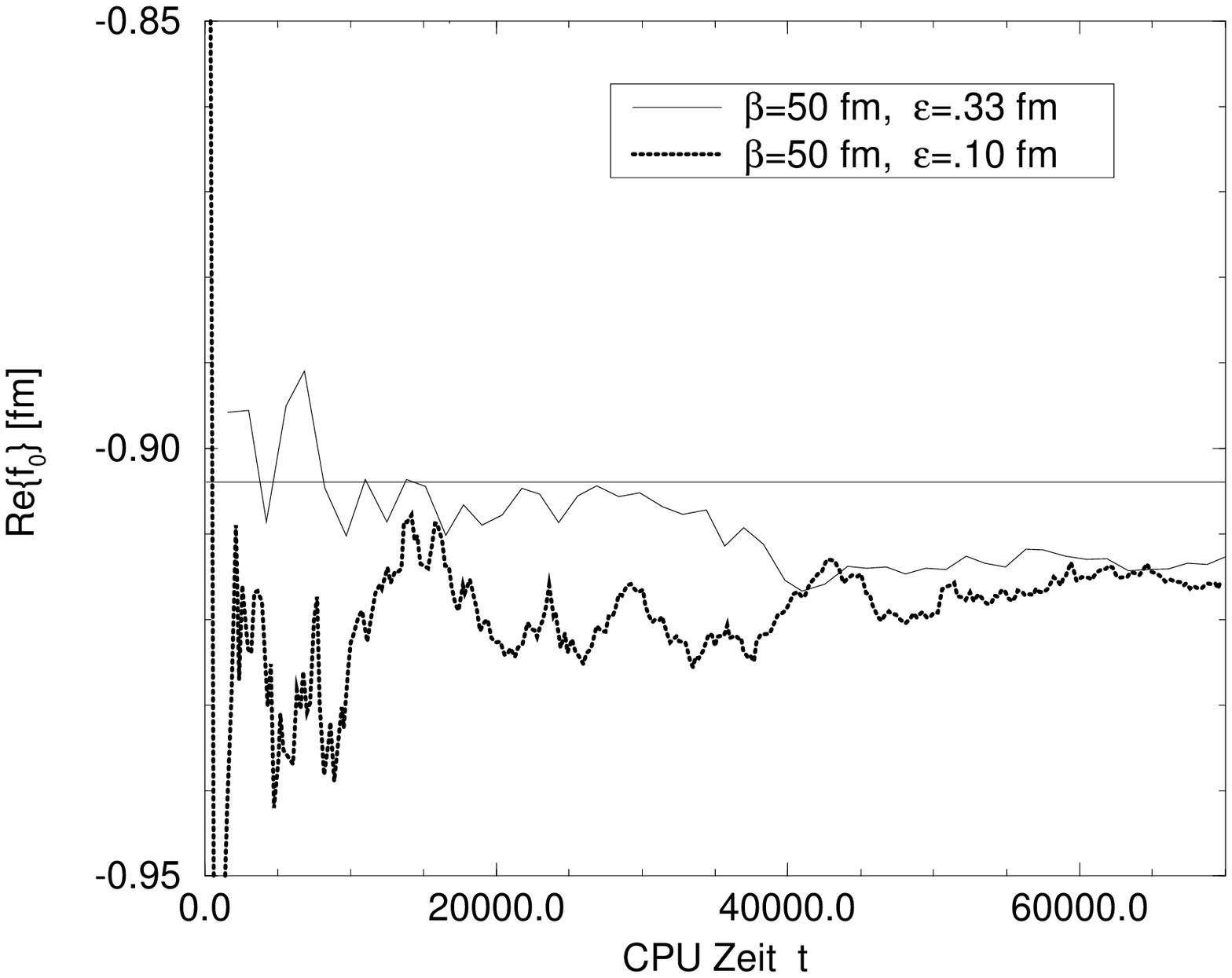}}
  \epsfysize=9.5cm
  \epsfxsize=11.5cm
  \centerline{\epsffile{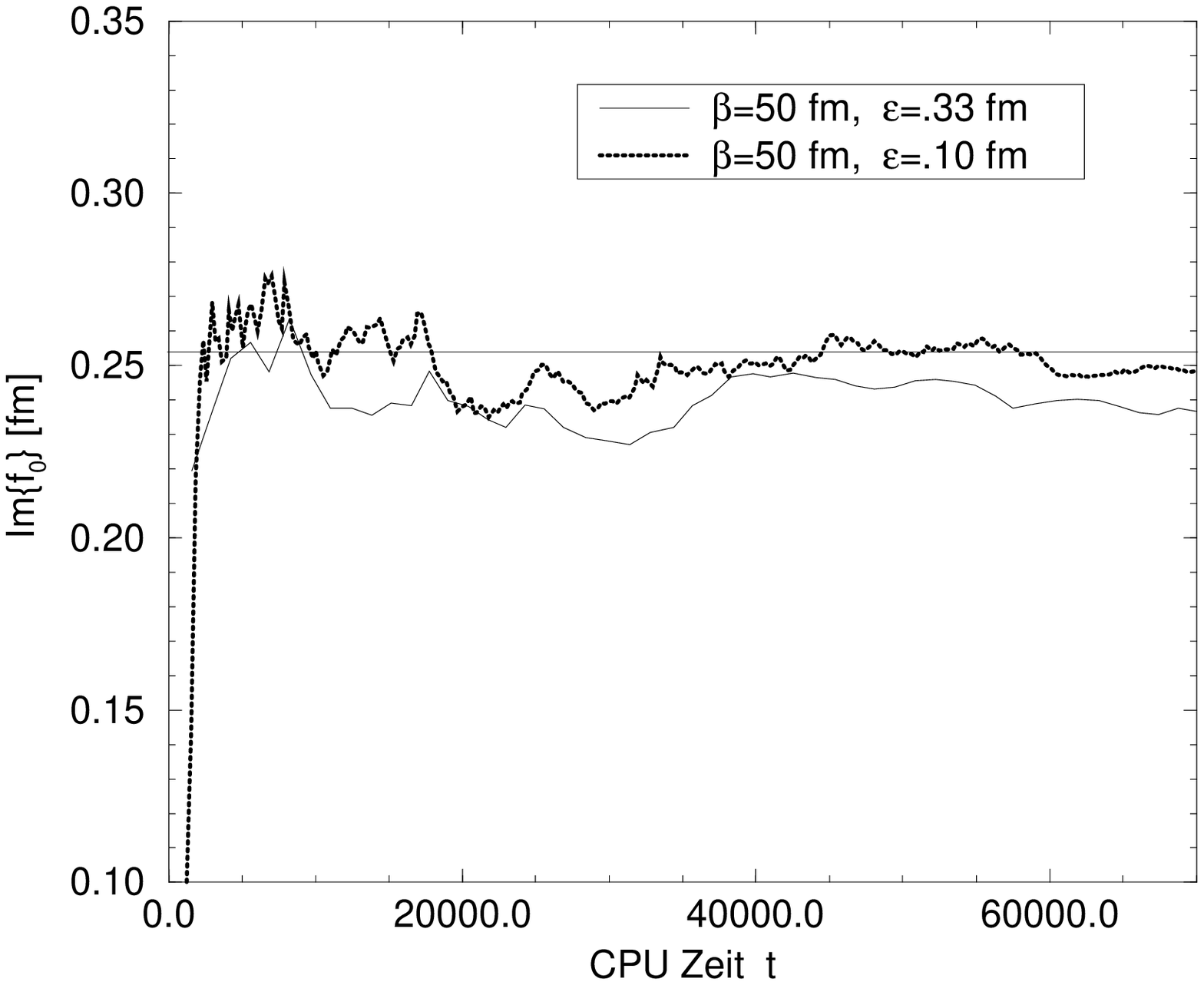}}
  \center{\parbox{11cm}{ 
  \caption{ Scattering length for complex potential $V_c =(2.3,-7.0)
	\;\mbox{fm}^{-1}$, obtained from the 'exact' hybrid algorithm. 
	    Horizontal line: exact result.
	  \label{hyb_fig2} }}}
\end{figure}

\begin{figure}[p]
  \epsfysize=9.5cm
  \epsfxsize=11.5cm
  \centerline{\epsffile{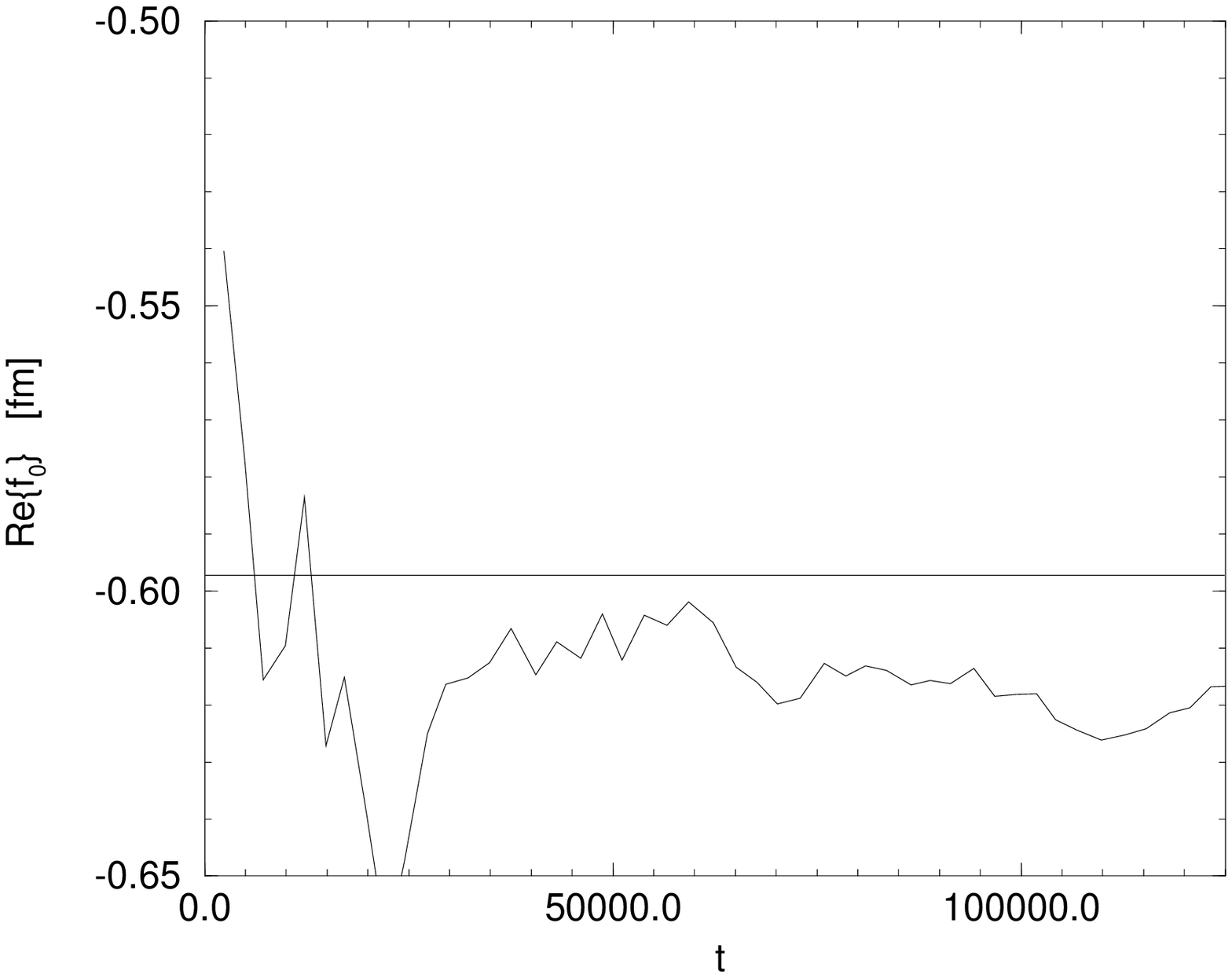}}
  \epsfysize=9.5cm
  \epsfxsize=11.5cm
  \centerline{\epsffile{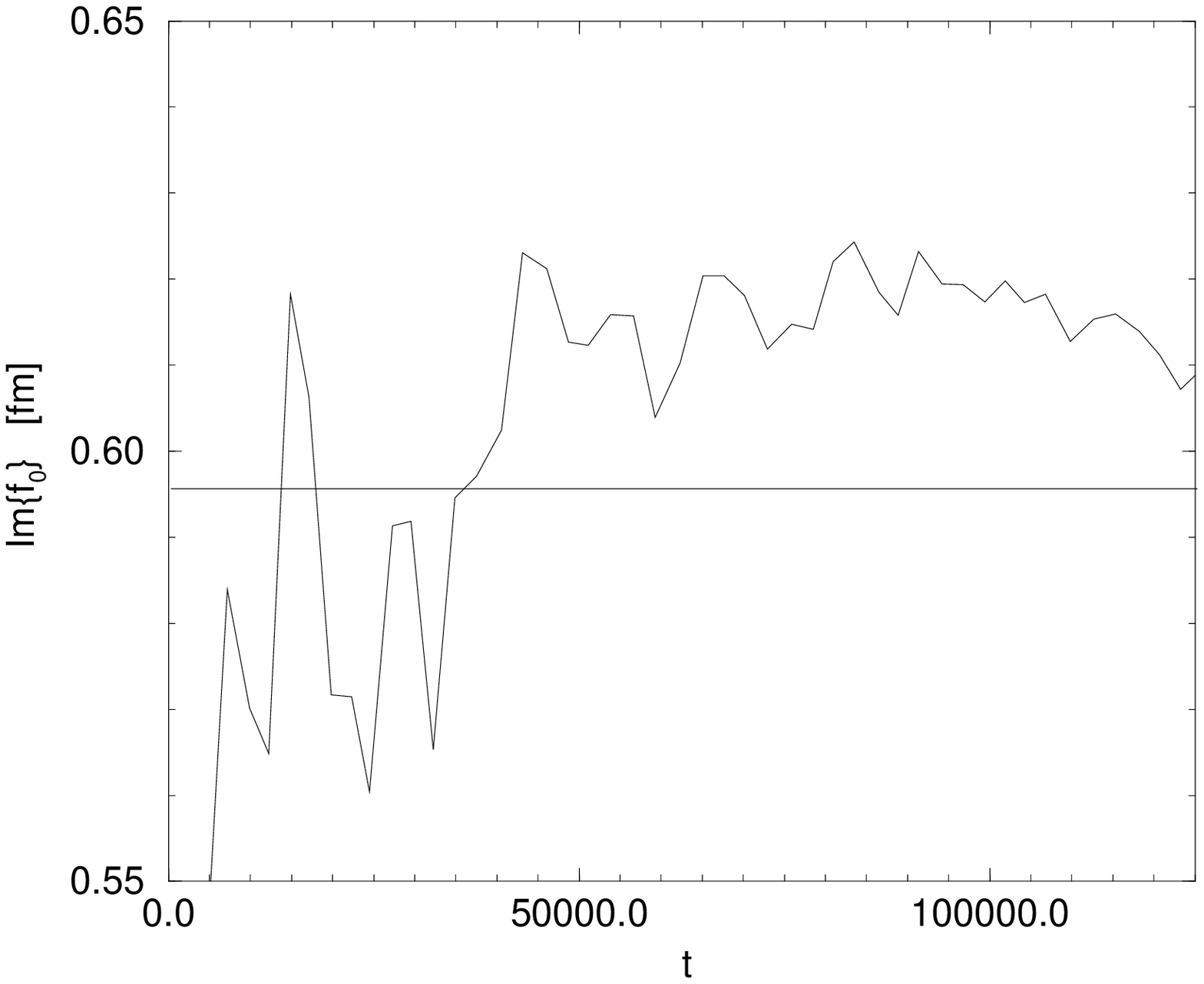}}
  \center{\parbox{11cm}{ 
  \caption{ Scattering length for complex potential 
            $V_c =(-0.5,-1.5) \;\mbox{fm}^{-1}$, 
            obtained from the 'exact' hybrid algorithm; 
	    $\beta=100\;\mbox{fm}$, $\varepsilon=0.33\;\mbox{fm}$.
	    Horizontal line: exact result.
	  \label{hyb_fig3} }}}
\end{figure}

\begin{figure}[p]
  \epsfysize=9.5cm
  \epsfxsize=11.5cm
  \centerline{\epsffile{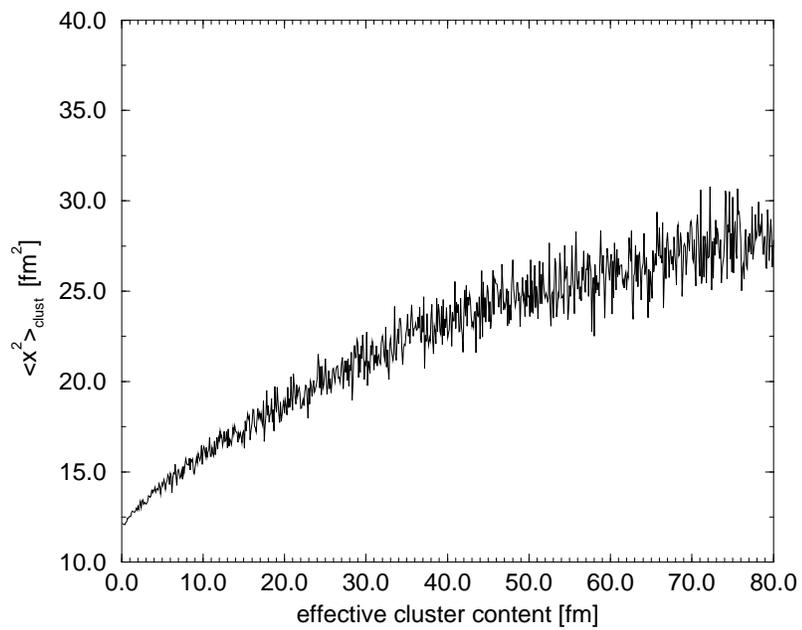}}
  \center{\parbox{11cm}{ 
  \caption{Mean square distance to the origin as function of 
  effective cluster size for $\beta=80\,\mbox{fm}\,,\;\varepsilon=0.10\,
  \mbox{fm}\,,\; V_c=(2.3,-7.0)\;\mbox{fm}^{-1}\,$.
\label{eval}}}}
\end{figure}

\begin{figure}[p]
  \epsfysize=9.5cm
  \epsfxsize=11.5cm
  \centerline{\epsffile{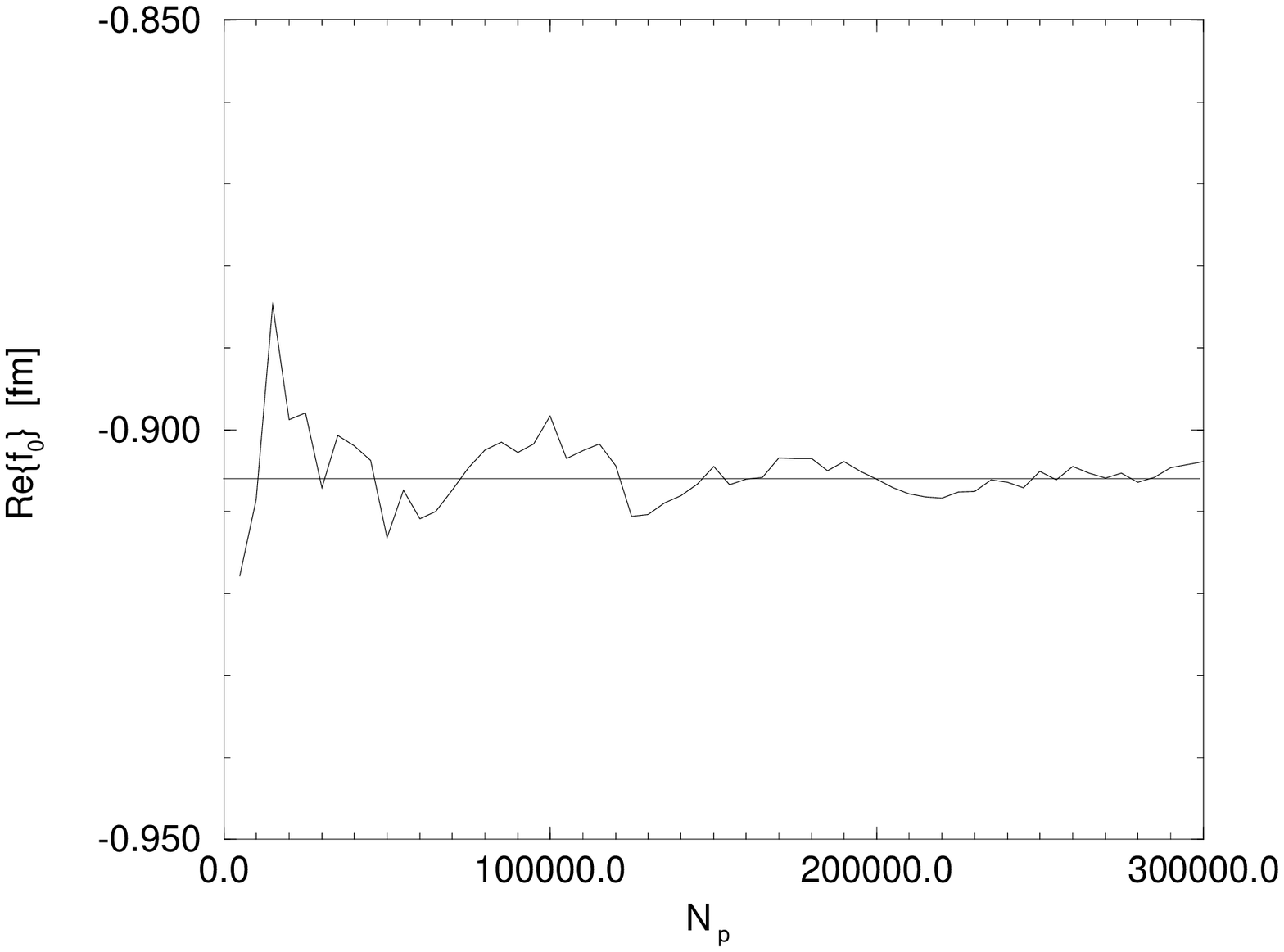}}
  \epsfysize=9.5cm
  \epsfxsize=11.5cm
  \centerline{\epsffile{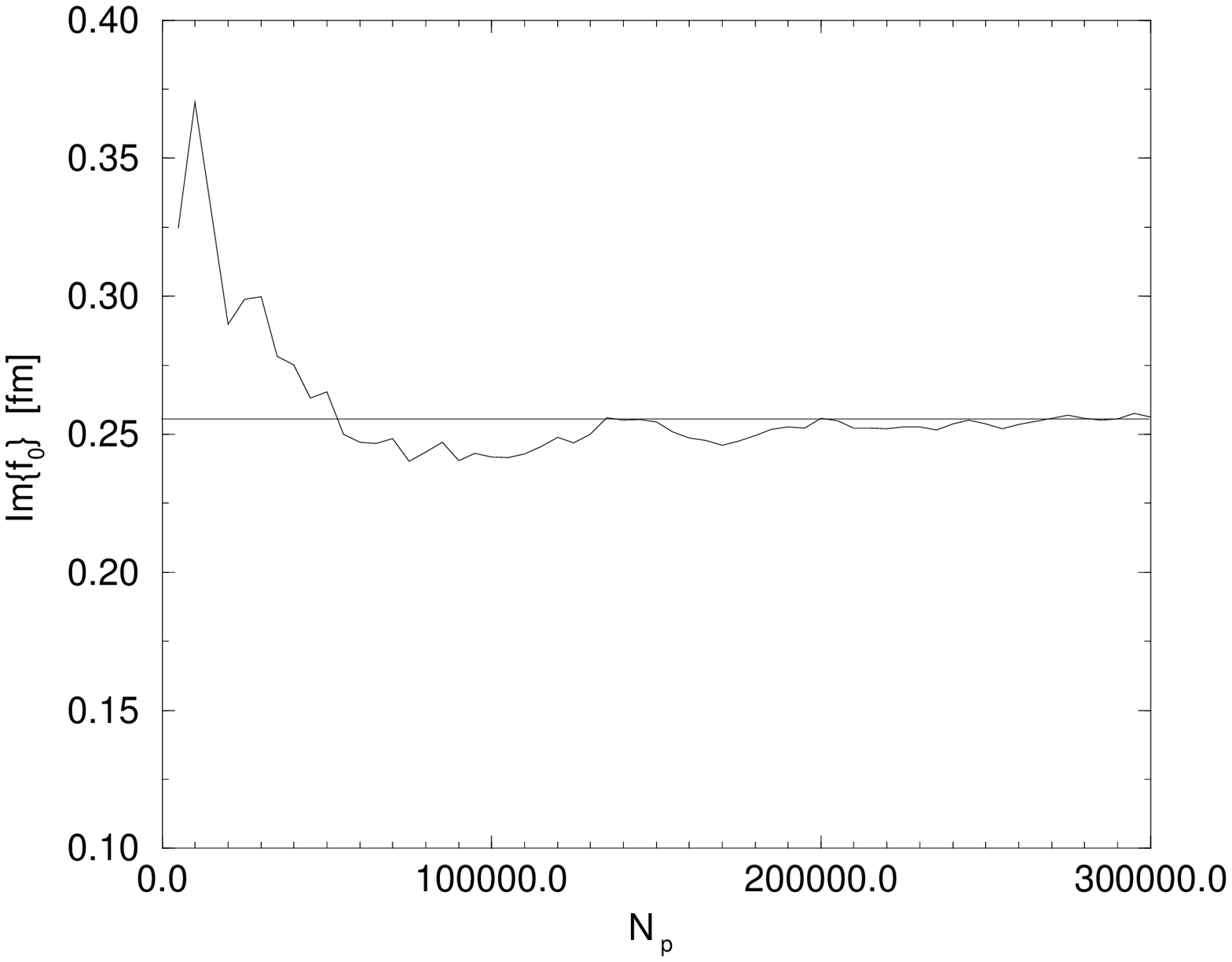}}
  \center{\parbox{11cm}{ 
  \caption{Scattering length $f_0$ as function of the number of paths, $N_p$,
	   generated by the cluster algorithm for 
  	   $\;\;V_c=(2.3,-7.0)\;\mbox{fm}^{-1}$, $\beta=80\;\mbox{fm},\;
	   \varepsilon=0.10\;\mbox{fm}$. Solid line: exact result.
\label{wo_fig1}\label{wo_fig2}}}}
\end{figure}

\begin{figure}[p]
  \epsfysize=9.5cm
  \epsfxsize=11.5cm
  \centerline{\epsffile{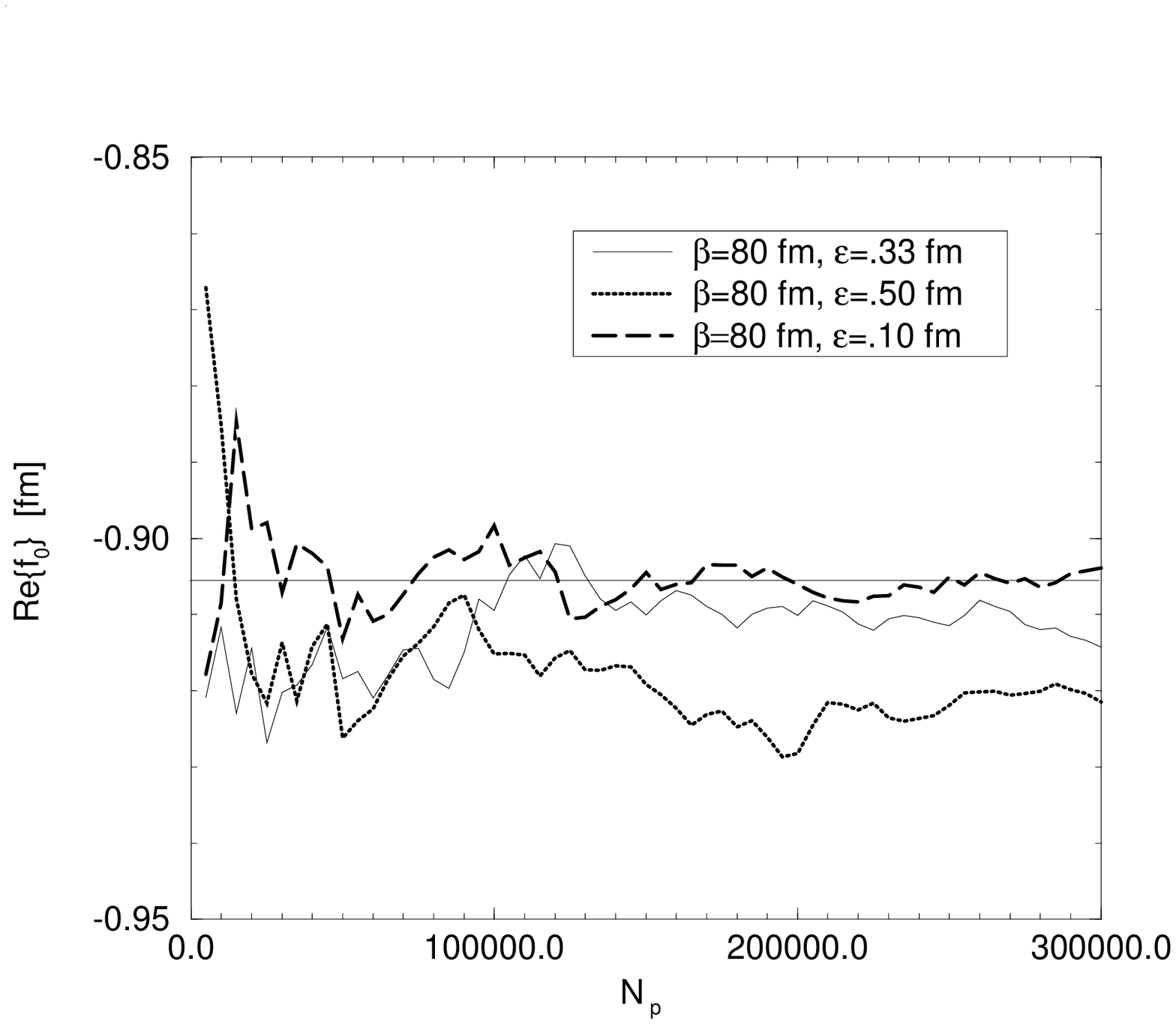}}
  \epsfysize=9.5cm
  \epsfxsize=11.5cm
  \centerline{\epsffile{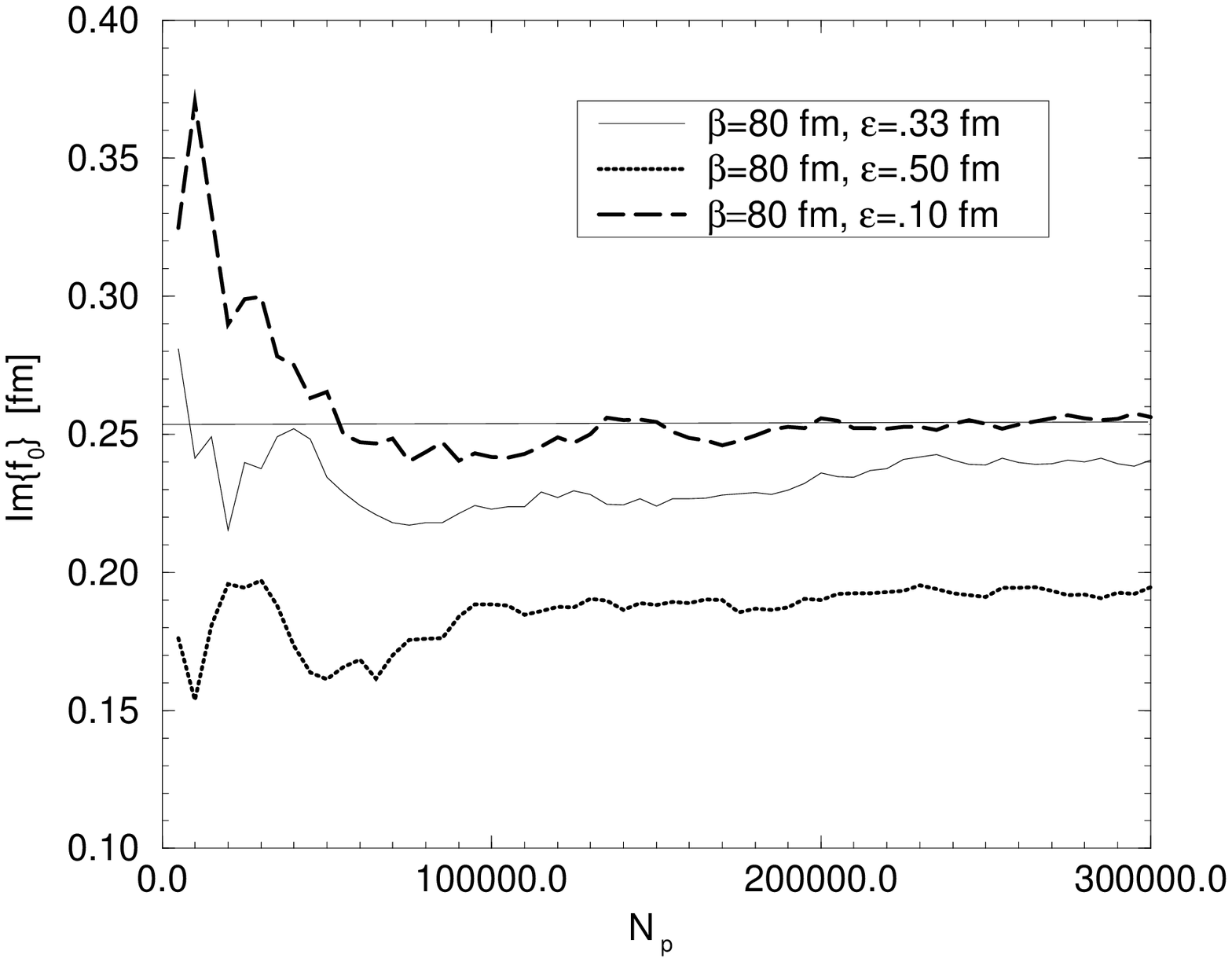}}
  \center{\parbox{11cm}{ 
  \caption{Scattering length $f_0$ as function of the number of paths, $N_p$,
	   generated by the cluster algorithm for $\;\;V_c=(2.3,-7.0)\;
	   \mbox{fm}^{-1}$, $\beta=80\;\mbox{fm}$ and different 
	   discretizations, $\varepsilon$. Solid line: exact result.
\label{wo_fig3}\label{wo_fig4}}}}
\end{figure}

We first show in Fig. \ref{wo_fig2} results for a potential 
with a repulsive real part.
The parameters are the same as in Fig. \ref{hyb_fig1}, where it was shown that
with the hybrid method a discretization of the projection times $\beta$ down to 
$\varepsilon = 0.25\;\mbox{fm}$ could not achieve convergence to the correct 
answer in both real and imaginary part. 
The results were obtained with the cluster algorithm for a
refined discretization of $\varepsilon = 0.1\;\mbox{fm}$. There is now a rapid 
convergence to the correct result for the scattering amplitude. For the hybrid 
algorithm such a fine discretization would have been extremely time consuming
and out of the question for an extension to many-body targets. In 
contrast, the calculations in Fig. \ref{wo_fig2} took less than 60 minutes with 
the cluster method. (All calculations discussed in this section were preceeded 
by 50000 equilibration steps before starting with the evaluation of the 
expectation value of the modified observable; the quoted computation times 
always include this initial equilibration.)
That the refined discretization was essential for the accurate result in Fig.
\ref{wo_fig2} can be seen in Fig. \ref{wo_fig4}, where different choices for
$\varepsilon$ are compared. Comparison of Figs. \ref{hyb_fig1} and \ref{wo_fig4}
shows that the influence of the discretization of the projection time is equal
for both methods; a calculation with the same projection time $\beta$ and 
discretization  $\varepsilon$ yields results of comparable quality for both the
hybrid method used in Fig. \ref{hyb_fig1} and the cluster method.

\begin{figure}[p]
  \epsfysize=9.5cm
  \epsfxsize=11.5cm
  \centerline{\epsffile{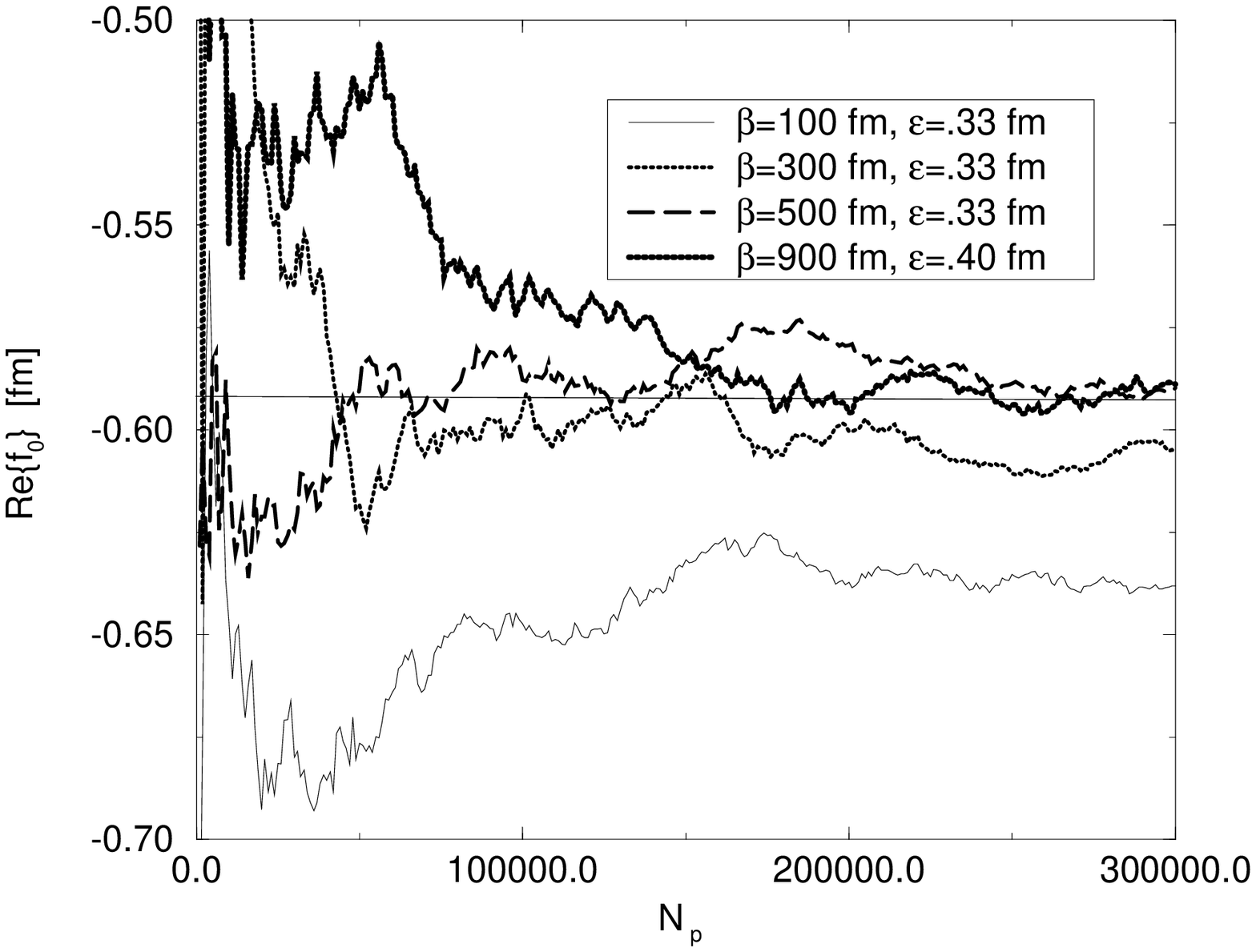}}
  \epsfysize=9.5cm
  \epsfxsize=11.5cm
  \centerline{\epsffile{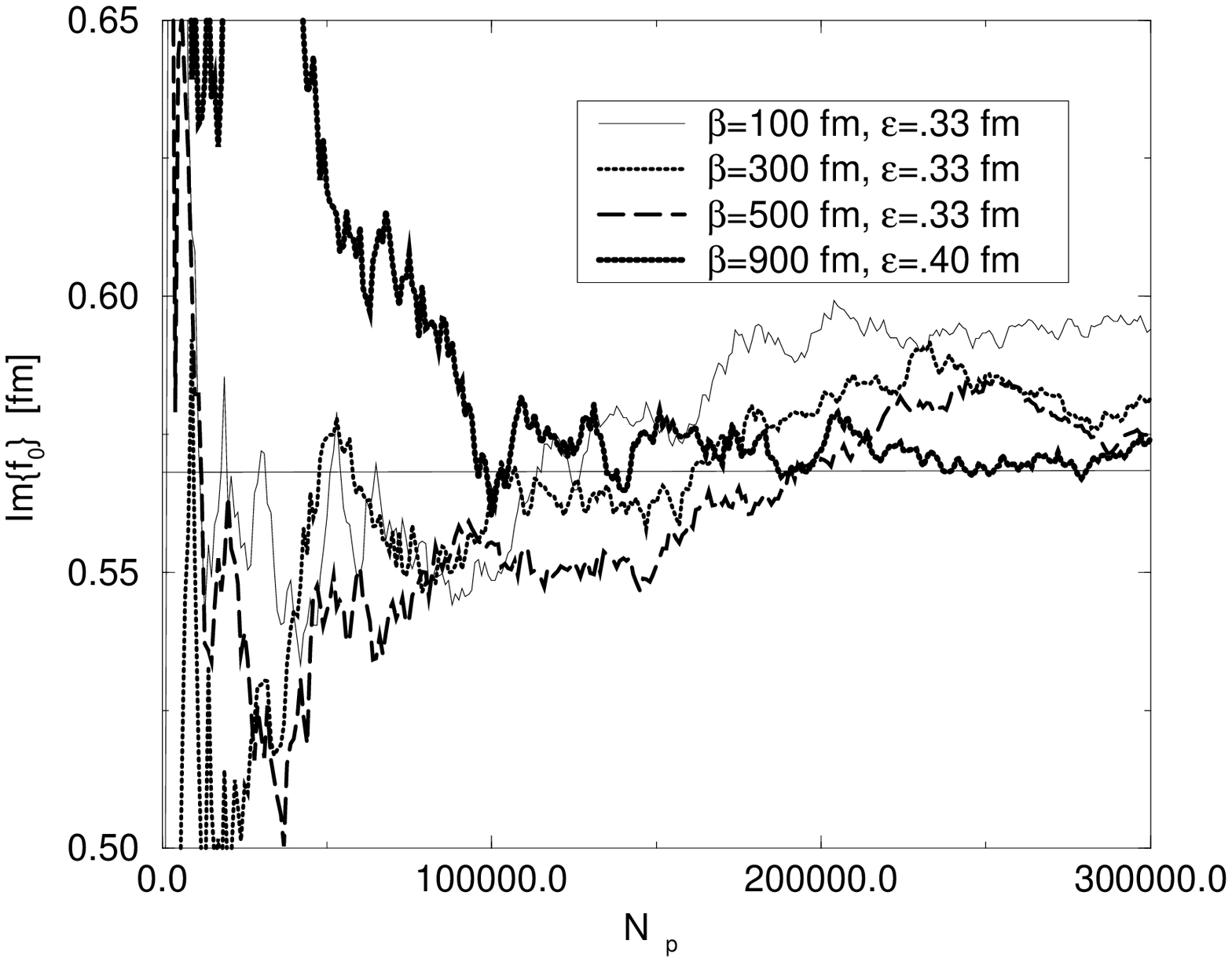}}
  \center{\parbox{11cm}{ 
  \caption{Scattering length $f_0$ as function of the number of paths, $N_p$,
	   generated by the cluster algorithm for $\,\,V_c=(-0.5,-1.5)\;
	   \mbox{fm}^{-1}$, different discretizations, $\varepsilon$, and
	   projection times, $\beta$. Solid line: exact result.
\label{wo_fig10}\label{wo_fig11}}}}
\end{figure}

To right away also test the limits of the cluster method, we used attractive 
potentials which, as we have seen before, put higher demands on the stochastic 
algorithms. The same qualitative results as for the repulsive potential in Figs.
\ref{wo_fig2} and \ref{wo_fig4} were also found for weakly attractive 
potentials as long as the strength does not exceed 
$U_0 = - 0.3 \;\mbox{fm}^{-1}$. 
However, for potentials with a stronger attractive part the necessary 
projection time increases rapidly. This can be seen in Fig. \ref{wo_fig11}, 
which shows the dependence of the accuracy on $\beta$. With the 
efficient cluster method one can afford to go to projection times as 
high as $900 fm$ to achieve good agreement 
with the exact result; a coarser discretization than for the repulsive 
potential could be used. The computation time with the cluster 
method was about 120 minutes for the $\beta=900\,\mbox{fm}$ case. 
Calculations with such a long projection time 
are impossible with the methods of Section \ref{LAHYB} within the limits 
we set for computation time. The results shown in Fig. \ref{hyb_fig3} for 
the same attractive potential and with a projection time 
$\beta = 200\;\mbox{fm}$ required 180 minutes and show 
no signs of convergence to the exact result. 

\begin{figure}[p]
  \epsfysize=9.5cm
  \epsfxsize=11.5cm
  \centerline{\epsffile{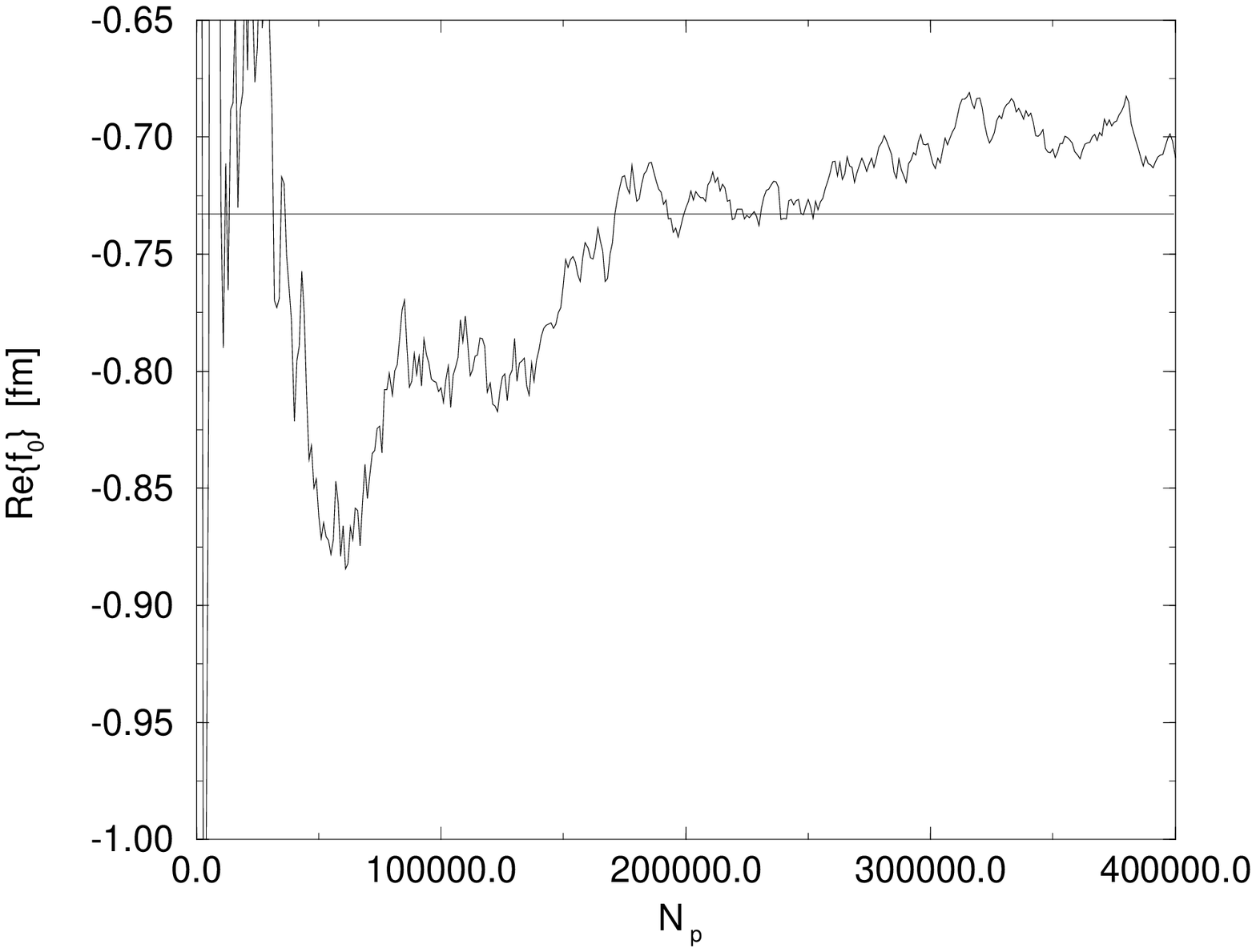}}
  \epsfysize=9.5cm
  \epsfxsize=11.5cm
  \centerline{\epsffile{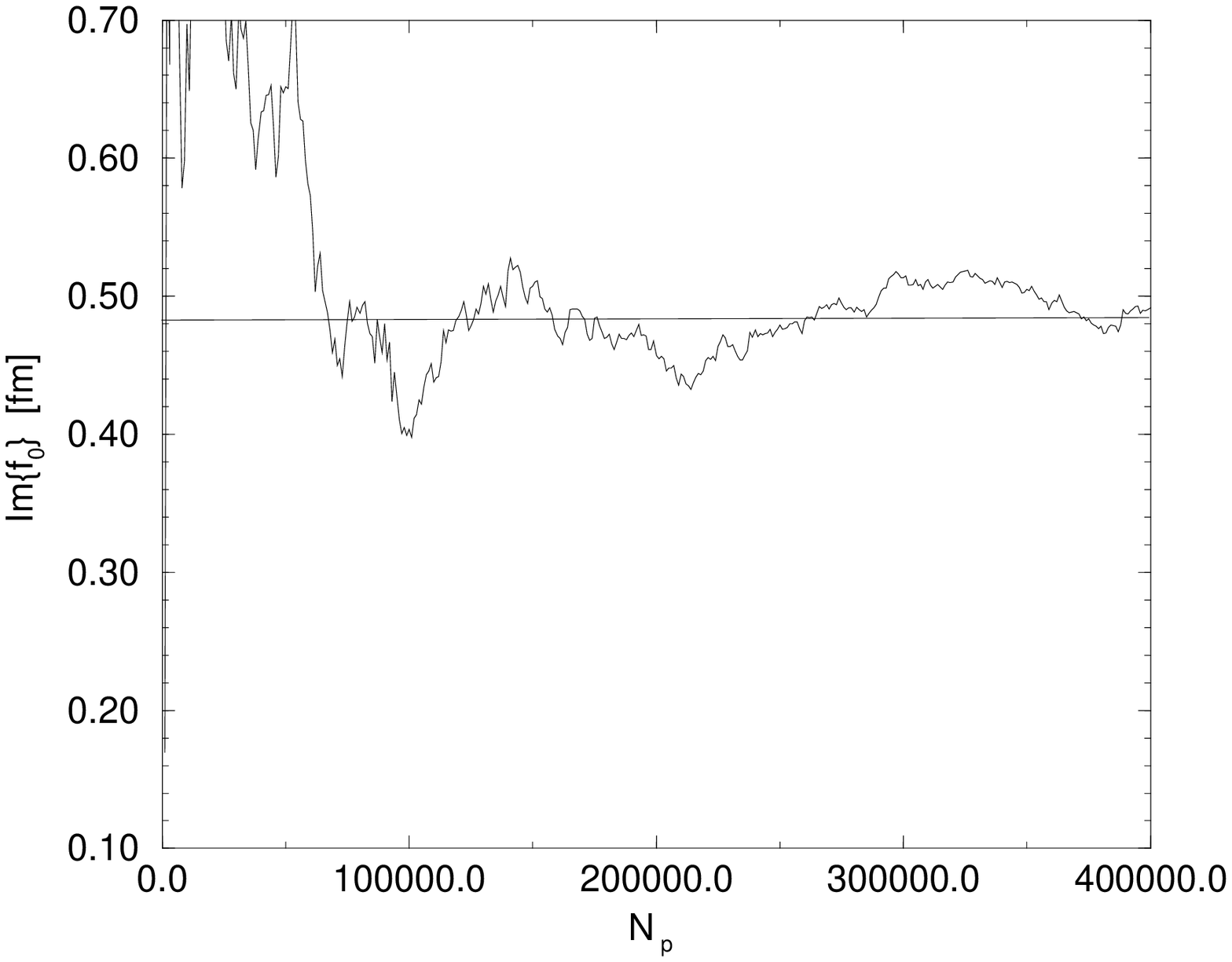}}
  \center{\parbox{11cm}{ 
  \caption{Scattering length $f_0$ as function of the number of paths, $N_p$,
	   generated by the cluster algorithm for $\;V_c=(-0.75,-2.25)\;
	   \mbox{fm}^{-1},\;\beta=900\;\mbox{fm},\;\varepsilon=0.50\;
	   \mbox{fm}$. Solid line: exact result.
\label{wo_fig22}\label{wo_fig23}}}}
\end{figure}

Closer inspection of Fig. \ref{wo_fig11} shows that even for the best result
some fluctuation persist. This feature is even more evident in 
Fig. \ref{wo_fig23}, where the attractive part of $-0.75\;\mbox{fm}^{-1}$ 
required a very long projection time, resulting in a computation time of 
180 minutes. As discussed above, for high projection times the middle of 
the path moves out of the potential region. Reflections occur increasingly 
on  planes at large distances from the origin, causing the endpoints of 
the path to move out of the potential region. This leads a low acceptance, 
slowing down the speed at which uncorrelated paths are generated and 
fluctuations remain in the result.

\begin{figure}[p]
  \epsfysize=9.5cm
  \epsfxsize=11.5cm
  \centerline{\epsffile{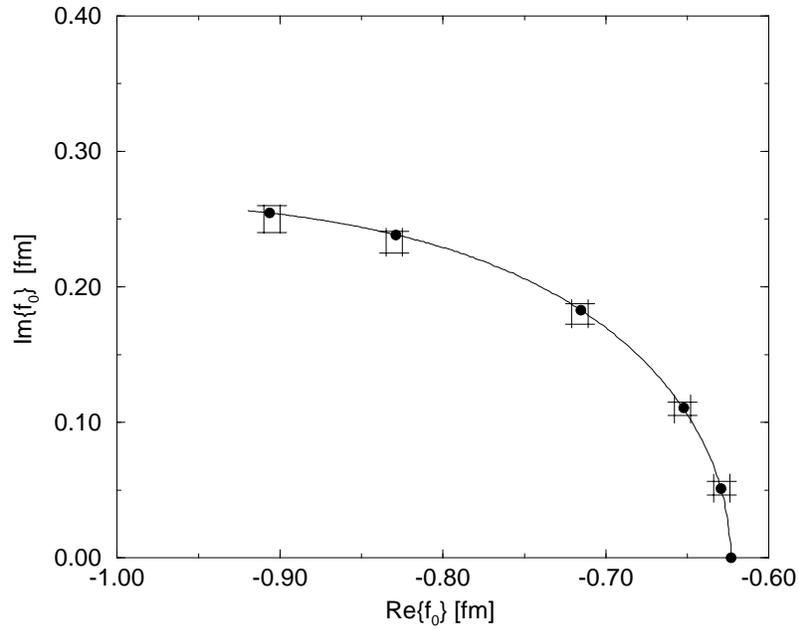}}
  \epsfysize=9.5cm
  \epsfxsize=11.5cm
  \centerline{\epsffile{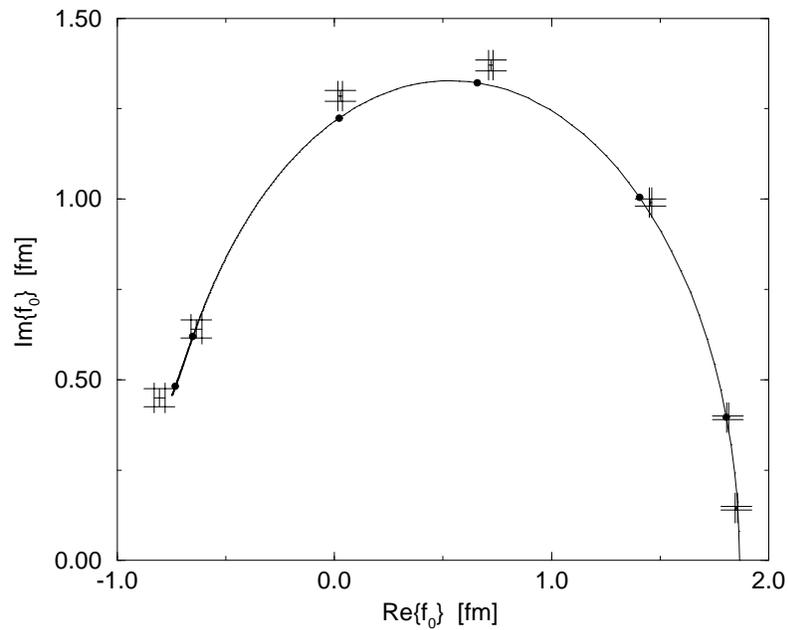}}
  \center{\parbox{11cm}{ 
  \caption{Scattering length for complex potentials with different absorptive
	   parts given in Table 1; absorptive part grows in absolute
	   value from right to left. Upper figure: attractive real part; 
	   lower figure: repulsive real part.
\label{wkoord_fig1}
\label{wkoord_fig2}
}}}
\end{figure}

To give a general impression of the performance of the cluster algorithm,
we show in Fig. \ref{wkoord_fig1} a set of results for attractive and repulsive 
potentials with varying imaginary parts. The corresponding parameters are 
listed in Table \ref{tabwolff2}. 
After the equilibration, 300 000 new paths were generated in all 
cases to evaluate
the observable. The error-bars in Fig. \ref{wkoord_fig1} are obtained from the 
results of the last 50 000 steps. 
The solid curve shows the exact result obtained by numerically solving the 
Schr\"odinger equation with a Runge-Kutta method. The computation time for the 
repulsive potentials was 45 minutes, while the attractive potentials 
require a longer projection time $\beta$ and took ca. 60 minutes each. 
The overall performance is quite good. The only significant discrepancies 
occur for the attractive
potential, which is close to having a bound state and requires longer 
projection times. The examples with a weak imaginary potential
could actually have been carried out in about half the computing time, since 
the scattering wave function of the full problem is quite similar to that
with the real part of the potential only (which is used to generate the
paths) and the convergence is fast. 

\begin{figure}[p]
  \epsfysize=9.5cm
  \epsfxsize=11.5cm
  \centerline{\epsffile{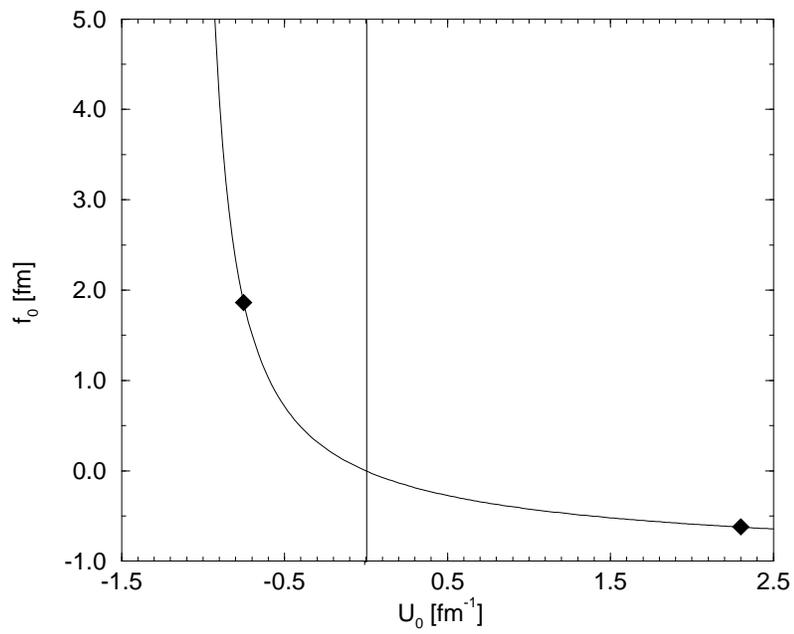}}
  \center{\parbox{11cm}{ 
  \caption{Scattering length $f_0$ as function of strength parameter 
  $U_0$. Range of real simulation potentials used for path generation
  indicated by diamonds.
\label{strreal_fig1}}}}
\end{figure}

The strength of the real simulation potentials we have used here varied from 
$U_0 = -0.75\;\mbox{fm}^{-1}$ to $2.3\;\mbox{fm}^{-1}$. To put this range into
perspective, we show in Fig. \ref{strreal_fig1} the corresponding exact scattering 
length as a function of the strength parameter $U_0$. The scattering length changes
substantially and grows rapidly as the real part of the scattering potential
approaches $U_0 = - 1.2\;\mbox{fm}^{-1}$ where a bound state appears and 
$f_0$ diverges.

In summary, the performance of the cluster algorithm is superior to that
of any of the other algorithms discussed before. While the hybrid algorithms
can be improved by finer discretization, the necessary computation time
grows roughly quadratically with the number of points \cite{SLen94} and becomes 
unreasonably large. In contrast, the increase in computation time is
only linear for the cluster algorithm, which allows one to obtain results
quite efficiently even for the difficult cases of attractive potentials which the
hybrid algorithm cannot manage anymore. In cases where both hybrid and cluster
methods could be applied, the achieved accuracy was comparable. 
To further improve the succesful cluster method and understand its performance in
more detail, one would need to better understand the decorrelation of paths in
this method. Further studies should also be directed towards optimizing the 
orientation of the reflection plane to yield a uniform acceptance of reflections 
of both the middle segments and end points of a path.

\section{Summary}

The path integral approach offers an opportunity to solve exactly the 
quantum mechanical scattering of a hadron from a composite, many-body target.
The standard methods, which have been used for decades, involve 
several approximations concerning both the reaction mechanism between 
projectile and target as well as the internal target dynamics.
Recent studies have shown that due these approximations one misses
important physics, such as the trapping of the projectile by several
scattering centers.

The possibility to apply the path integral formalism to low energy scattering
has been pointed out several years ago. However, little has been done 
since then to actually use it in practice and to investigate which computational
method is best suited for its application. The path integral approach 
is of course a common tool for the study of localized, bound systems. But 
when considering scattering one is faced with new aspects. 
An obvious difference is that in scattering the wavefunction is not
normalizable and the spectrum of the Hamiltonian continous. Another important
feature that has to be dealt with  by the numerical methods is the presence 
of two very different length scales. One is provided by the size of the target 
or the relatively short range of the interaction between projectile and the 
target constituents. The other scale, which is much larger, is introduced 
by the exterior region into which the paths can drift off with increasing 
physical time.

In order to assess the efficiency of the different methods 
for applying the path integral approach to scattering, we have tested them
for a situation where the exact answer can still be obtained by standard
methods: the scattering of a particle from a potential. 
As the idea is to find a method that can be extended to the much more 
extensive and time consuming calculations for a many-body target,
the desired computational method has to be fast and numerically stable. 
Keeping in mind the applications to systems with absorptive channels,
{\it e.g.} for calculating the antiproton - nucleus scattering length,
we have used a complex potential. Due to the presence of a complex action,
the usual probability interpretation of the distribution of paths is not 
possible. We therefore used the fact that one can choose a different real
action to generate the paths and to accordingly rearrange the matrix-element
of the observable one evaluates. In our approaches, the imaginary part of 
the potential is always contained in such a 'modified observable'. In the
action used for generating paths we used here for simplicity the real part 
of the scattering potential; in other applications
it is of course possible to make another choice suggested by the 
physics of the system.

We compared three different methods to generate paths for the
evaluation of the observable. The first is the random walk method,
where paths are built up point for point. In contrast, the Langevin algorithm,
with its extensions to the Langevin algorithm with a kernel and the hybrid 
algorithm, deals with the updating of entire paths, rather than stepping 
through each path in the physical time. These two methods, which had previously 
been applied to bound systems, showed systematic difficulties when applied
to scattering. We could trace the difficulties of these methods,
which have been used extensively for bound states, to the different 
physical situation in scattering and tested some modifications
and hybrid methods. The insights gained from these studies finally this
led us to propose the 'cluster algorithm' for scattering. It deals not with
individual points of a path but instead with clusters of varying sizes.

For the random walk method we considered both path generation by the
simple free diffusion method as well as the method of path replication. 
Both yielded rather inaccurate results, mainly due to the fact that only
few of the generated paths actually contributed significantly to the
integral. Some improvement was obtained by the 'time-dependent guided
random walk', but large fluctuations in the observable persisted, in
particular for attractive potentials, which require a large projection time
and put higher demands on numerical stability.

While the random walk sequentially generates the elements of a path, 
the Langevin algorithm updates an entire path. As discretization errors
become important, the generated paths were treated as proposals for a
Metropolis procedure. Within the computational limits we considered, this
procedure was seen to be quite time-consuming and yielded only 
unsatisfactory results. This was traced in part to the small number of 
uncorrelated paths produced in this fashion. To speed up the algorithm,
a kernel was introduced into the stochastic differential equation, which 
allows one to decouple the long-ranged and short-ranged modes
and substantially lowers the 'decorrelation time' between independent
paths. This yielded very stable results, but only over a limited range of 
potential parameters. As the matrix algebra makes this method more
time consuming, its application to more complicated scattering problems
does not look promising. A similar performance was found when combining
the Langevin method with deterministic methods to the 'hybrid algorithm'.
This was shown to lead to a substantial gain in the
the decorrelation time. Nevertheless, the need for refined discretization
also makes this method rather time consuming, especially when 
attractive potentials are considered.

Instead of dealing with individual points when updating a path, the
'cluster algorithm', which we finally proposed, treats groups of points 
as the new degrees of freedom. When generating a new path from a given 
one, these
clusters are reflected collectively on a randomly oriented plane. Due to the
procedure for forming and reflecting these clusters, this was seen to 
lead to an updating in terms of large scale clusters in the exterior region,
while small scale clusters are used in the potential region, allowing
an accurate probing of the details of the potential. This algorithm was 
found to be very stable and efficient, able to handle all but the 
very strongly attractive potentials. Another aspect that makes this method 
look very promising for application to nuclear targets is the reflection 
procedure. It allows a path to move quickly over larger distances,
{\it e.g.} between regions that are separated by repulsive potential barriers.
If point-by-point updating were used, this would be extremely time consuming.

For none of the three types of algorithms have we tried
to optimize specific computational aspects, such as the matrix manipulations
for the Langevin algorithm with a kernel. Individual improvements are
clearly possible and will cut down the computation time in each case.
However, we believe that the qualitative differences among the
different approaches which we have found are quite general and that
the cluster approach is the one that should be pursued first when
considering scattering from nuclear targets.

\section*{Acknowledgements}

\noindent
We would like to thank Prof. F. Lenz for initiating this work and for his 
continued interest. Discussions with Prof. U. Wiese were the stimulus for the 
developement of the cluster method. J.H.K. thanks the Institute f\"{u}r
Theoretische Physik III for its hospitality during several extended stays.
The work of H.R.M. and S.L. was supported in part by the DFG Graduiertenkolleg 
'Starke Wechselwirkung' Erlangen-Regensburg and by the Bundesministerium f\"ur
Bildung und Forschung (BMBF).
The work of J.H.K. is part of the research program of the Foundation for
Fundamental Research of Matter (FOM) and the National Organisation for
Scientific Research (NWO).

\pagestyle{empty}
\newpage
\begin{table}\centering
\section*{Tables and Figures}
\begin{tabular}[t]{{|r@{.}l||}*{7}{r@{.}l|}}
\hline
  \multicolumn{2}{|c||}{\mbox{$U_0$}}
  &\multicolumn{14}{c|}{$W_0$}\\ 
\hline
    2&{30}& -0&{50}&  -1&{15}& -2&{30}& -4&{60}& -7&{00}& -&-& -&-\\

    -0&{75}& -0&{02}&  -0&{05}& -0&{15}& -0&{30}& -0&{50}& -1&{50}& -2&{25}\\
\hline
\end{tabular}
\caption{Potential parameters (in $\mbox{fm}^{-1}$) used for Fig. 19.
\label{tabwolff2}}
\end{table}

\newpage

\newpage

\newpage

\newpage

\newpage

\newpage

\newpage

\newpage

\newpage

\newpage

\newpage

\newpage

\newpage

\newpage

\newpage

\newpage

\newpage

\newpage

\newpage

\end{document}